\documentclass[a4paper]{aa}
\usepackage{lscape}
\usepackage{graphicx}
\usepackage{latexsym,amsmath,amssymb}
\usepackage{natbib}
\bibpunct{(}{)}{;}{a}{}{,}

\def \xmm {XMM-Newton}
\def \src {GX\,13$+$1}

\def \nh {N${\rm _H}$}

\def \hcm {\hbox {\ifmmode $ atom cm$^{-2}\else atom cm$^{-2}$\fi}}
\def \arcmin {\hbox{$^\prime$}}
\def \arcsec {\hbox{$^{\prime\prime}$}}
\def \deg {$^{\circ}$}
\def \chisq {$\chi ^{2}$}
\def \rchisq {$\chi_{\nu} ^{2}$}
\def \approxgt{\mathrel{\hbox{\rlap{\lower.55ex \hbox {$\sim$}}
        \kern-.3em \raise.4ex \hbox{$>$}}}}
\def \approxlt{\mathrel{\hbox{\rlap{\lower.55ex \hbox {$\sim$}}
        \kern-.3em \raise.4ex \hbox{$<$}}}}

\newcommand{\mc}{\multicolumn}

\newcommand {\fetfive} {\ion{Fe}{xxv}}
\newcommand {\fetsix} {\ion{Fe}{xxvi}}
\newcommand {\ka} {K$\alpha$}
\newcommand {\kb} {K$\beta$}
\newcommand {\catwenty} {\ion{Ca}{xx}}

\newcommand {\fetone} {\ion{Fe}{xxi}}

\newcommand {\arseventeen} {\ion{Ar}{xvii}}
\newcommand {\sfifteen} {\ion{S}{xv}}
\newcommand {\ssixteen} {\ion{S}{xvi}}
\newcommand {\sithirteen} {\ion{Si}{xiii}}
\newcommand {\sifourteen} {\ion{Si}{xiv}}

\newcommand {\mgtwelve} {\ion{Mg}{xii}}
\newcommand {\nitseven} {\ion{Ni}{xxvii}}
\newcommand {\niteight} {\ion{Ni}{xxviii}}

\newcommand {\mntfive} {\ion{Mn}{xxv}}
\newcommand {\crtfour} {\ion{Cr}{xxiv}}

\def\ang {$\rm\AA$}

\def\ergcmsec{\hbox{erg cm$^{-2}$ s$^{-1}$}}
\def\countsec{\hbox{count s$^{-1}$}}

\def \mxb {MXB\,1658$-$298}
\def \bigdip {X\,1624$-$490}
\def \twelve {XB\,1254$-$690}
\def \grs {GRS\,1915+105}

\def \thirteen {4U\,1323$-$62}

\def \nhwarmabs {$N{\rm _H^{warmabs}}$}

\def \xiunit {\hbox{erg cm s$^{-1}$}}
\def \logxi {$\log(\xi)$}

\def \countsec{\hbox{counts s$^{-1}$}}

\newcommand {\egau} {$E_{\rm gau}$}

\newcommand {\ktdbb} {$kT_{\rm dbb}$}
\newcommand {\ktbb} {$kT_{\rm bb}$}

\newcommand {\ew} {$EW$}
\newcommand {\ews} {$EW$s}

\def \nh {$N{\rm _H}$}
\def \nhabs {$N{\rm _H^{abs}}$}

\def \xiunit {\hbox{erg cm s$^{-1}$}}
\def \logxi {$\log(\xi)$}
\def \xil {$\xi$}

\newcommand {\sigmav} {$\sigma_{\rm v}$}
\newcommand {\kms} {km~s$^{-1}$}

\def \kbb {$k_{\rm bb}$}
\def \kdbb {$k_{\rm dbb}$}
\def \kgau {$k_{\rm gau}$}

\begin{document}

\title{\xmm\ observations of \src: correlation between photoionised absorption and broad line emission}

\author{M. D{\'i}az Trigo\inst{1} \and L. Sidoli\inst{2} \and L. Boirin\inst{3}  \and A.N. Parmar\inst{4}}
\institute{
ESO, Karl-Schwarzschild-Strasse 2, D-85748 Garching bei M\"unchen, Germany
              \and
       INAF-IASF, via Bassini 15, I-20133 Milano, Italy
\and
      Observatoire Astronomique de Strasbourg, 11 rue de l'Universit\'e,
      F-67000 Strasbourg, France
       \and
	ESAC, P.O. Box 78, E-28691 Villanueva de la Ca\~nada, Madrid, Spain    
}

\date{Received ; Accepted:}

\authorrunning{D{\'i}az Trigo et al.}

\titlerunning{\src\ }

\abstract{We analysed data from five \xmm\ observations of \src\
to investigate the variability of the photo-ionised absorber present
in this source. We fitted EPIC and RGS spectra obtained from the
``least-variable" intervals with a model consisting of disc-blackbody
and blackbody components together with a Gaussian emission feature at
$\sim$6.55--6.7~keV modified by absorption due to cold and photo-ionised
material. We found a significant correlation between the hard, $\sim$~6--10~keV, flux, 
the ionisation and column density of the absorber and the equivalent width of the broad iron line. We interpret 
the correlation in a scenario in which 
a disc wind is thermally driven at large, $\sim$\,10$^{10}$~cm, radii and the broad line results from reprocessed emission in the wind and/or hot atmosphere. The breadth of the emission line is naturally explained by a combination of scattering, recombination and fluorescence processes. 
We attribute the variations in the absorption and emission along the orbital period to the view of different parts of the wind, possibly located at slightly different inclination angles. 
We constrain the inclination of \src\ to be between 60 and 80\deg\ from the presence of strong absorption in the line of sight, that obscures up to 80\% of the total emission in one observation, and the absence of eclipses.  We conclude that the presence of a disc wind and/or a hot atmosphere can explain the current observations of narrow absorption and broad iron emission features in neutron star low mass X-ray binaries as a class.
} \maketitle

\section{Introduction}
\label{sect:intro}

\src\ is a type I X--ray burster \citep{gx13:fleischman85aa, gx13:matsuba95pasj} with a bright persistent X--ray emission.  It is
associated with an evolved late-type K5~{\sc iii} star, located at a
distance of 7$\pm$1~kpc \citep{Bandyopadhyay1999}. \src\ is the most
luminous ``atoll'' source known to date, with an intermediate
luminosity ($\sim$0.1 Eddington luminosity) between the so-called
``Z'' Low Mass X--ray Binaries (LMXBs) and all other ``atoll''
sources.  Its nature has been debated in the literature since it shows
some features from Z sources, like 57-69 QPOs \citep{Homan1998},
persistent radio emission \citep{Grindlay1986} and high accretion rate,
and some proper from atoll sources (e.g. the path in the colour-colour
diagram).  The investigation of the aperiodic variability of bright
persistent X--ray sources led \citet{Reig2003} to propose a new
classification for LMXBs, depending on the rms amplitude and the slope
of their power density spectra.  In their scheme, \src\ is the only
``atoll'' source which has been classified as a type I source (lower
rms, rms$<$20\%, and flatter power spectra, $\alpha<0.9$), a class
which includes all the ``Z'' sources.  \citet{gx13:schnerr03aa} proposed to
explain the source peculiar spectral and power spectral properties
with the presence of a relativistic jet.  \citet{Paizis2006} first
detected a hard X--ray tail in the \src\ average hard X--ray spectrum
($>$20 keV), summing together all the available INTEGRAL observations.
They found a correlation for ``Z'' sources and bright ``GX'' atolls
(like \src), between the X-ray hard tail (40-100 keV) and their radio
luminosity, likely indicating that hard tails and energetic electrons
causing the radio emission may have the same origin, most likely the
Compton cloud located inside the neutron star (NS) magnetosphere.

\citet{gx13:corbet03apj} found a modulation of 24.065$\pm$0.018~d in
RXTE ASM data spanning 7~years and suggested that this may be the
orbital period of the system. A similar modulation was found by
\citet{gx13:bandyopadhyay02apj} after performing infrared photometry
of the source. The 24~day period has been recently confirmed from both
the K-band and additional X-ray observations \citep{gx13:corbet10apj}.
Other NIR observations \citep{Froning2008} show that the accretion disc is the dominant
emitter in \src\ ($\sim$3/4 of NIR flux comes from the disc at the
time of those observations).

\citet{gx13:ueda01apjl} discovered narrow resonant absorption
lines near 7~keV from \src\ using ASCA. The same features were studied with \xmm\
by \citet{gx13:sidoli02aa} who found a complex of features
identified with resonant scattering from the K$\alpha$ and
K$\beta$ lines of He- and H-like iron (Fe\,{\sc xxv} and Fe\,{\sc
xxvi}), H-like calcium (Ca\,{\sc xx}) K$\alpha$ as well as the
presence of a deep ($\tau \sim 0.2$) Fe~{\sc xxv} absorption edge
at 8.83~keV. The absorption lines were superposed on a broad
emission feature whose energy and breadth were poorly determined,
partly due to the presence of the deep Fe K$\alpha$ features.
$Chandra$ HETGS observations \citep{gx13:ueda04apj} revealed that
the features are blue-shifted indicating an outflowing plasma with
a velocity of $\sim$400~km~s$^{-1}$. 

\src\ is unusual in that all the other LMXB systems that exhibit
prominent \fetfive\ and \fetsix\ features are dipping sources
\citep[see Table~5 of ][]{1916:boirin04aa}. Dipping sources are
LMXB systems that are observed from close to the plane of the
accretion. The dips 
are believed to be caused by
periodic obscuration of the central X-ray source by structure
located in the outer regions of a disc
\citep{1916:white82apjl}. The depth, duration and spectral
evolution of the dips varies from source to source and often from cycle
to cycle. The lack of an orbital modulation of the absorption
features outside of the dips indicates that the absorbing plasma
is located in a cylindrical geometry around the compact object.
The important
role that photo-ionised plasmas play in LMXBs was recognised by
\citet{1323:boirin05aa} and \citet{ionabs:diaz06aa} who were able
to model the changes in {\it both} the narrow
X-ray absorption features and the continuum during the dips from all the bright
dipping LMXB observed by XMM-Newton by an increase in the column
density and a decrease in the amount of ionisation of a
photo-ionised absorbing plasma. Since dipping sources are
normal LMXBs viewed from close to the orbital plane, this implies
that photo-ionised plasmas are common features of LMXBs. Outside
of the dips, the properties of the absorbers do not vary strongly
with orbital phase suggesting that the ionised plasma 
has a cylindrical geometry with a maximum column density close to
the plane of the accretion disc.

Since \src\ exhibits deep Fe absorption features similar to the
dipping LMXBs, this suggests that \src\ may be a high-inclination LMXB
and that as such the overall X-ray continuum may be similarly
affected.  To investigate the origin of the variability of the
absorption and the presence of dipping activity, we successfully
applied for five observations of \src\ with \xmm, spaced by at least
one day. In this paper, we report on this monitoring program. We
include results from the RGS detectors and the optical monitor, which
were not reported for the early 2000 observations.

\section{XMM-Newton observations}
\label{sec:observations}

The XMM-Newton Observatory \citep{xmm:jansen01aa} includes three
1500~cm$^2$ X-ray telescopes each with an EPIC 
(0.1--15~keV) at the focus. Two of the EPIC imaging
spectrometers use MOS CCDs \citep{xmm:turner01aa} and one uses pn CCDs
\citep{xmm:struder01aa}. The RGSs \citep[0.35--2.5~keV,][]{xmm:denherder01aa} 
are located behind two of the
telescopes. In addition, there is a co-aligned 30~cm diameter 
Optical/UV Monitor telescope \citep[OM,][]{xmm:mason01aa}, 
providing simultaneously coverage with the X-ray instruments.
Data products were reduced using the Science Analysis
Software (SAS) version 10.0. The EPIC MOS cameras were not used during the
observation in order to allocate their telemetry to the EPIC pn camera
and avoid Full Scientific Buffer in the latter. We present here the analysis 
of EPIC pn data, RGS data from both gratings and OM data.

\begin{table*}
\begin{center}
\caption[]{XMM-Newton observations of \src. Results from the first
three observations are reported in \citet{gx13:sidoli02aa}. $T$ is the
total EPIC pn exposure time and $C$ is the pn 0.6--10~keV total
count rate after dead time correction. In all cases the EPIC 
pn thin filter was used. For obs~5 only RGS and OM data are available.}
\begin{tabular}{cclcccc}
\hline \noalign {\smallskip}
Obs  & Observation & \mc{3}{c}{Observation Times (UTC)} & $T$  & $C$   \\
Num & ID   & \mc{2}{c}{Start}  & End & (ks) & (s$^{-1}$) \\
        &        & (year~month~day)& (hr:mn) & (hr:mn) \\
\hline \noalign {\smallskip}
1 & 0122340101 & 2000 March 30 & 14:10 & 16:58 & 6.0 & 674 \\
2 & 0122340901 & 2000 April 01 & 04:29 & 07:54 & 8.1 & 781 \\
3 & 0122341001 & 2000 April 01 & 08:51 & 11:16 & 4.6 & 690 \\
4 & 0505480101 & 2008 March 09 & 15:48 & 22:50 & 22.9 & 843 \\
5 & 0505480701 & 2008 March 11 & 19:00 & 23:10 & 12.0 & --  \\
6 & 0505480201 & 2008 March 11 & 23:10 & 27:16 & 12.4 & 909 \\
7 & 0505480301 & 2008 March 22 & 02:20 & 07:15 & 15.3 & 818 \\
8 & 0505480501 & 2008 March 25 & 23:01 & 26:59 & 11.9 & 761 \\
9 & 0505480401 & 2008 September 5 & 21:49 & 26:48 & 15.5 & 746 \\
\noalign {\smallskip} \hline \label{tab:obslog}
\end{tabular}
\end{center}
\end{table*}

\begin{figure*}[!ht]
\includegraphics[angle=0.0,width=0.23\textheight]{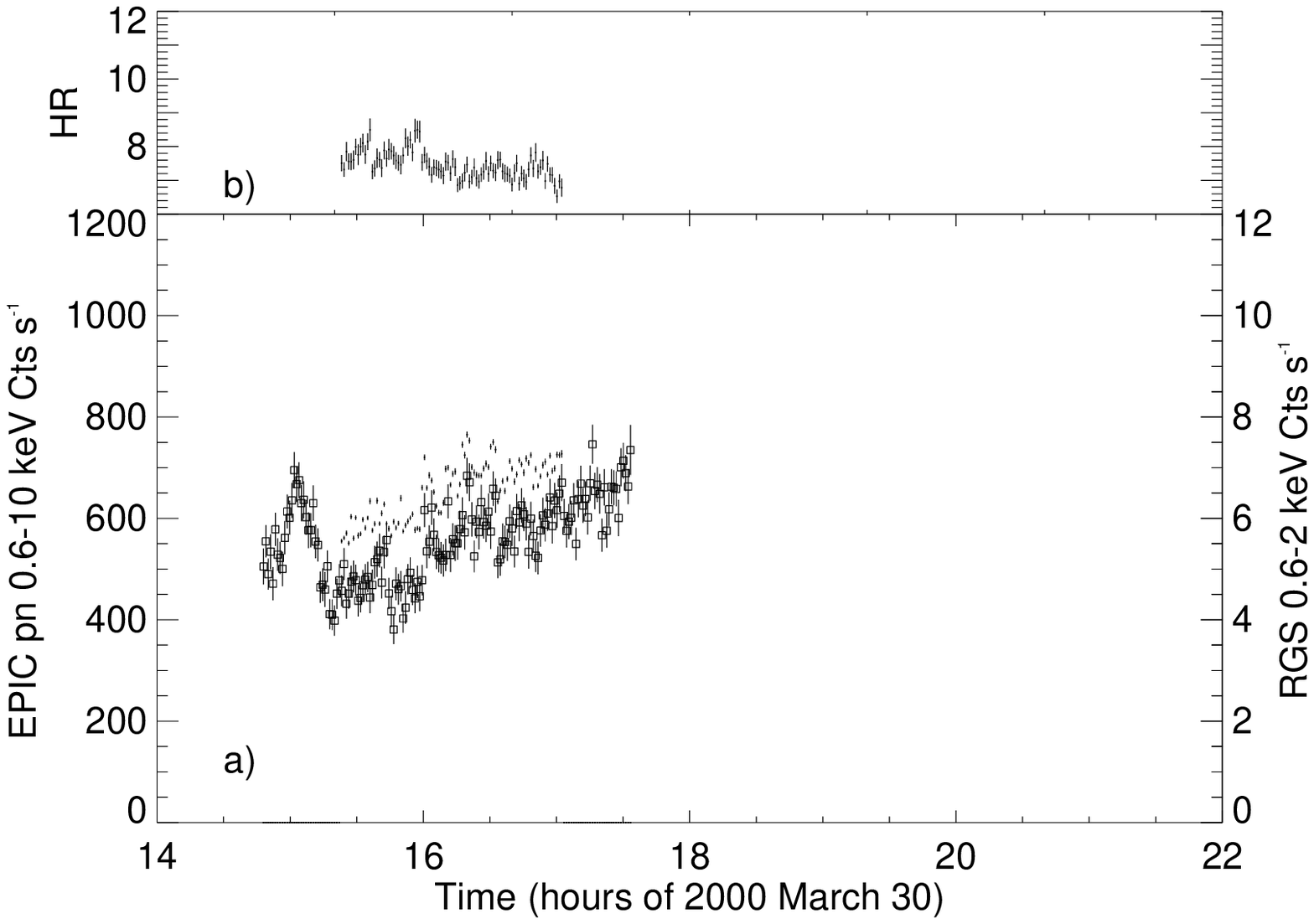}
\includegraphics[angle=0.0,width=0.23\textheight]{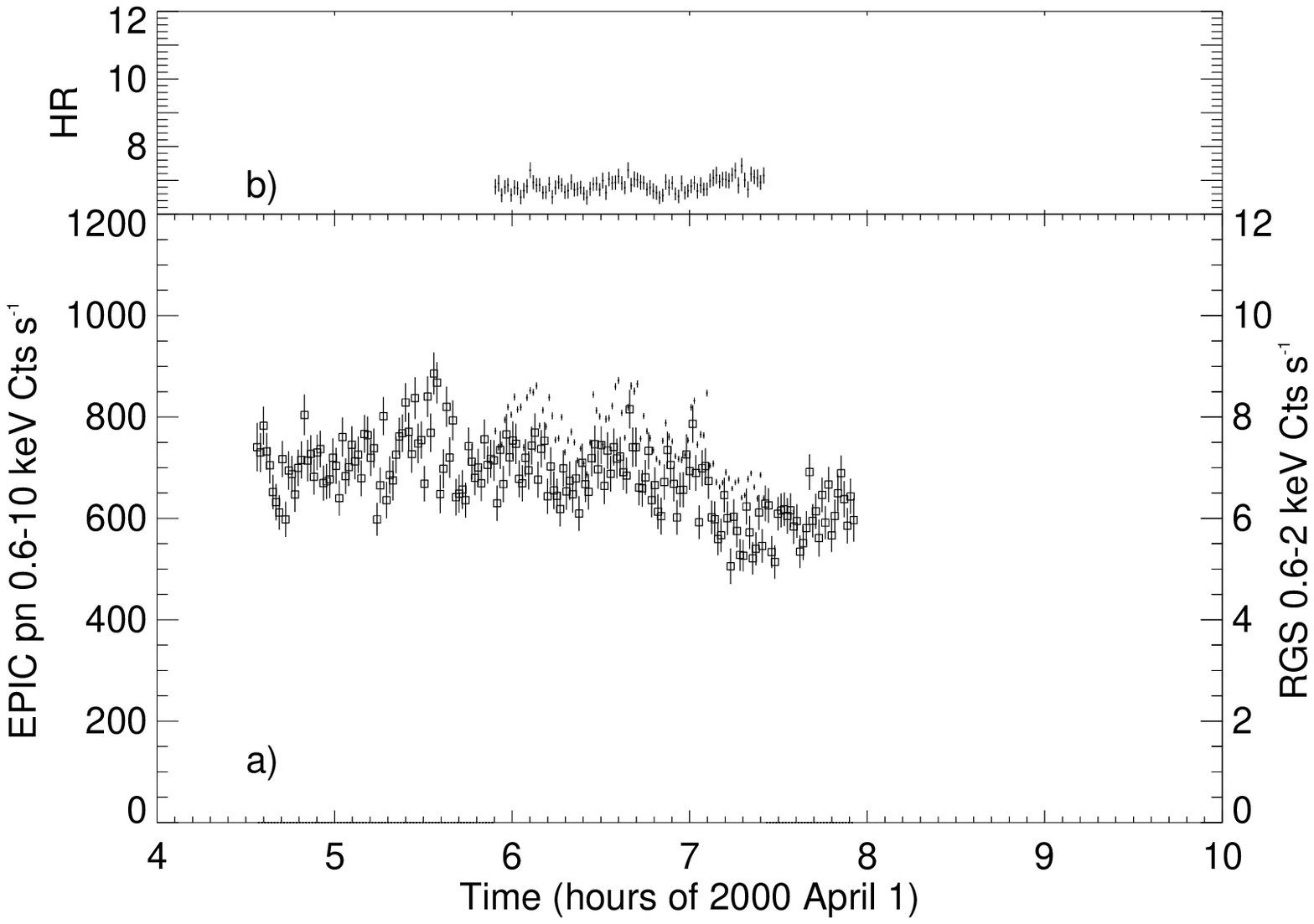}
\includegraphics[angle=0.0,width=0.23\textheight]{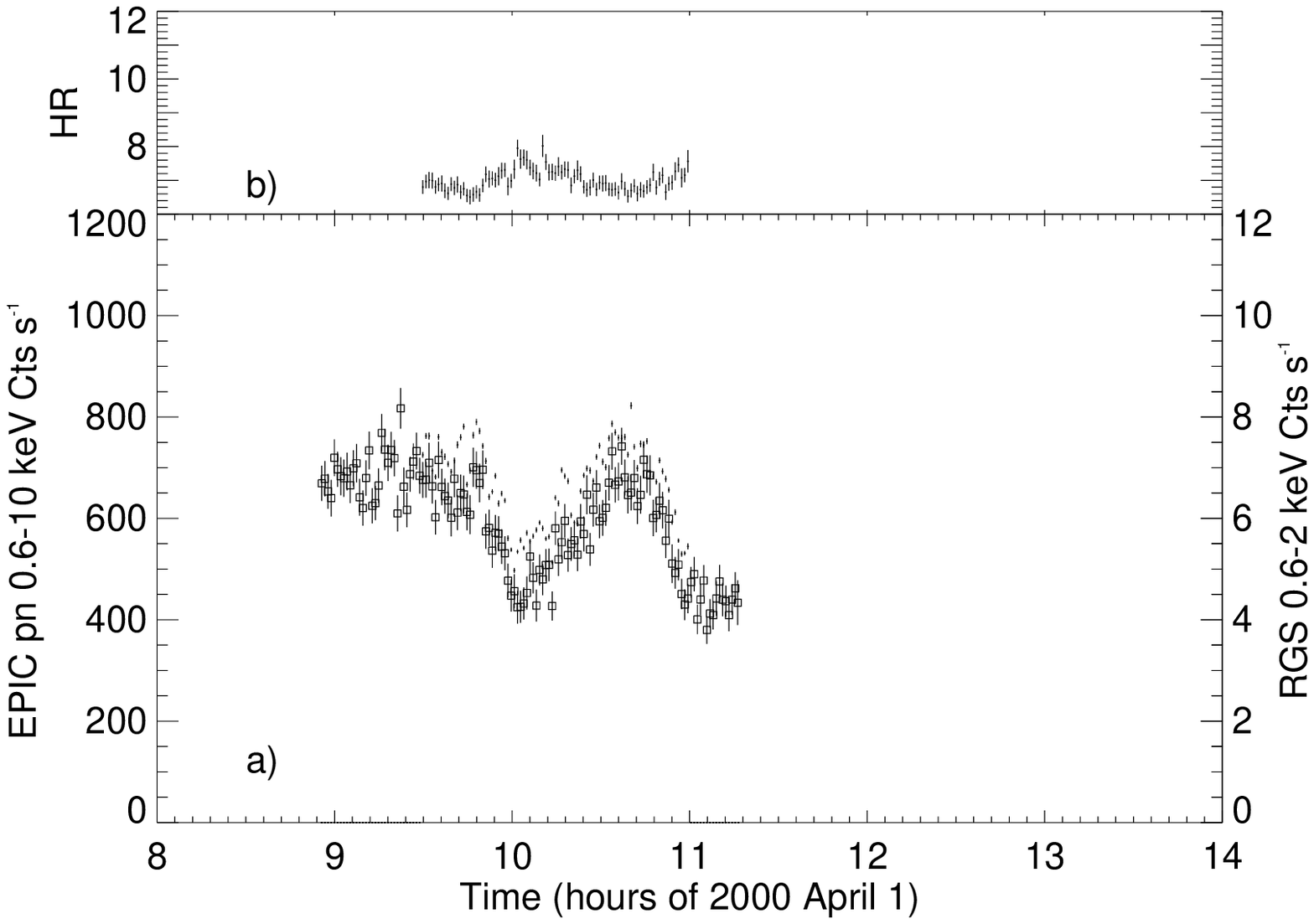}
\includegraphics[angle=0.0,width=0.23\textheight]{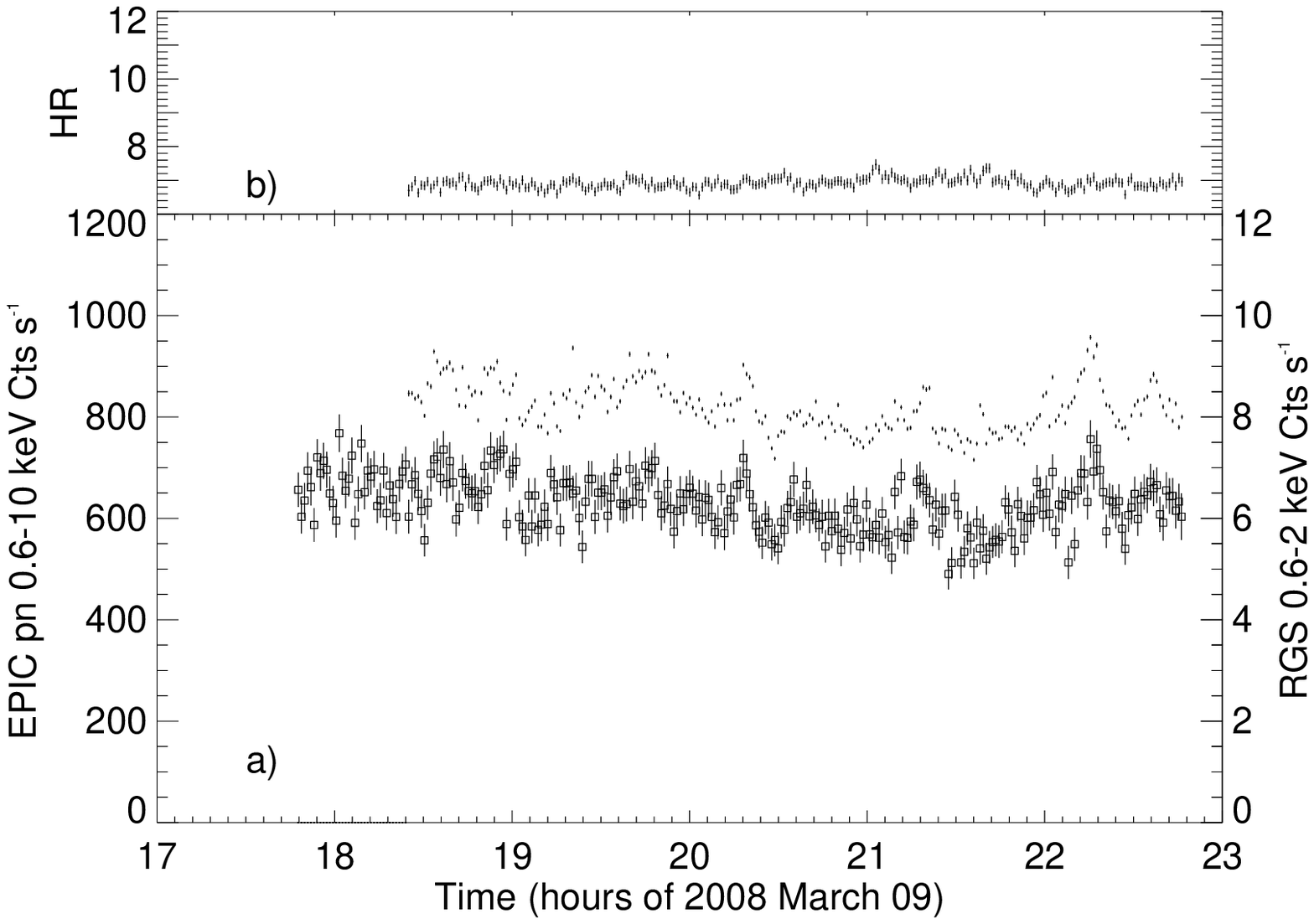}
\includegraphics[angle=0.0,width=0.23\textheight]{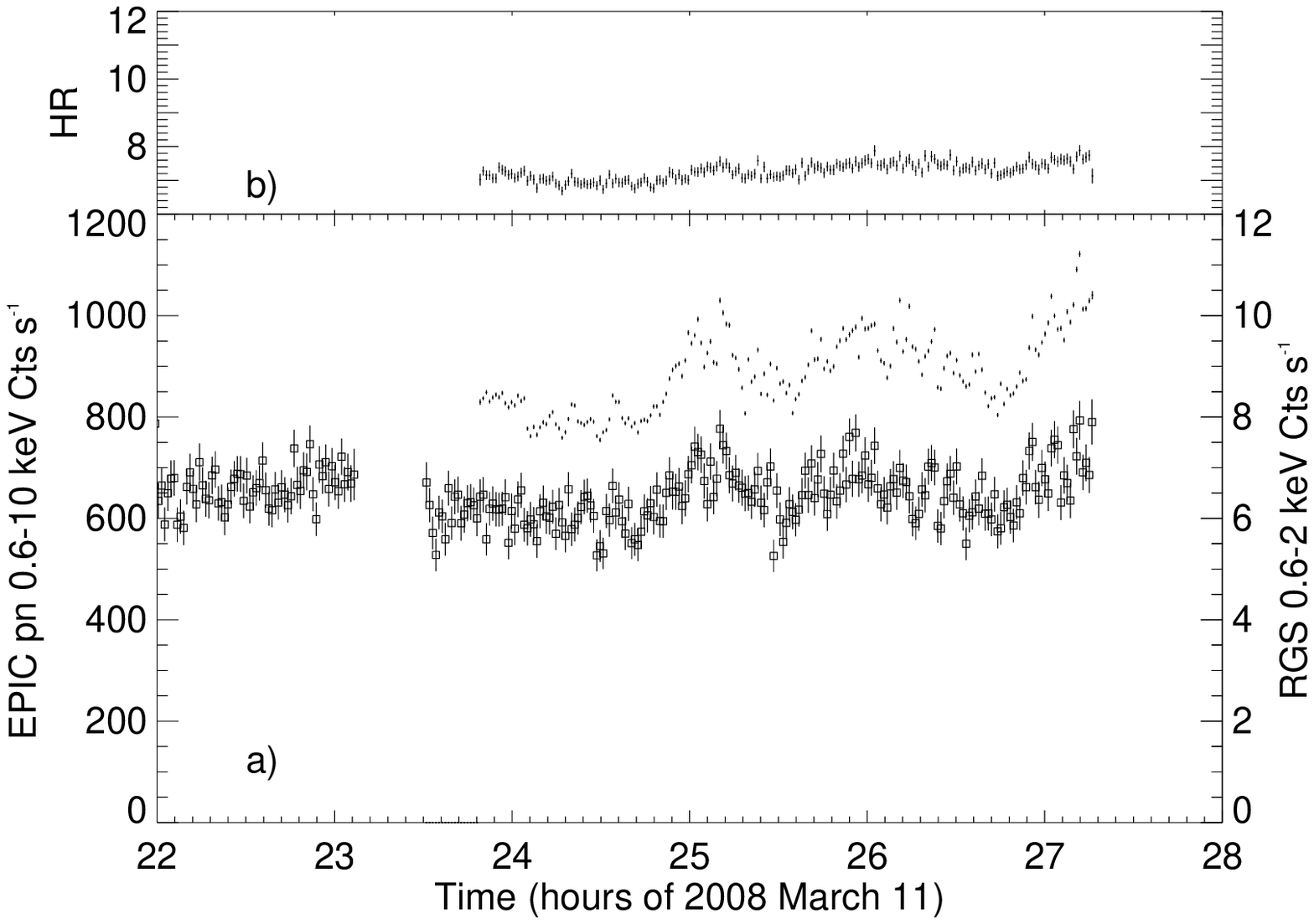}
\includegraphics[angle=0.0,width=0.23\textheight]{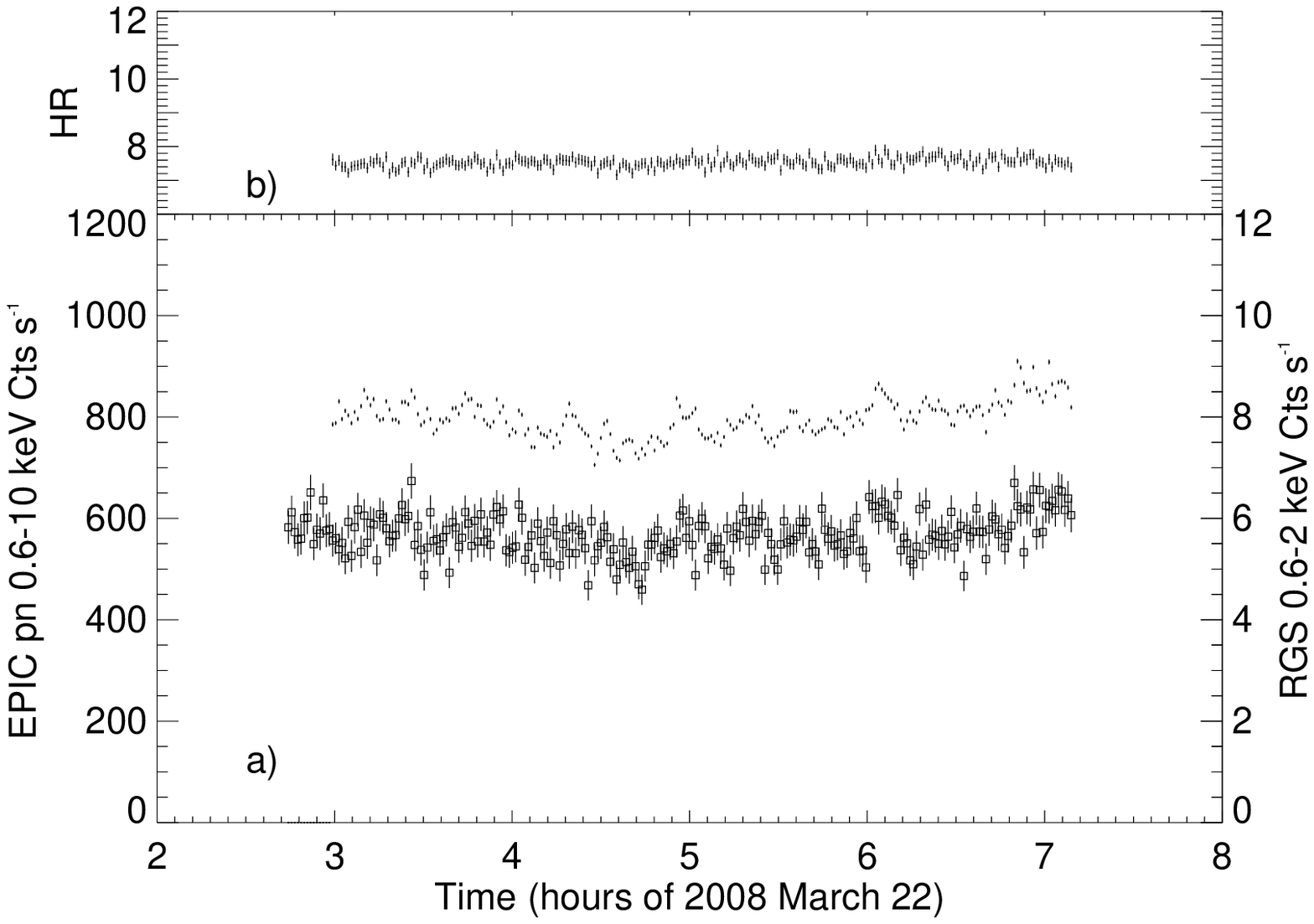}
\includegraphics[angle=0.0,width=0.23\textheight]{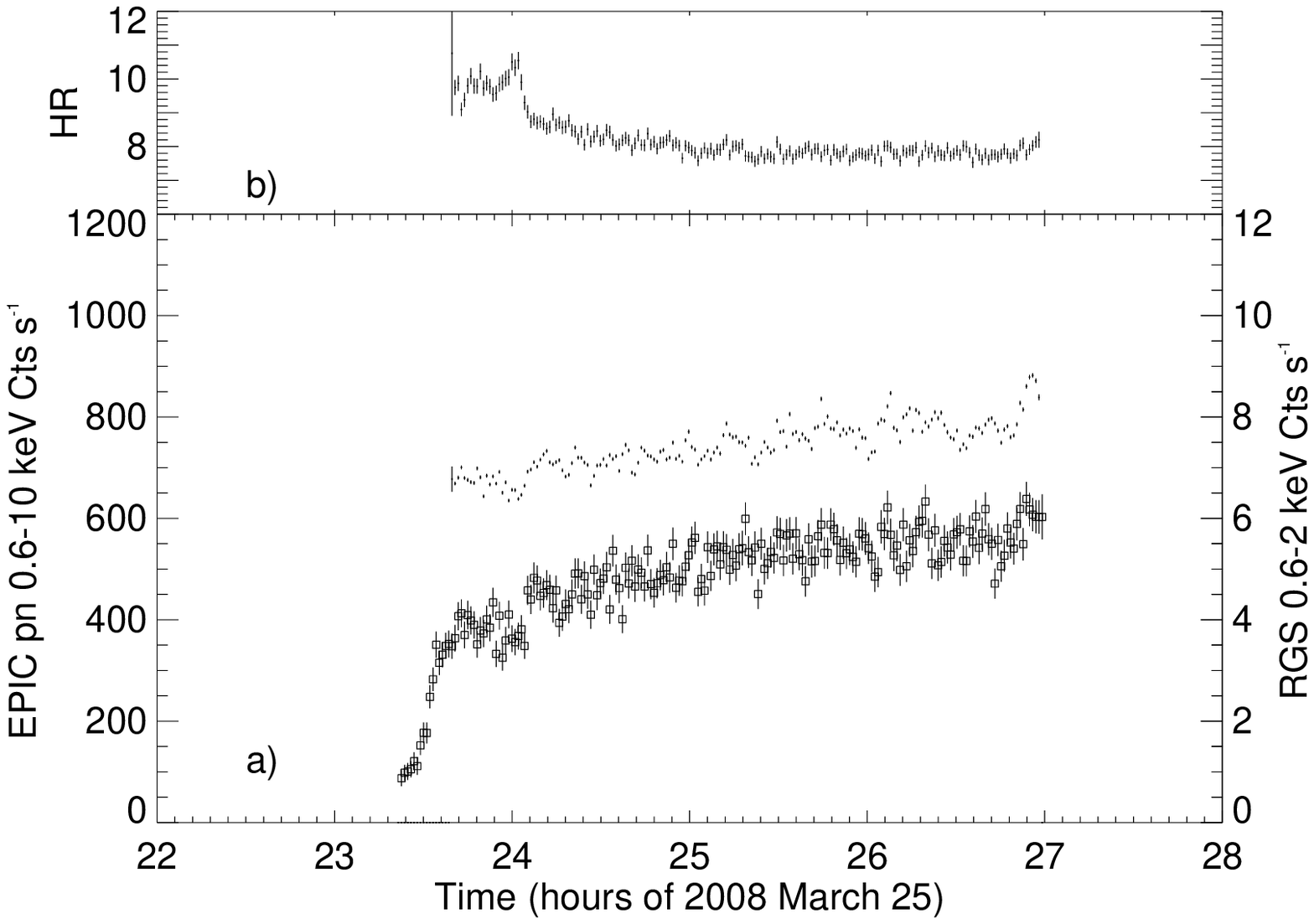}
\hspace{0.23cm}
\includegraphics[angle=0.0,width=0.23\textheight]{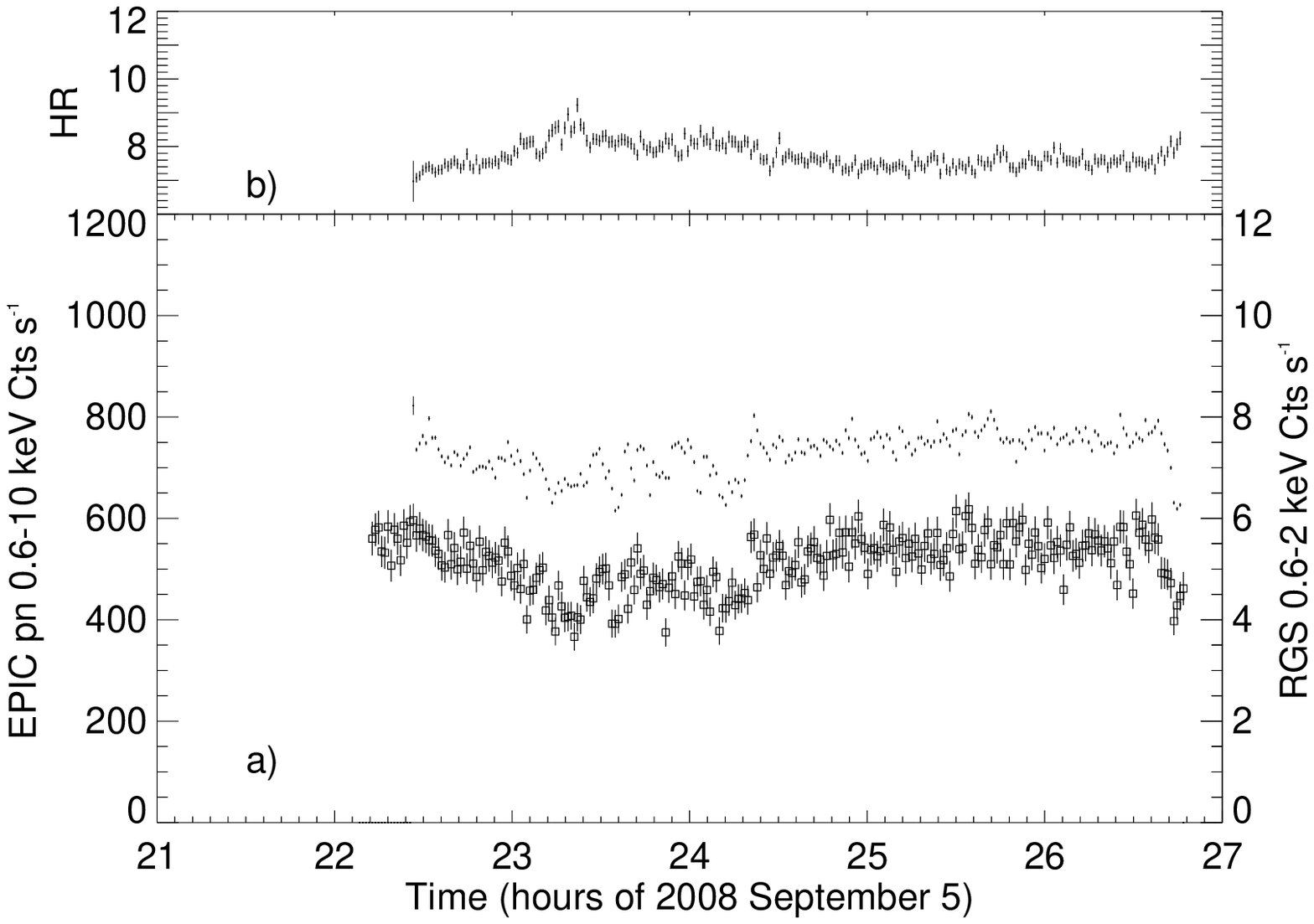}
\vspace{0.cm}
\caption{a) 0.6--10 keV EPIC pn (dots) and 0.6--1.8 keV RGS (squares)
background subtracted light curves for each \src\ observation with a
binning of 64~s. The left (right) axes show the scale for EPIC pn
(RGS). b) Hardness ratio (counts in the 2--10~keV band divided by
those between 0.6--2~keV) for the EPIC pn light curves.}
\label{fig:lightcurves}
\end{figure*}

Table~\ref{tab:obslog} is a summary of the XMM-Newton observations. We
used the EPIC pn in timing mode. In this mode only one CCD chip is
operated and the data are collapsed into a one-dimensional row
(4\farcm4) and read out at high speed, the second dimension being
replaced by timing information. This allows a time resolution of
30~$\mu$s. 
We used the SAS task {\tt epfast} on the event files to correct for a 
Charge Transfer Inefficiency (CTI) effect which has been observed in EPIC 
pn timing mode when high count rates are present\footnote[1]{More 
information about the CTI correction can be 
found in the {\it EPIC status of calibration and data analysis} and 
in the Current Calibration File (CCF) release note {\it Rate-dependent CTI correction for EPIC-pn timing modes}, by
Guainazzi et al. (2009), at http:$\slash\slash$xmm.esac.esa.int$\slash$external$\slash$xmm$\_$calibration}.
Ancillary response files were generated using the SAS task
{\tt arfgen} following the recommendations of the {\it XMM-Newton SAS
User guide} for piled-up observations in timing mode, whenever applicable. 
Response matrices were generated using the SAS task {\tt rmfgen}.

Light curves were generated with the SAS task
{\tt epiclccorr}, which corrects for a number of effects like
vignetting, bad pixels, PSF variation and quantum efficiency and
accounts for time dependent corrections within a exposure, like dead
time and GTIs.

The SAS task {\tt rgsproc} 
was used to produce calibrated RGS event lists, spectra, and response
matrices.
We also chose 
the option {\it keepcool}=no to discard single
columns that give signals a few percent below the values expected from
their immediate neighbours. Such columns are likely to be important
when studying weak absorption features in spectra with high
statistics. We used the SAS task {\tt rgsbkgmodel} to compute model background spectra from RGS background templates. We generated RGS light curves with the SAS task 
{\tt rgslccorr}. 

The OM was operated in image+fast mode. The U, UVW1 and UVM2
filters were used. In this
mode the instrument produces images of the entire 17\arcmin\ x
17\arcmin\ FOV with a time resolution between 800 and 5000~s and event
lists with a time resolution of 0.5~s from a selected 11\arcsec\ x
11\arcsec\ region. The SAS task {\tt omfchain} was used to extract
light curves of \src\ from the high time resolution fast mode data. We
used a time sampling of 100~s in the curve extraction to improve the
signal-to-noise ratio. 

\subsection{Pile-up and X-ray loading in the EPIC pn camera}

The count rate in the EPIC pn was close to, or above, the
800~\countsec\ level, at which X-ray loading and pile-up effects
become significant. 

Pile-up occurs when more than one photon is read in a pixel during a
read-out cycle. This causes photon loss, pattern
migration from lower to higher pattern types and hardening of the spectrum,
because the charge deposited by more than one photon is added up before being read 
out\footnote[2]{See {\it XMM-Newton Users Handbook}
for more information on pile-up.}. 
In addition, when high
count rates are present, the offset map calculated at the beginning of
a exposure may be contaminated by X-ray events from the source, the
so-called ``X-ray loading''. As a consequence pattern migration from
higher to lower pattern types and a shift to lower energy for all the
events associated with the contaminated pixel occur\footnote[3]{More 
information about X-ray loading can be 
found in the {\it PN X-ray loading investigation results}
calibration document by Smith (2004) at 
http:$\slash\slash$xmm.esac.esa.int}.
Since both X-ray loading and pile-up cause significant spectral distortion,
we investigated in detail their presence before extracting the spectra. 

We used the SAS task {\tt
epatplot}, which utilizes the relative ratios of single- and
double-pixel events which deviate from standard values in case of
significant pile-up, as a diagnostic tool in the pn camera timing mode
data and found that all the spectra were affected by pile-up. Next, we
extracted several spectra selecting single and double timing mode
events (patterns 0 to 4) but different spatial regions for the
source. Source events were first extracted from a 62\arcsec\ (15
columns) wide box centred on the source position (Region~1). Next we
excluded 1, 3, 5 and 7 columns from the centre of Region~1
(Regions~2--5) and extracted one spectrum for each of the defined
regions. 
We obtained spectra free of pile-up once events within the inner 5 columns were
excluded. 

Next, we inspected the pn offset maps to diagnose the presence of X-ray
loading. For each observation, we calculated the so-called ``residual'' offset 
map by subtracting the offset map of a nearby observation taken with closed 
filter from the offset map of the observation. We found that all the observations 
showed effects of X-ray loading in at least the inner 5 columns of the PSF. 
X-ray loading could be significant in the inner 7 columns, but the noise fluctuations 
in the individual rows for outer columns are of the order of the level of excess in the 6th and 7th columns.
Therefore, we extracted the final spectra after excluding events from the central 5 
columns.

\subsection{Background subtraction}
\label{subsec:bkg}
In the EPIC pn timing mode, there are no source-free background
regions, since the PSF of the telescope extends further than the
central CCD boundaries. The central CCD has a field of view of 
13\farcm6 $\times$ 4\farcm4 in the pn. In timing mode, the largest
column is the one in which the data are collapsed into one-dimensional
row. Therefore, the maximum angle for background extraction is 2\arcmin, 
compared to 5\arcmin\ for imaging modes. Since \src\ is very bright, its
spectrum will not be significantly modified by the
``real'' background which contributes less than 1\% to the total count
rate in most of the bandwidth. Conversely, subtracting the background
extracted from the outer columns of the central CCD will modify the
source spectrum, since the PSF is energy dependent and the source
photons scattered to the outer columns do not show the same energy
dependence as the photons focused on the inner columns.
Therefore, we chose not to 
subtract the ``background'' extracted from 
the outer regions of the central CCD \citep[see also][]{gx339:done10mnras,ng10aa}. We expect a contribution from the background to the total count rate of 
more than 1\% below 1.7~keV. Therefore, below this energy  we used only the RGS spectra, for which background templates were available.

\section{Light curves}

\subsection{EPIC pn and RGS light curves}
\label{sec:x-lc}

Figure~\ref{fig:lightcurves} shows 0.6--10~keV EPIC pn and 0.6--1.8~keV
RGS light curves of five \xmm\ observations of \src\ performed
in 2008 together with three observations performed in 2000, for comparison, 
with a binning of 64~s.

The EPIC pn light curves are remarkably similar to the simultaneous RGS 
light curves despite the different energy band. We note that the lower RGS count
rate with respect to the pn count rate in the 2008 observations compared to the 2000 observations is 
due to the inoperative CCDs 7 (RGS1) and 4 (RGS2) since early after the beginning of the mission.
A high variability is present in all
the observations. The average count rate changes significantly from one 
observation to the next one, with values between $\sim$700 and 900~\countsec. 
The count rate variability is not associated to significant changes in the spectral 
hardness for obs~4 and 6. However, obs~1, 3, 8 and 9 show spectral hardening 
associated with a decrease of count rate.
Obs~6 shows the highest count rate, with 
peaks of up to $\sim$1100~\countsec\ in the EPIC pn camera. Obs~8 shows
the lowest count rate, with less than 1~\countsec\ at the beginning of the 
observation in the RGS, compared to $\sim$6~\countsec\ at the end of the 
observation. This behaviour could be related to dipping activity. 
Obs~9 shows a decrease of count rate at $\sim$23-24.5~h 
which is related to spectral hardening and could be again an indication for
dipping behaviour in \src. 

\begin{figure}[!ht]
\includegraphics[angle=0,width=0.49\textwidth]{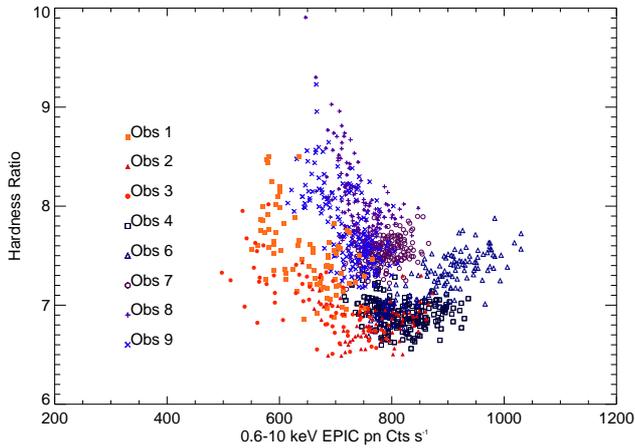}
\caption{Hardness ratio (2--10~keV/0.6--2~keV counts) versus
0.6--10~keV EPIC pn count rate for all the observations. Data from
obs~1 to 3 are shown as filled squares, filled triangles and filled circles, 
respectively. Data from obs~4 and 6 to 9 are shown as
open squares, open triangles, open circles, crosses and crooked crosses, 
respectively. Each point corresponds to a binning of 64~s.}
 \label{fig:colours2}
\end{figure}

We next examined the hardness ratio (counts in the 2--10~keV band divided
by those between 0.6--2~keV) as a function of 0.6--10~keV count rate
for all the observations (Fig.~\ref{fig:colours2}). The largest
hardness ratios correspond to the intervals of low count rate present
in obs~1, 3, 8 and 9. This behaviour is typical of deep dipping states of dippers.
A detailed comparison of Fig.~\ref{fig:colours2}
 with Fig.~4 from \citet{1323:boirin05aa} shows that  indeed obs~1, 3, 8 and 9 occupy the same parameter space in the 
 figure as the deep dipping state of \thirteen, while obs 2, 4 and 7 resemble the shallow dipping states and 
 obs~6 the persistent state.
We analyse in detail the origin of these states in Sect.~\ref{simultaneous_fits}. 

\subsection{Light curves from other X-ray missions}
\label{sec:lc-superperiodicity}

In order to investigate the variability of \src\
further, we extracted from the High Energy
Astrophysics Science Archive Research Center (HEASARC) light
curves from previous \src\ X-ray observations 
with the maximum time resolution available. We did not perform a 
systematic search of dips in all the \src\ archival observations. Instead, 
we looked for the signature of dipping episodes in observations
where a ``bi-modal" behaviour of the count rate with hardness ratio
had been reported in the literature \citep[]{gx13:stella85conf,
gx13:schnerr03aa}. 
Our goal was to determine if dipping behaviour may have been not 
recognized in previous observations of \src\ and to see if such
dipping behaviour, if existent, is associated with a certain state 
of the source.

\begin{figure}[!ht]
\includegraphics[angle=0,width=0.49\textwidth]{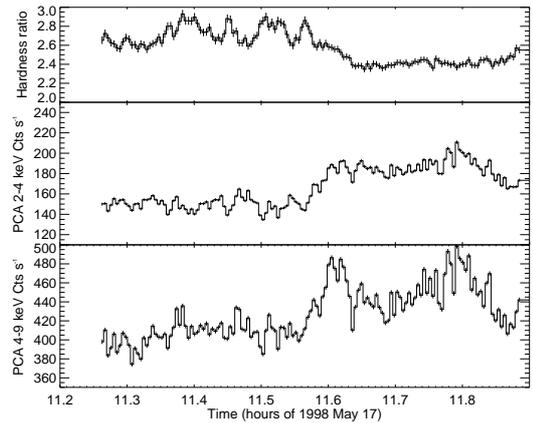}
\caption{2--4 keV (middle) and 4--9 keV (bottom) RXTE PCA
background subtracted light curves for an archival \src\ observation with a
binning of 16~s. {\it Top:} Hardness ratio (counts in the 4--9~keV band divided by
those between 2--4~keV) for the RXTE PCA light curves shown in the middle and bottom panels.}
\label{fig:archive}
\end{figure}

Details on the observations of \src\ from archival X-ray observations 
which may show dipping behaviour are given in Table~\ref{tab:dips}. 
The light curve corresponding to the RXTE observation is shown in 
Fig.~\ref{fig:archive}. The light curve from the EXOSAT archival observation
is shown in Fig.~1 from \citet{gx13:stella85conf}. In both observations
a decrease of count rate, which is more pronounced in the soft band, is visible. 
The decrease in the 1--4~keV EXOSAT (2--4~keV RXTE) light curves is $\sim$25--30\% and 
is accompanied by spectral hardening.  

\begin{table}
\begin{center}
\caption[]{X-ray archival observations of \src\ with potential dip states.}
\begin{tabular}{lcl}
\hline \noalign {\smallskip}
Date & Mission & Reference  \\
\hline \noalign {\smallskip}
1983 September 22 & EXOSAT & \citet{gx13:stella85conf} \\
1998 May 17 & RXTE & \citet{gx13:schnerr03aa} \\
\noalign {\smallskip} \hline \label{tab:dips}
\end{tabular}
\end{center}
\end{table}

Finally we examined the \src\ RXTE All-Sky Monitor (ASM) 1.5--12~keV
light curves spanning more than 13~years and with a time resolution of 
96~s to search for any characteristic signature at the time of appearance
of dips. We corrected the ASM light curves to the solar
system barycentre using the FTOOL task {\tt faxbary} and folded the
light curves into segments between 0.1 and 24~days 
duration, but clear dipping activity was not detected. 
A modulation of the ASM data has been already extensively searched and
it is reported in the literature \citep{gx13:corbet03apj, gx13:corbet10apj}. Therefore, we did 
not look further for periodic behaviour in the ASM data. 

\section{X-ray spectra}
\label{sec:spectra}

We extracted EPIC pn and RGS spectra for each observation.  We
rebinned the EPIC pn spectra to over-sample the $FWHM$ of the energy
resolution by a factor 3 and to have a minimum of 25 counts per bin,
to allow the use of the $\chi^2$ statistic. To account for systematic
effects we added a 0.8\% uncertainty to each spectral
bin after rebinning (we note that this is equivalent to add $\sim$2\% 
uncertainty to each spectral bin before rebinning). 
We used the RGS spectra with two different binnings: in 
Sect.~\ref{subsec:pn-RGS} we rebinned
the RGS spectra 
to have a minimum of 25 counts per bin, to be able to
consistently use the $\chi^2$ statistic for both EPIC pn and RGS
spectra. In Sect.~\ref{subsec:RGS} we rebinned the RGS spectra to
over-sample the $FWHM$ of the energy resolution by a factor 3 to be
sensitive to narrow features and we used the C-statistic \citep{cash79apj}. We performed
spectral analysis using XSPEC \citep{arnaud96conf} version 12.6.0q. Since there were no photons
in the RGS spectra below $\sim$\,0.8~keV, we used the RGS spectra in the energy 
interval 0.8--1.8~keV. We used the pn spectra between 1.7 and 10~keV in order
to exclude energy bins for which we expect the spectrum to be affected by background
(see Sect.~\ref{subsec:bkg}). 
To 
account for absorption by neutral gas we used the {\tt
tbabs} XSPEC model with solar abundances \citep{anders89}.
Spectral uncertainties are given at 90\%
confidence ($\Delta$\chisq = 2.71 for one interesting parameter), and
upper limits at 95\% confidence.

We extracted intervals of low variability for all the observations, using as a
criterion the hardness ratio and the count rate of the observations.
For obs~7 there are no
significant variations of hardness ratio or count rate during the
observation. Therefore we only consider one interval for this
observation. We
divided obs~4, 6, 8 and 9 in two intervals each (``high'' and ``low'') and
examined the spectra separately. A quick look shows that while for
obs~4, 8 and 9 both intervals show significant spectral changes, the two
intervals of obs~6 are only different in the normalisation
factors. Therefore we continued the analysis for obs~6 only with one
interval. We continued the analysis for obs~4, 8 and 9 only with the ``high'', less variable, intervals, for which
we could accumulate a sufficient, $>$\,5~ks, net exposure time.

\subsection{EPIC and RGS spectral analysis}
\label{subsec:pn-RGS}

\subsubsection{Phenomenological model}
\label{model1}

We first examined the spectra of the low-variability interval separately for each
observation. We combined the RGS spectra (RGS1 and RGS2, order
 1) and the EPIC pn spectrum and fitted them with a model consisting of a disc blackbody 
 and a blackbody components modified by neutral material. A constant factor, fixed to 1 for
the pn spectrum, but allowed to vary for each RGS spectrum, was
included multiplicatively in order to account for cross-calibration
uncertainties. 
\begin{figure*}[!ht]
\includegraphics[angle=0,width=0.33\textwidth]{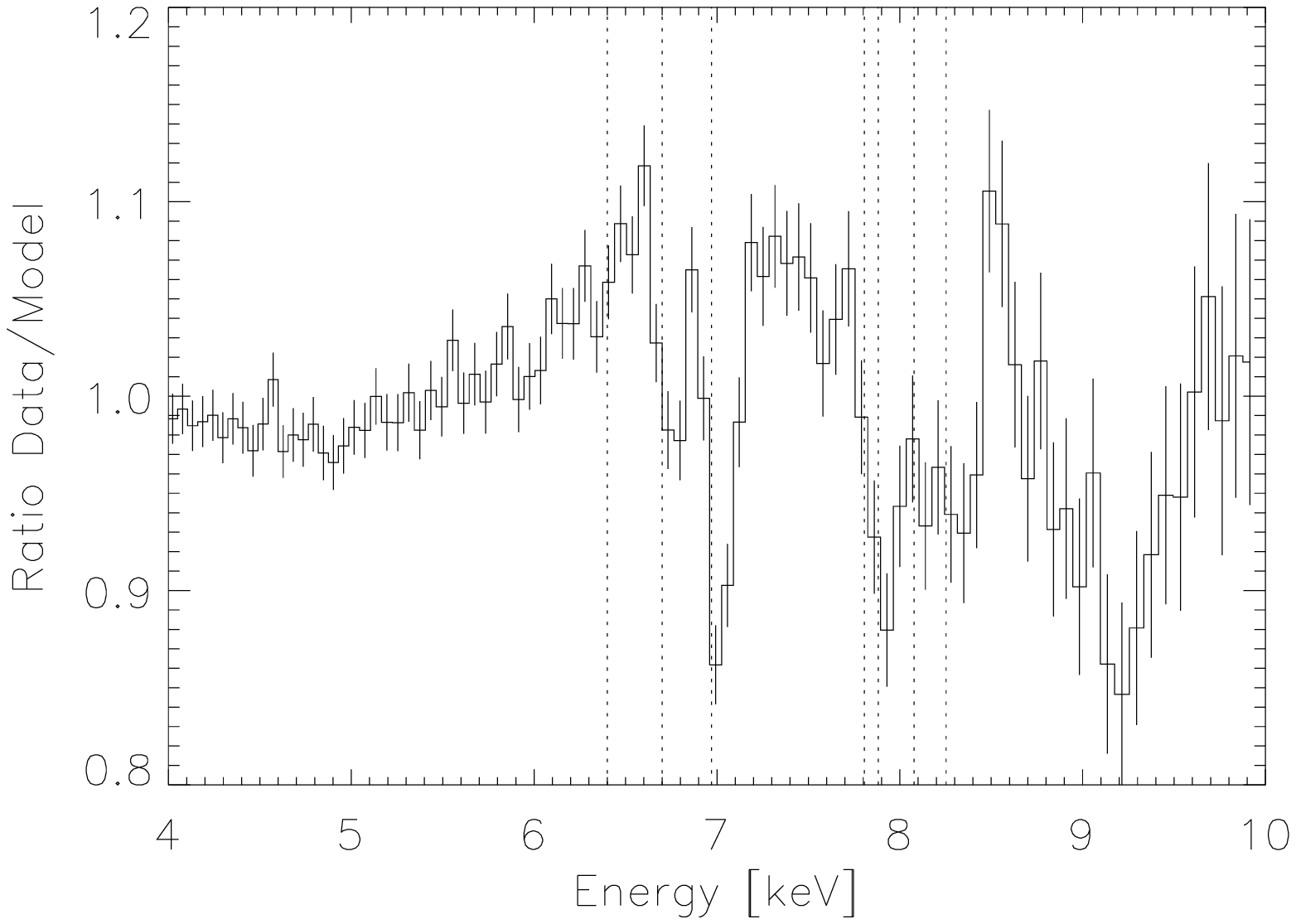}
\includegraphics[angle=0,width=0.33\textwidth]{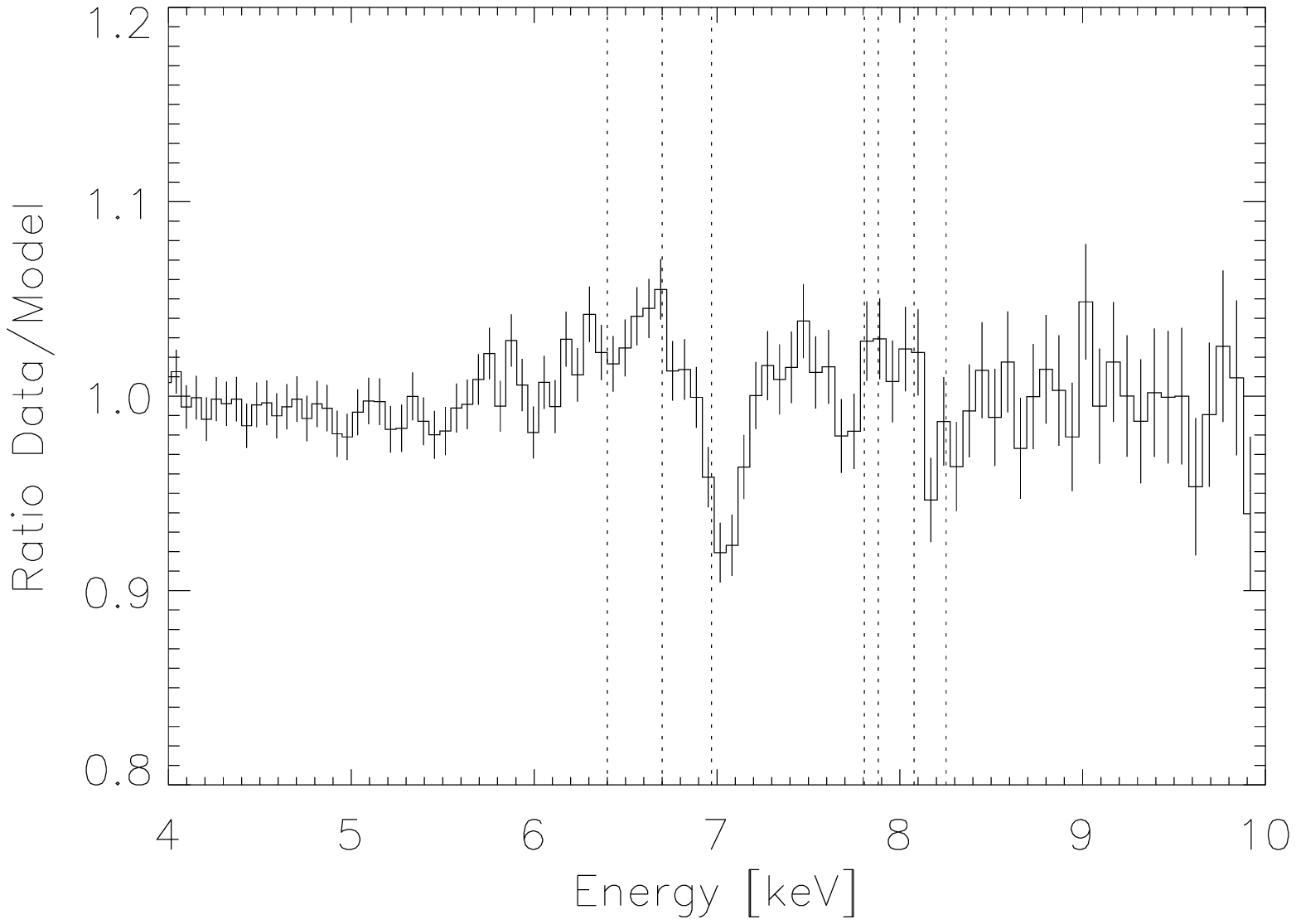}
\includegraphics[angle=0,width=0.33\textwidth]{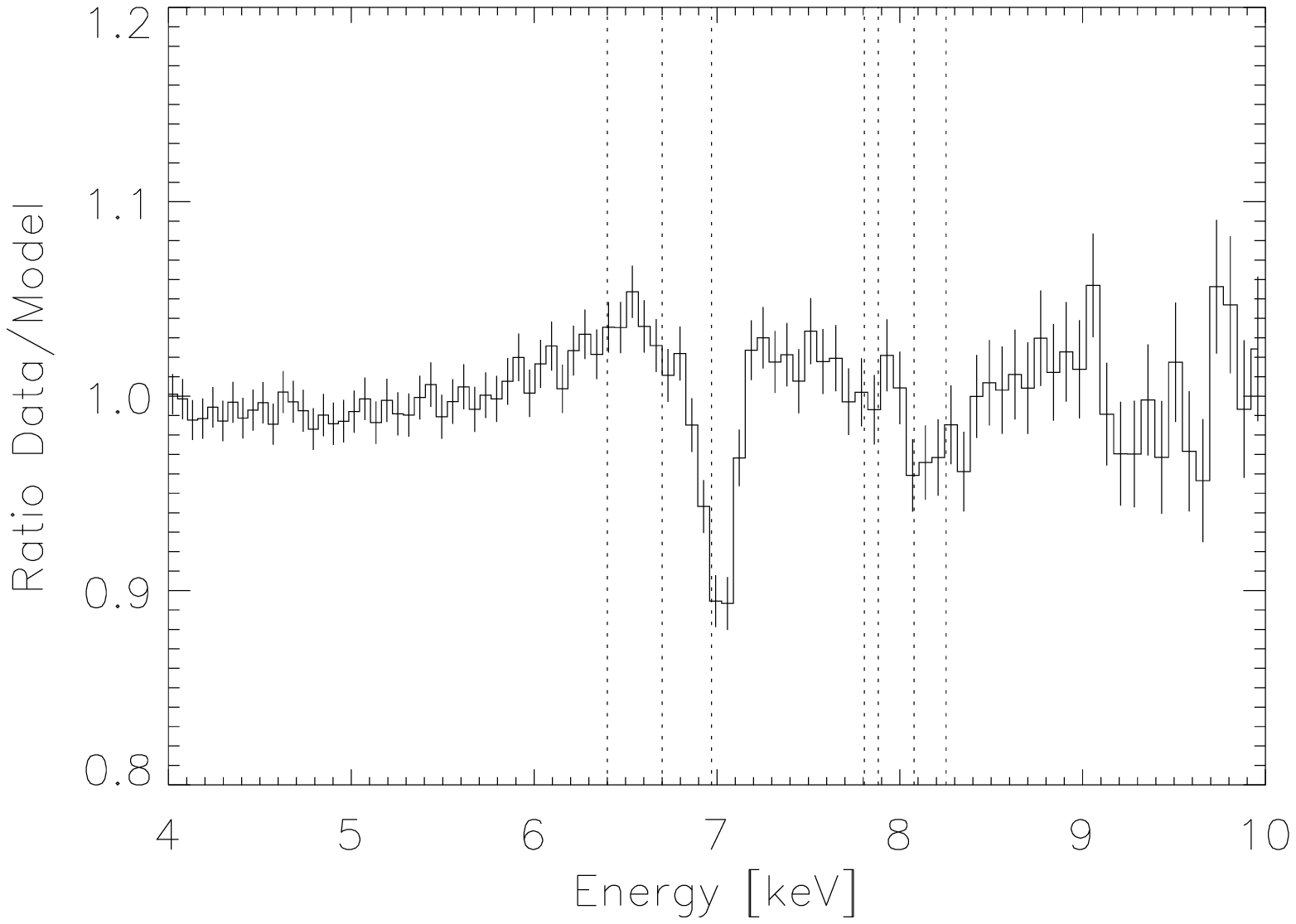}
\includegraphics[angle=0,width=0.33\textwidth]{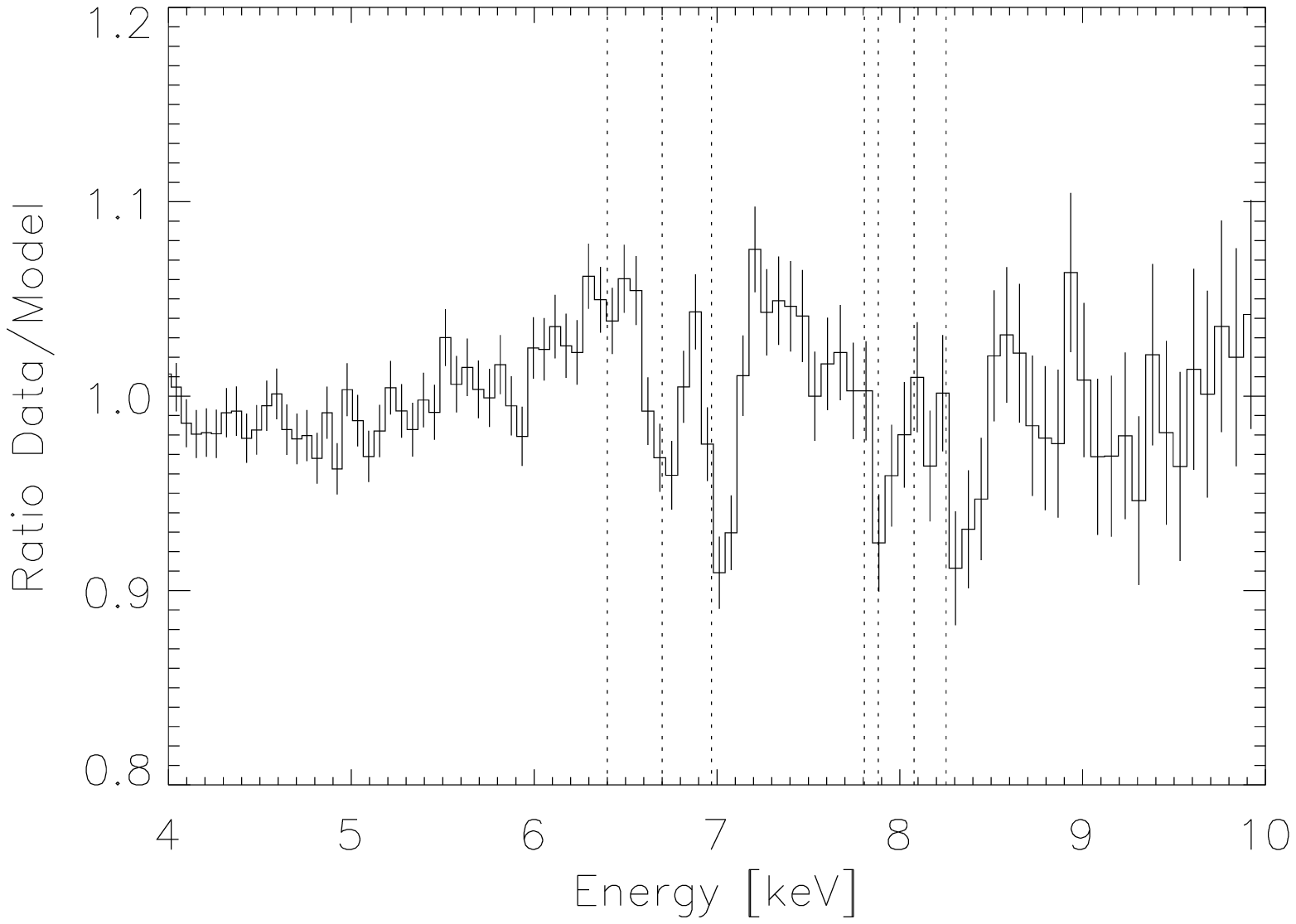}
\includegraphics[angle=0,width=0.33\textwidth]{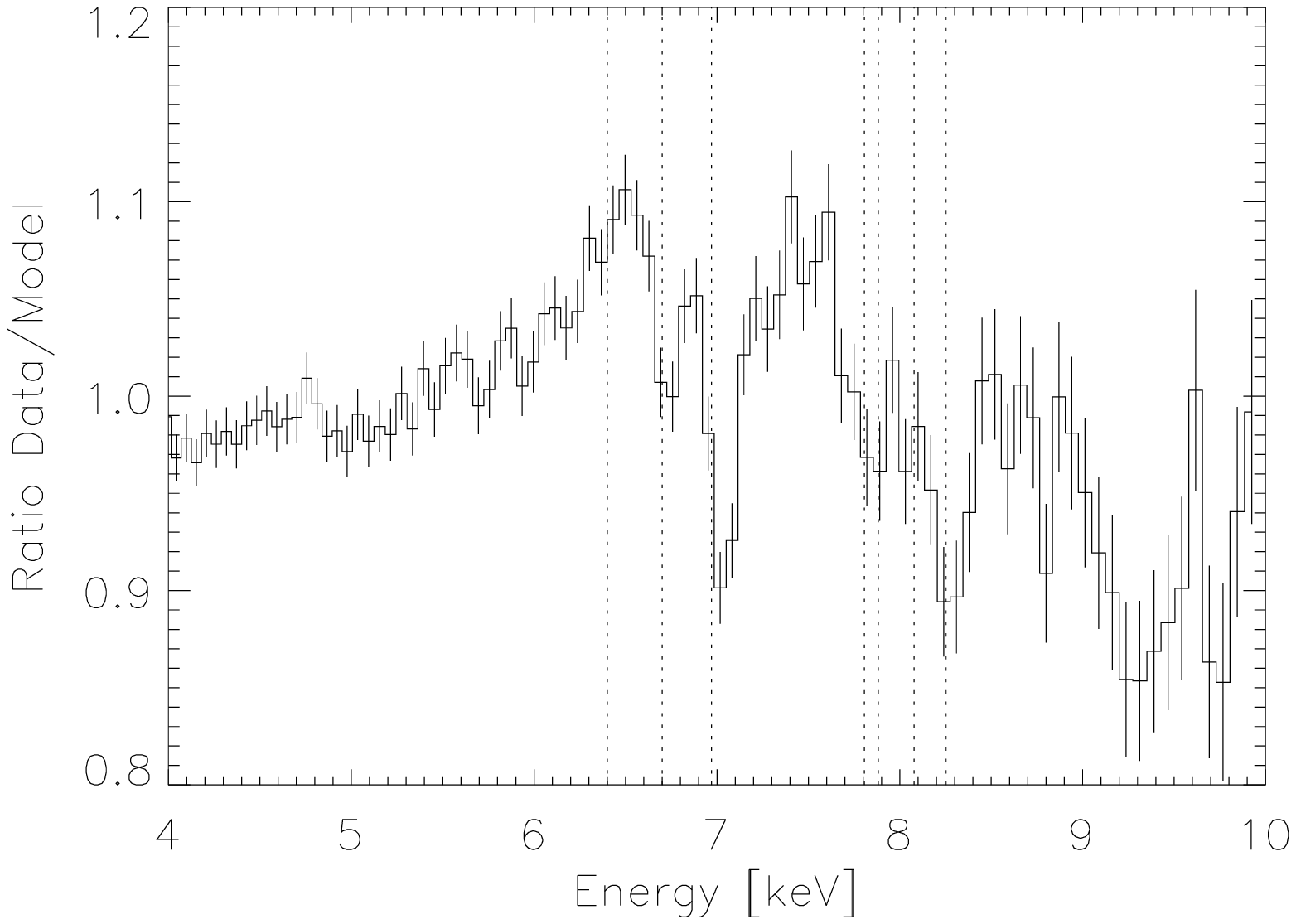}
\caption{Ratio of the data to the continuum model at the Fe~K region for obs~4 (upper-left), 6 (upper-middle), 7 (upper-right), 8 (lower-left) and 9 (lower-middle). The dotted vertical lines indicate from left to right the rest energy of the transitions of neutral Fe, \fetfive\ \ka, \fetsix\ \ka, \nitseven\ , \fetfive\ \kb, \niteight\ and \fetsix\ \kb. } 
\label{fig:rgspndiscrepant}
\end{figure*}

The residuals showed strong absorption lines near 7~keV
superposed on a broad emission feature and weak absorption features at 
$\sim$1.8 and 2.3~keV. The latter are most likely due to residual calibration 
uncertainties (after using the SAS task
{\tt epfast}), due to an incorrect application of the CTI 
correction in the EPIC pn camera when high count rates are present\footnote[4]{We note that while the use of {\tt epfast} is recommended by the EPIC calibration team, the recent discovery that the count-rate dependency of the CTI may be caused by X-ray loading implies that the {\tt epfast} correction may improve, but not completely solve, the CTI calibration deficiency (for more details, see ``Evaluation of the spectral calibration accuracy in EPIC-pn fast
modes" at http:$\slash\slash$xmm.esac.esa.int$\slash$external$\slash$xmm$\_$calibration).}.
We modeled such features with two Gaussian absorption 
components and do not discuss them further. However, we note that
these features, if due to residual uncertainties of the CTI correction, indicate
that the energy gain could be compromised in the whole energy band.
Detailed plots of the residuals from the best-fit continuum model around the Fe~K~region are 
shown in Fig.~\ref{fig:rgspndiscrepant}. Absorption features from highly ionised species of iron, such as 
\fetfive\ and \fetsix\ \ka\ and \kb\ are evident. A broad iron
emission line is present in all the observations. Both the absorption and emission features show significant
variations among observations, with obs~6 and 7 showing the most ionised species in absorption and the
weakest line in emission. Obs~4 and 9 show the strongest line in emission and a prominent absorption edge at $\sim$9~keV, that we identify with absorption from \fetfive\ and \fetsix.

\begin{figure}[!ht]
\hspace{-0.8cm}
\includegraphics[angle=0,width=0.5\textwidth]{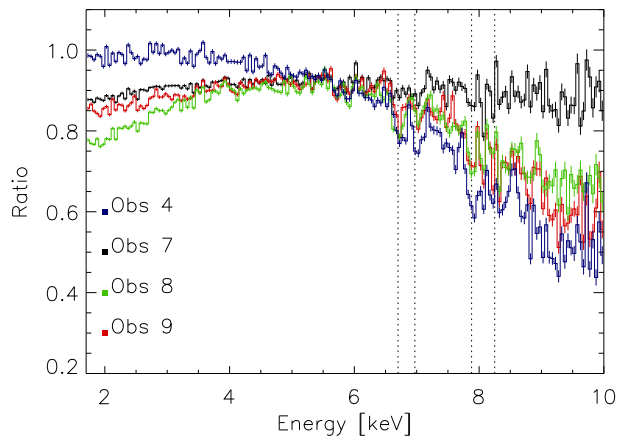}
\caption{Ratio of EPIC pn \src\ spectra from obs~4, 7, 8 and 9 with respect to 
obs~6.} 
\label{fig:ratio}
\end{figure}

We first evaluated the effect of the continuum in the absorption and emission features in a model-independent
manner calculating the ratio among the spectra for all the observations with respect to obs~6 (see Fig.~\ref{fig:ratio}). 
Obs~7 shows a similar spectral shape to obs~6 and a flux lower by $\sim$10\% in all the energy band. Above $\sim$6~keV, obs~4, 8 and 9 show a significant deficit of photons with respect to obs~6 and 7. Below $\sim$5~keV, obs~8 and 9 show the lowest fluxes. 
Interestingly, obs~4 shows the lowest 6-10~keV flux of all the observations but the highest 1.5-5~keV flux together with obs~6. 
This analysis indicates that the hard, $\sim$\,6--10~keV, flux could be correlated with the ionisation of the absorption features and the significance of the broad emission line. As the hard flux increases from obs~4 to obs~9, 7 and 6 the significance of the broad emission line decreases and the absorption features become weaker and reveal an increase of ionisation in the plasma. Obs~8 shows an intermediate hard flux and broad emission line, but the less ionised absorption features, indicating that the flux below $\sim$5~keV may play an important role in the ionisation of the absorbing plasma.
 
In order to quantify the relation between the significance of the emission and absorption features and the changes in the continuum, 
we next included in the model Gaussian absorption and emission features and absorption edges 
to account for the residuals near 7~keV. 
The total model consisted of disc blackbody and blackbody components, 
one Gaussian emission feature at $\sim$6.6~keV and four Gaussian absorption features at the energies of \fetfive\ and \fetsix\ \ka\ and \kb, 
modified by photo-electric absorption from neutral material and by absorption edges at the energies of \fetfive\ and \fetsix\
({\tt tbabs*edge$_1$*edge$_2$*(diskbb+bbodyrad+gau$_1$+gau$_2$+gau$_3$\\+gau$_4$+gau$_5$)}, hereafter Model~1).
The fits with Model~1 were acceptable, with \rchisq\ between 1.0 and 1.2
for $\sim$530--700 d.o.f. for obs~4--9. The parameters of the Gaussian emission and absorption features and edges for the best-fit with
this model are given in Table~\ref{tab:bestfit-gau}. 

\begin{figure}[!ht]
\includegraphics[angle=0,width=0.48\textwidth]{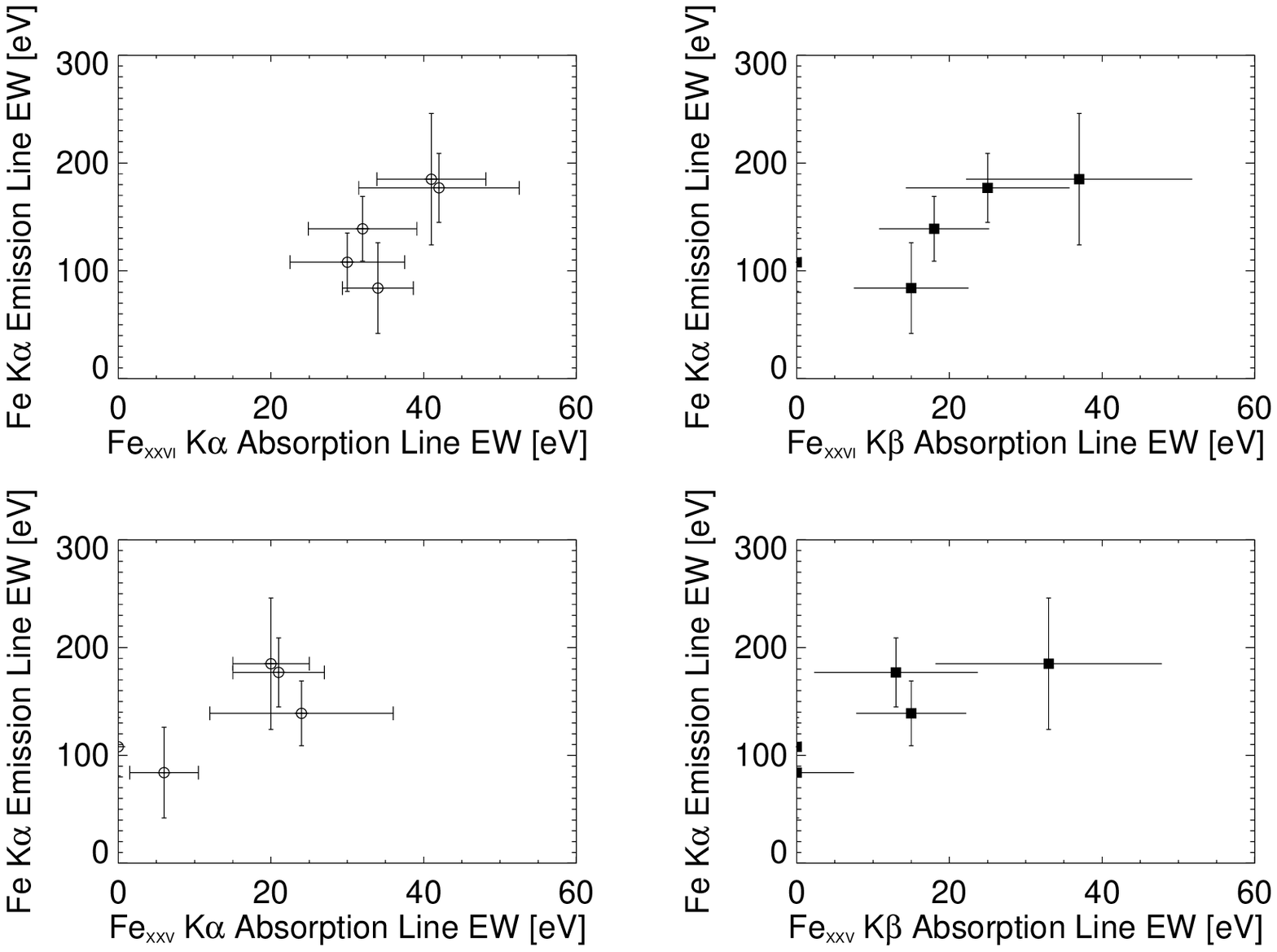}
\caption{\ew\ of the broad emission line with respect to the narrow absorption lines of \fetsix\ \ka\ (upper-left), \fetsix\ \kb\ (upper-right), \fetfive\ \ka\ (lower-left) and \fetfive\ \kb\ (lower-right).} 
\label{fig:lines}
\end{figure}

We can infer some properties of the absorber 
from the equivalent width (\ew) of the absorption features, the depth of the edges and the predominance of a given ion. For example, the depth of the \fetfive\ edge indicates
that obs~4 and 9 have a column density of ionised plasma significantly larger than obs~6--8. The ratio of the \fetfive\ and \fetsix\ \ka\ to \kb\ lines in obs~4  shows saturation of
the lines, indicating most likely a column density above 10$^{23}$~atom~cm$^{-2}$ \footnote[5]{The exact column density at which the absorption lines reach saturation depends on the velocity broadening of the lines \citep[see e.g. Fig.~1 of ][]{tombesi11apj}.}. 
The ratio of the \ew\ of the \fetfive\ to \fetsix\ lines determines the ionisation of the plasma and shows that obs~6 and 7 have a very
ionised plasma and that obs~8 has the least ionised plasma. Confirmation of the ionisation is given by the presence of low significance absorption features at 
$\sim$1.88~keV in obs 8 and 9, that we identify with \sithirteen, and at $\sim$2.62~keV in obs~4, 8 and 9, that we identify with \ssixteen. Obs~8 shows also small features
at $\sim$2.0, 2.5 and 4.2~keV that we identify with \sifourteen, \sfifteen\ and \catwenty.
Taking the above into account we can also examine the relation between emission and absorption lines. The \ew\ of the broad emission line is correlated with the \ew\ of the narrow absorption lines (see Fig.~\ref{fig:lines}). Obs~6 and 7 show the smallest emission line and the thinnest absorbing plasma. Obs 4 and 9 show instead the largest
emission lines together with the largest column density of ionised plasma. The \ew\ of the emission and absorption lines for obs 8 are intermediate between 
obs~4 and 9 on one side and obs~6 and 7 on the other side. Finally, the column density of the ionised plasma increases as the plasma becomes less ionised for
obs~4, 6, 7 and 9. This is expected since obs~6 and 7 are likely to have a fraction of fully ionised plasma that reveals itself only through electron scattering. Instead, for obs~8 
the decrease in the column density with respect to obs~4 and 9 cannot be explained simply by a change in ionisation and indicates that at the time of obs~8 a smaller column density is present.  

\begin{table*}
\begin{center}
\caption[]{Parameters of the most significant absorption lines and edges and of the broad emission line detected in the EPIC pn spectra of obs~4--9. {\it f} indicates that a parameter
was fixed. The width of the absorption features was constrained to be $\approxlt$\,0.1~keV and of the broad emission feature to $\approxlt$\,1~keV. {\it p} indicates that the error of a parameter pegged at the limit imposed. {\tt gau$_1$} corresponds to the broad emission feature and {\tt gau$_2$}--{\tt gau$_5$} to narrow absorption features. The energy of the \fetfive\ feature of obs~7 was unconstrained due to its low significance. Therefore, we fixed its energy considering that it would show a similar blueshift as the \fetsix\ feature for the same observation. 
}
\begin{tabular}{lcccccc}
\hline \hline\noalign{\smallskip}
Observation No. & & 4 & 6 & 7 & 8 & 9 \\
\noalign{\smallskip\hrule\smallskip}
& Comp. & & & & & \\
Parameter & & & & & & \\
& {\tt gau$_1$} & & & & & \\
\multicolumn{2}{l}{ \egau\ {\small(keV)}} & 6.53\,$^{+0.11}_{-0.09}$ & 6.63\,$\pm$\,0.13 & 6.60\,$^{+0.10}_{-0.14}$ & 6.65\,$^{+0.08}_{-0.11}$ & 6.71\,$\pm$\,0.08 \\
\multicolumn{2}{l}{ $\sigma$ {\small(keV)}} & 0.72\,$^{+0.14}_{-0.16}$ & 0.74\,$\pm$\,0.19 & 0.62\,$^{+0.20}_{-0.09}$ & 0.61\,$\pm$\,0.18 & 0.60$^{+0.13}_{-0.11}$ \\
\multicolumn{2}{l}{ \kgau\ {\small(10$^{-3}$ ph cm$^{-2}$ s$^{-1}$)}} & 12\,$\pm$\,4 & 8\,$\pm$\,2 & 6\,$\pm$\,3 & 9\,$\pm$\,2  & 11\,$\pm$\,2 \\
\multicolumn{2}{l}{ \ew\ (eV)} & 185\,$\pm$\,62 & 108\,$\pm$\,27 & 84\,$\pm$\,42 & 139\,$\pm$\,31 & 177\,$\pm$\,32 \\
& {\tt gau$_2$} & & & & & \\
\multicolumn{2}{l}{ \egau\ {\small(keV)}} & 6.75\,$^{+0.02}_{-0.05}$ & -- & 6.73 (f) & 6.71\,$\pm$\,0.02 & 6.73\,$\pm$\,0.02 \\
\multicolumn{2}{l}{ $\sigma$ {\small(keV)}} & $<$\,0.04 & -- &  0.008\,$^{+0.002p}_{-0.008}$ & $<$\,0.07 & $<$\,0.05 \\
\multicolumn{2}{l}{ \kgau\ {\small(10$^{-4}$ ph cm$^{-2}$ s$^{-1}$)}} & 12\,$\pm$\,3 & -- & 4\,$\pm$\,3 & 16$^{+4}_{-8}$ & 14\,$\pm$\,4 \\
\multicolumn{2}{l}{ \ew\ (eV)} & 20\,$\pm$\,5 & -- & 6\,$\pm$\,5 & 24\,$^{+6}_{-12}$ & 21\,$\pm$\,6 \\
& {\tt gau$_3$} & & & & & \\
\multicolumn{2}{l}{ \egau\ {\small(keV)}} & 7.02\,$\pm$\,0.06 & 7.03\,$\pm$\,0.02 & 7.00\,$\pm$\,0.01 & 7.02\,$\pm$\,0.02  & 7.03\,$\pm$\,0.01\\
\multicolumn{2}{l}{ $\sigma$ {\small(keV)}} & $<$\,0.03 & 0.07\,$\pm$\,0.05 & $<$\,0.06 & $<$\,0.04 & $<$\,0.06 \\
\multicolumn{2}{l}{ \kgau\ {\small(10$^{-4}$ ph cm$^{-2}$ s$^{-1}$)}} & 23\,$\pm$\,4 & 20\,$\pm$\,5 & 22\,$\pm$\,3 & 18\,$\pm$\,4 & 24\,$^{+4}_{-6}$ \\
\multicolumn{2}{l}{ \ew\ (eV)} & 41\,$\pm$\,7 & 30\,$\pm$\,8 & 34\,$\pm$\,5 & 32\,$\pm$\,7 & 42\,$^{+7}_{-11}$ \\
& {\tt gau$_4$} & & \\
\multicolumn{2}{l}{ \egau\ {\small(keV)}}  & 7.91\,$^{+0.46}_{-0.15}$ & -- & -- & 7.91\,$\pm$\,0.05 & 7.84\,$^{+0.10}_{-0.06}$ \\
\multicolumn{2}{l}{ $\sigma$ {\small(keV)}} & 0.003\,$^{+0.007p}_{-0.003}$ & -- & -- & 0.005\,$^{+0.095p}_{-0.005}$ & 0.03\,$^{+0.07p}_{-0.03}$ \\
\multicolumn{2}{l}{ \kgau\ {\small(10$^{-4}$ ph cm$^{-2}$ s$^{-1}$)}} & 11\,$\pm$\,4 & -- & -- & 5\,$\pm$\,3 & 5\,$\pm$\,4 \\
\multicolumn{2}{l}{ \ew\ (eV)} & 33\,$\pm$\,12 & -- & -- & 15\,$\pm$\,9 & 13\,$\pm$\,10 \\
& {\tt gau$_5$} & & \\
\multicolumn{2}{l}{ \egau\ {\small(keV)}} & 8.24\,$^{+0.04}_{-0.06}$ & -- & 8.20\,$\pm$\,0.10 & 8.34\,$\pm$\,0.05 & 8.26\,$\pm$\,0.04 \\
\multicolumn{2}{l}{ $\sigma$ {\small(keV)}} & 0.1\,$^{+0p}_{-0.1p}$ & -- & 0.1\,$^{+0p}_{-0.1p}$ & $<$\,0.097 & 0.02\,$^{+0.08p}_{-0.02p}$ \\
\multicolumn{2}{l}{ \kgau\ {\small(10$^{-4}$ ph cm$^{-2}$ s$^{-1}$)}} & 10\,$\pm$\,4 & -- & 6\,$\pm$\,3 & 5\,$\pm$\,2 & 7\,$\pm$\,3 \\
\multicolumn{2}{l}{ \ew\ (eV)} & 37\,$\pm$\,15 & -- & 15\,$\pm$\,8 & 18\,$\pm$\,7 & 25\,$\pm$\,11 \\
& {\tt edge$_1$} & & & & & \\
\multicolumn{2}{l}{ E {\small(keV)}} & 8.83 (f) & 8.83 (f) & 8.83 (f) & 8.83 (f) & 8.83 (f) \\ 
\multicolumn{2}{l}{ $\tau$ } & 0.18\,$\pm$\,0.03 & $<$0.04 & $<$0.04 & $<$0.05 & 0.08\,$\pm$\,0.04 \\
& {\tt edge$_2$} & & & & & \\
\multicolumn{2}{l}{ E {\small(keV)}} & 9.28 (f) & 9.28 (f) & 9.28 (f) & 9.28 (f) & 9.28 (f) \\
\multicolumn{2}{l}{ $\tau$ } & $<$\,0.03 & 0.04\,$\pm$\,0.03 & 0.04\,$\pm$\,0.03 & $<$\,0.05 & $<$\,0.09 \\
\noalign {\smallskip}
\noalign {\smallskip}
\hline\noalign {\smallskip}
\multicolumn{2}{l}{\rchisq (d.o.f.)} & 1.08 (574) & 1.11 (694) & 1.03 (678) & 1.07 (530) & 1.18 (561) \\
\noalign{\smallskip\hrule\smallskip}
\noalign{\smallskip\hrule\smallskip}
\label{tab:bestfit-gau}
\end{tabular}
\end{center}
\end{table*} 

\subsubsection{Photoionised plasma model}
\label{sec:model2}
 
The strength of the absorption features shown in Table~\ref{tab:bestfit-gau} indicates that the continuum may be significantly 
affected by the absorbing plasma. Therefore, to quantify the above changes in a more physical manner, we substituted the absorption features
and edges by the component {\tt warmabs}, which models the absorption due to a photoionised
plasma in the line of sight. This component not only accounts
for the narrow absorption features evident near 7~keV but also modifies
the overall continuum shape at regions where the spectral resolution is not enough to resolve
individual features. We note that the {\tt warmabs} model
does not include Compton scattering, in contrast to the {\tt xabs} model in the SPEX package \citep{kaastra96}.
The contribution of Compton scattering is expected to be significant when modeling the changes
between persistent and dipping emission \citep{1323:boirin05aa, ionabs:diaz06aa}. Therefore 
we added the component {\tt cabs} to the model, which accounts for non-relativistic, optically-thin Compton scattering. We forced
the column density of the {\tt cabs} model to be equal to the column density of the {\tt warmabs} component multiplied by a factor of 1.21. The latter
factor accounts for the number of electrons per hydrogen atom for a material of solar abundances \citep{herx1:stelzer99}. The non-relativistic approximation
of {\tt cabs} will overestimate the scattering fraction at high energies, and therefore it should
not be used when broad band energy data is available. Moreover, the column density of the {\tt warmabs} component sets only 
a lower limit to the amount of Compton scattering since there may be fully ionised material which cannot be identified via line absorption but still
contributes to Compton scattering.

Our final model consisted of disc blackbody and blackbody components and
one Gaussian emission feature at $\sim$6.6~keV, modified by photo-electric 
absorption from neutral and ionised material
({\tt tbabs*cabs*warmabs*(diskbb+bbodyrad+gau)}, hereafter Model~2).

\begin{figure*}[!ht]
\includegraphics[angle=0,width=0.35\textwidth]{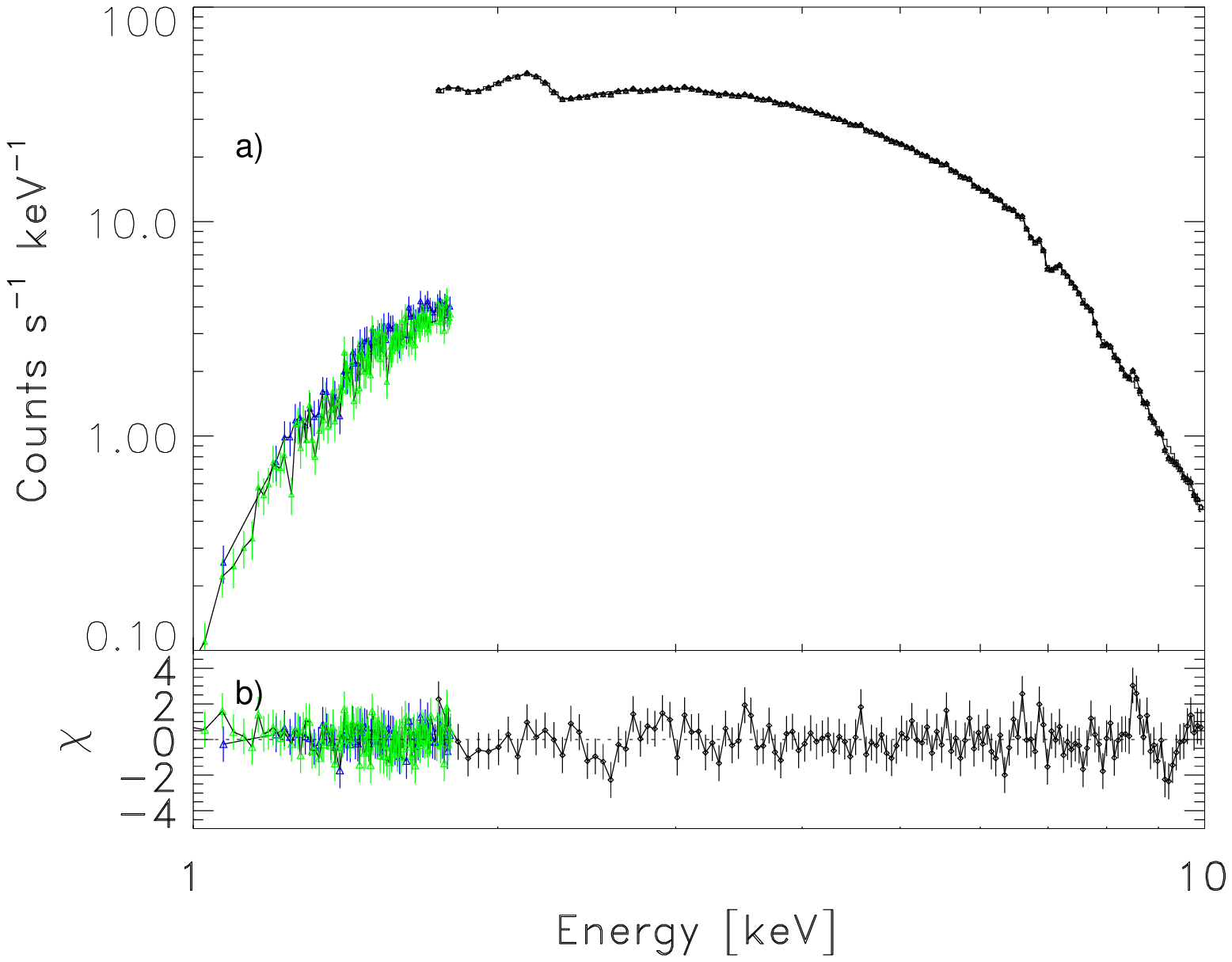}
\includegraphics[angle=0,width=0.35\textwidth]{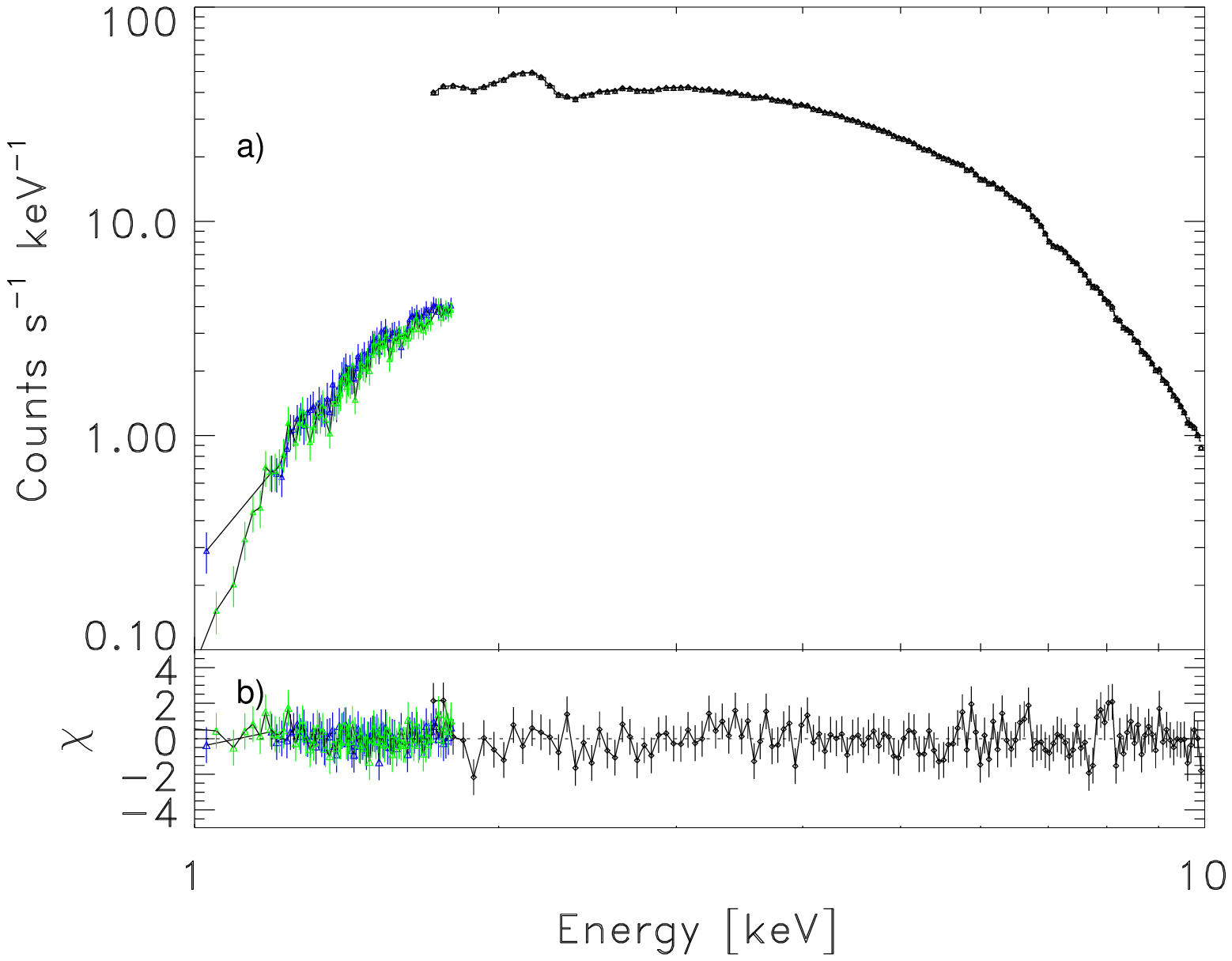}
\includegraphics[angle=0,width=0.35\textwidth]{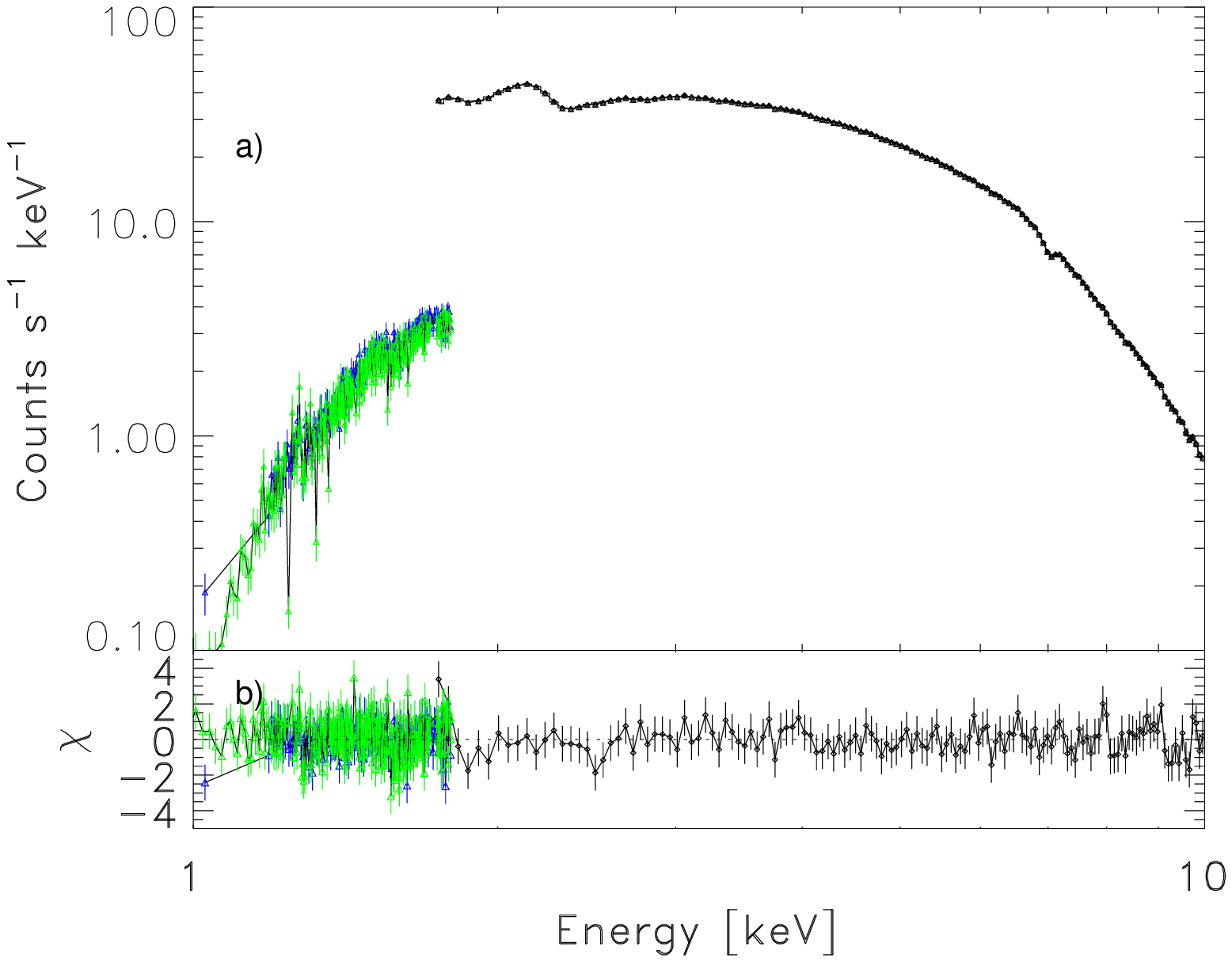}
\includegraphics[angle=0,width=0.35\textwidth]{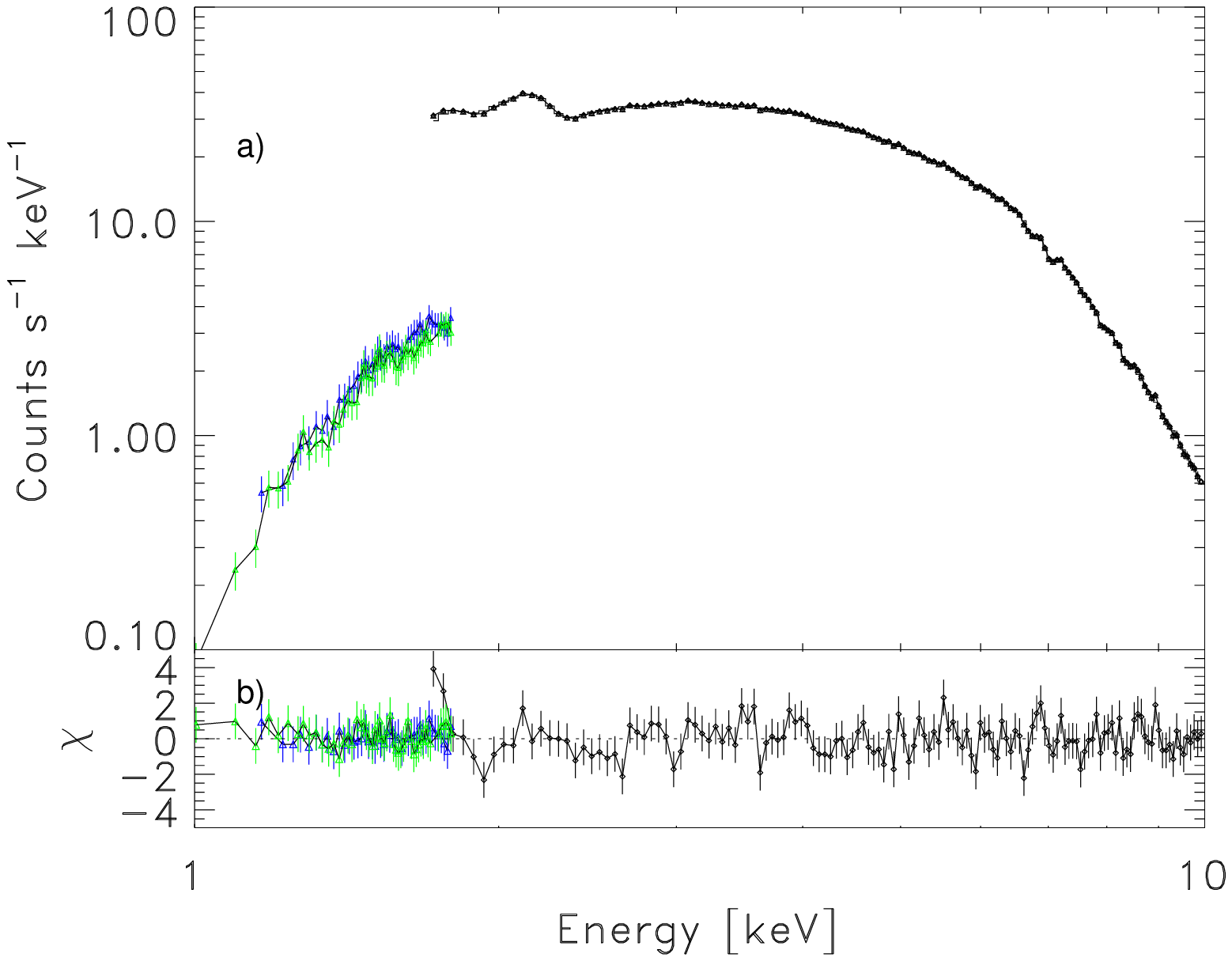}
\includegraphics[angle=0,width=0.35\textwidth]{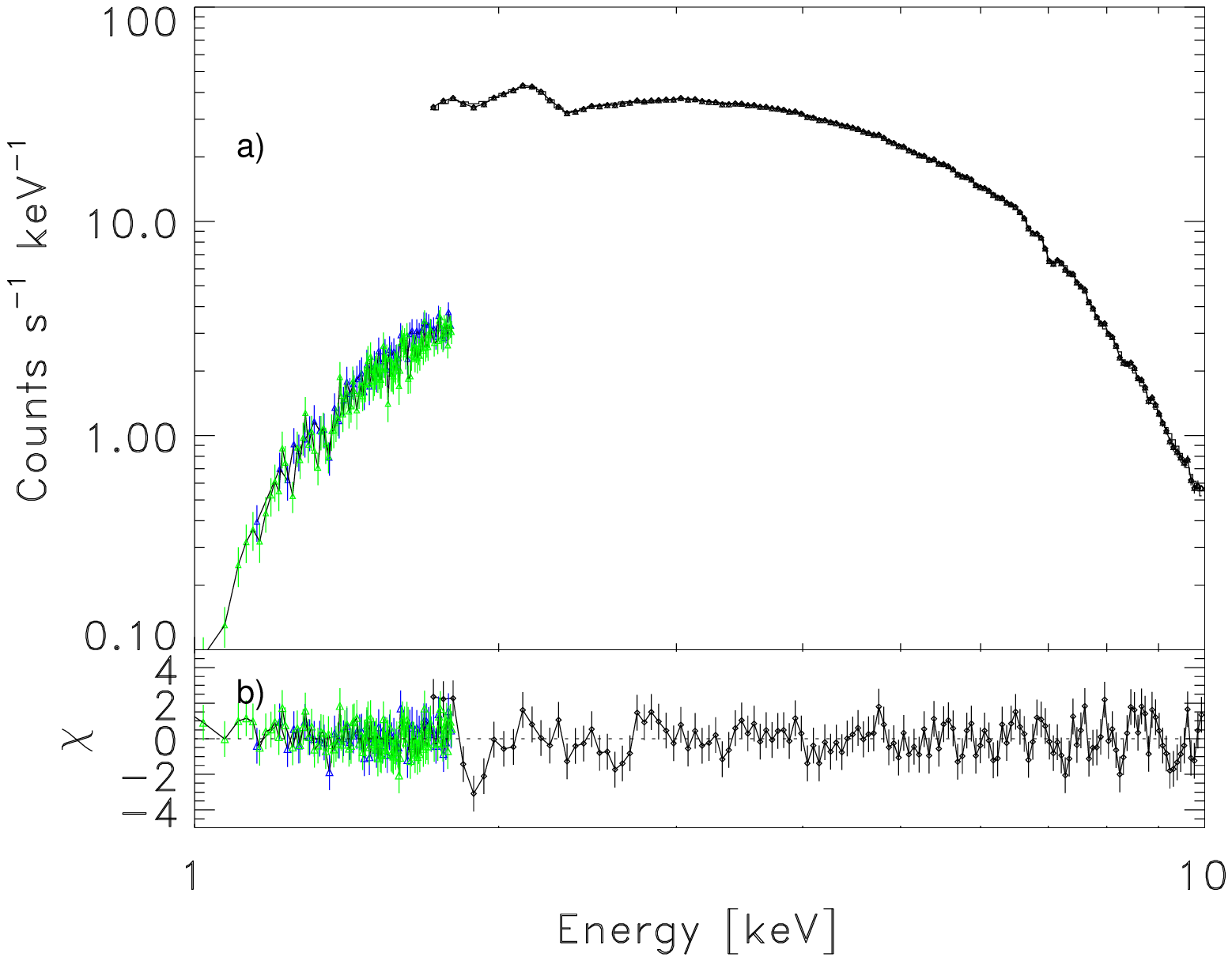}
\caption{Obs~4, 6 to 9 are shown from left to right and from up to
bottom.  (a) 1.7--10 keV EPIC pn (black), and 0.8--1.8~keV RGS1 (green)
and RGS2 (blue) \src\ spectra fitted with a disc-blackbody ({\tt
diskbb}), a blackbody ({\tt bbodyrad}) and a Gaussian ({\tt gau})
components modified by absorption from neutral ({\tt tbabs}) and
ionised ({\tt cabs*warmabs}) material. (b) Residuals in units of standard
deviation from the above model.}
\label{fig:rgspn}
\end{figure*}

\begin{figure*}[!ht]
\includegraphics[angle=0,width=0.35\textwidth]{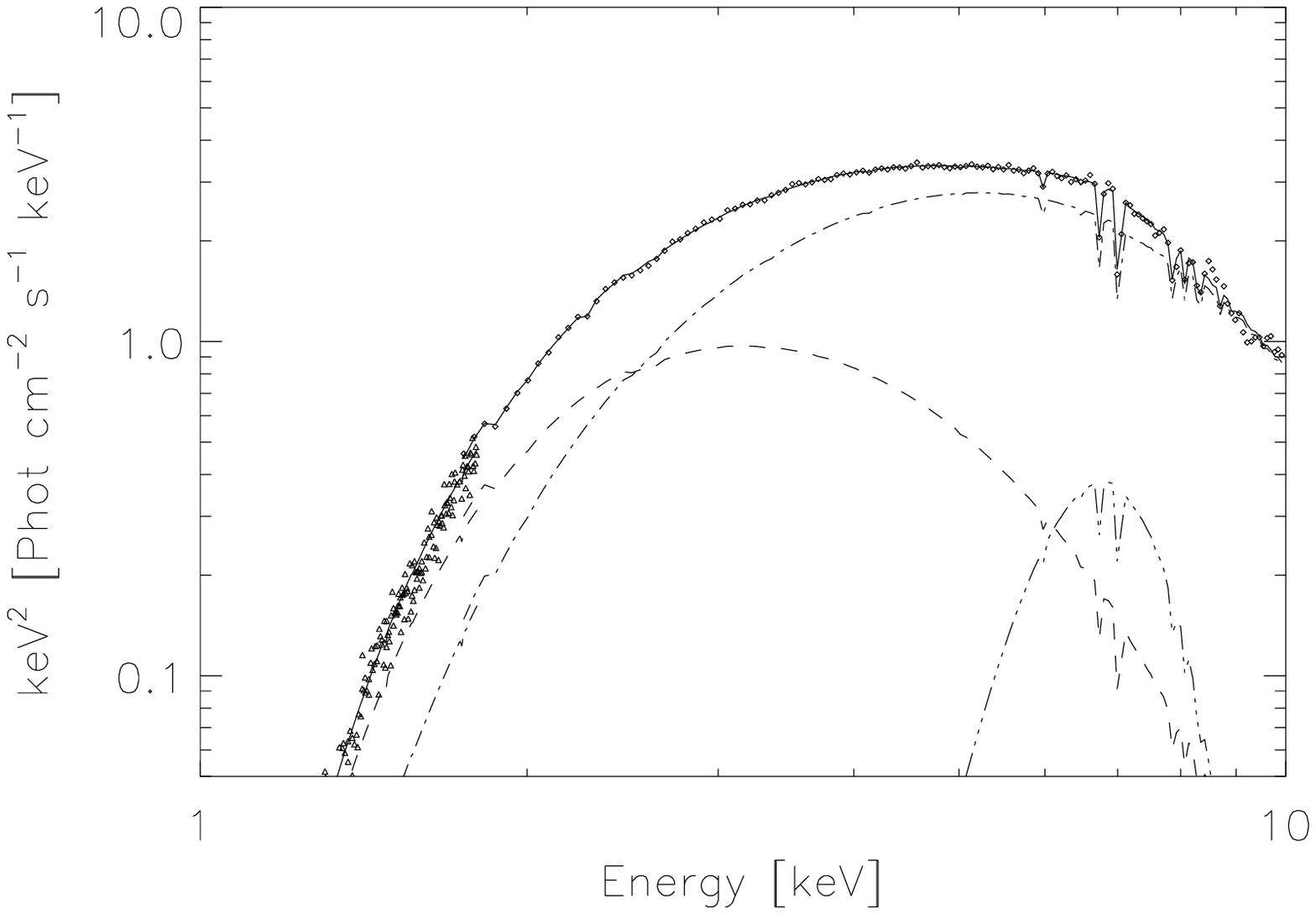} 
\includegraphics[angle=0,width=0.35\textwidth]{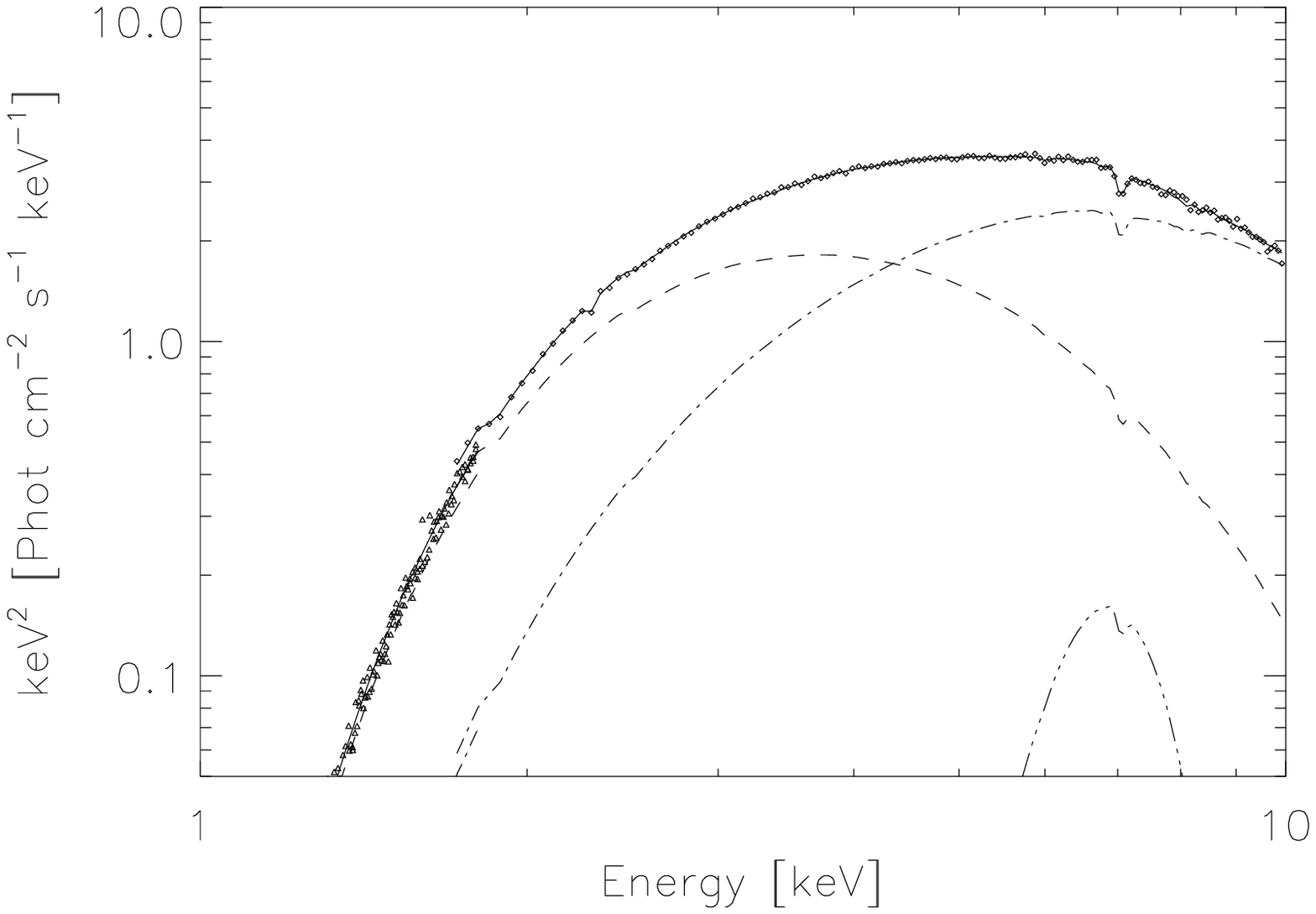}
\includegraphics[angle=0,width=0.35\textwidth]{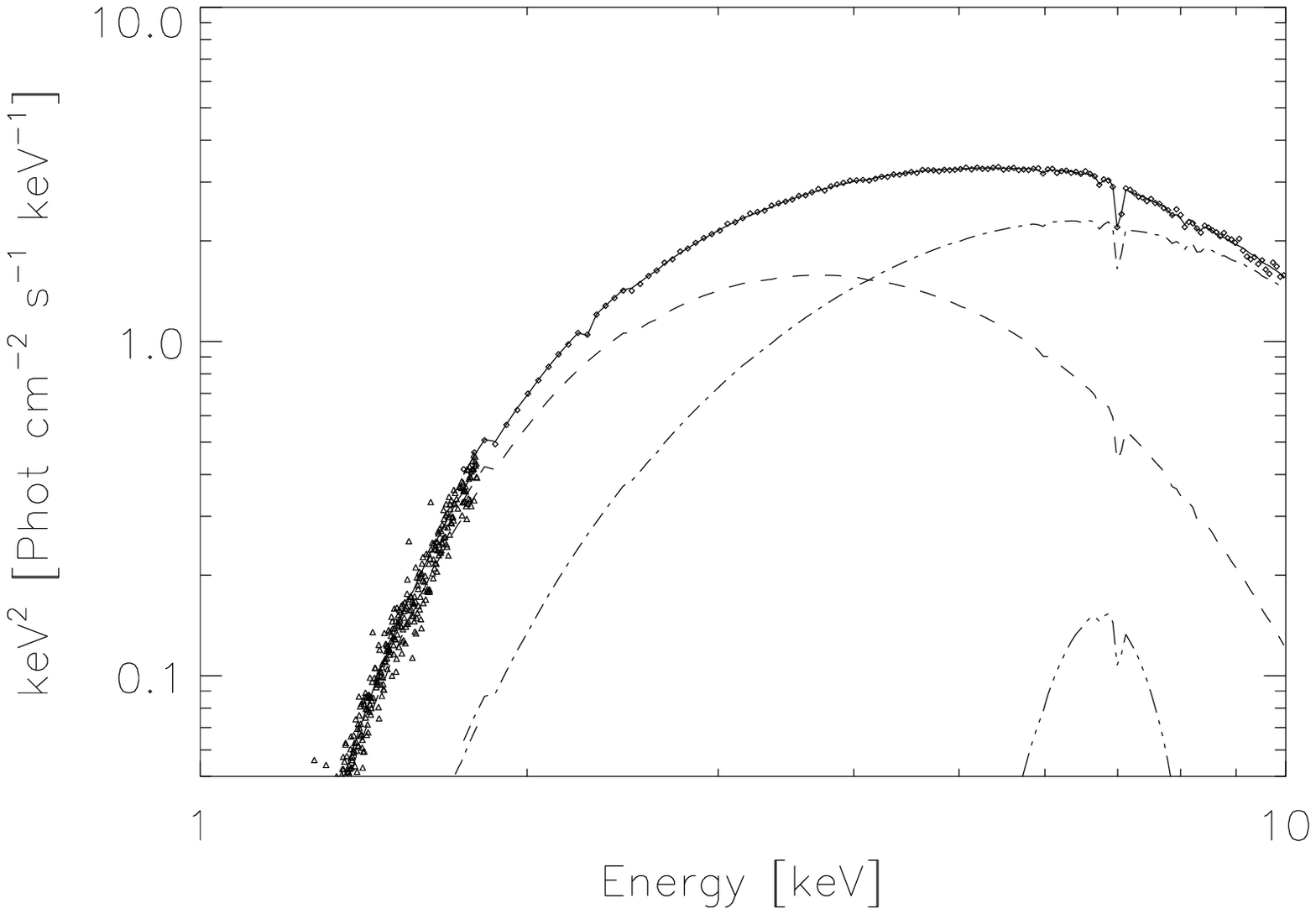}
\includegraphics[angle=0,width=0.35\textwidth]{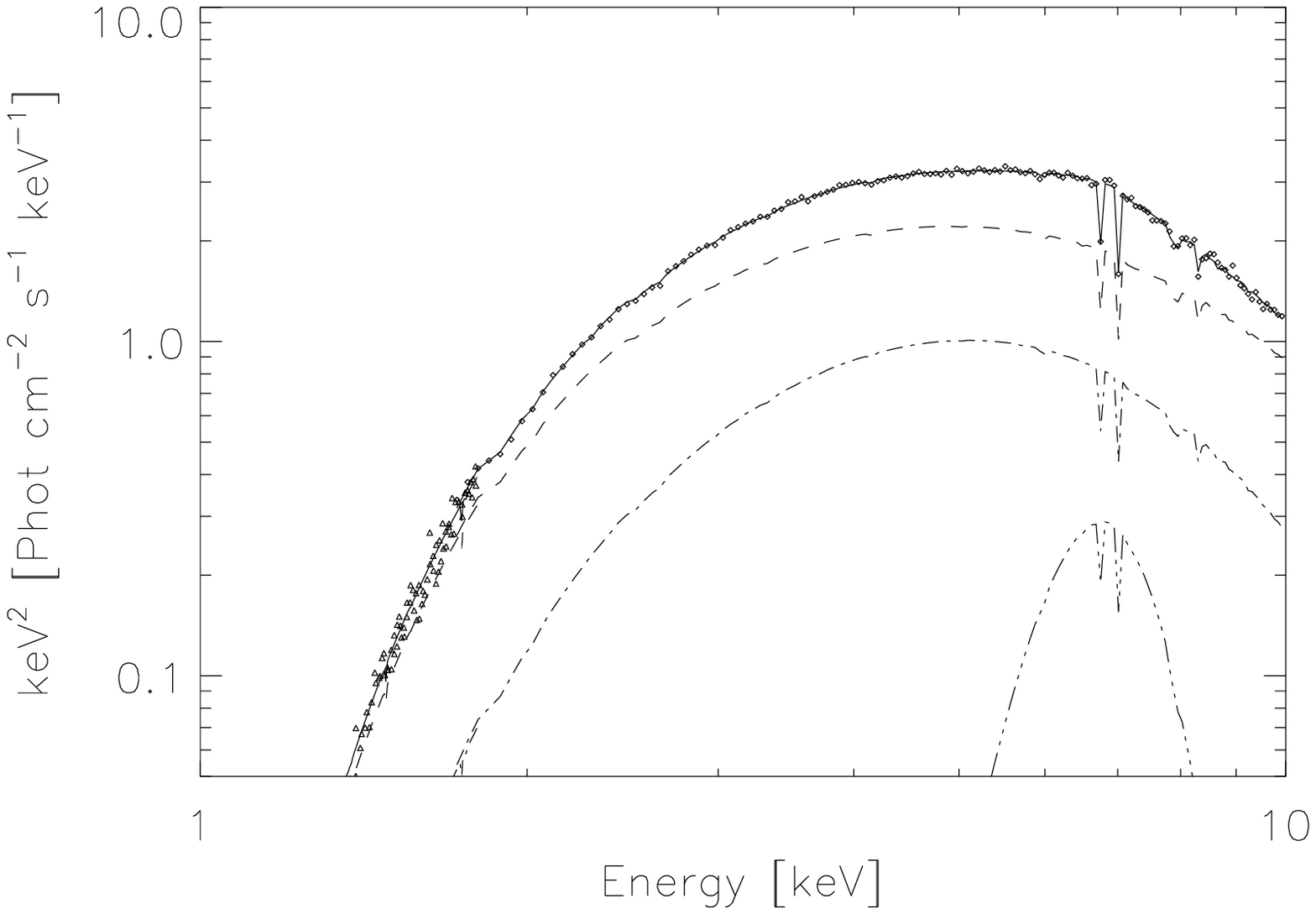}
\includegraphics[angle=0,width=0.35\textwidth]{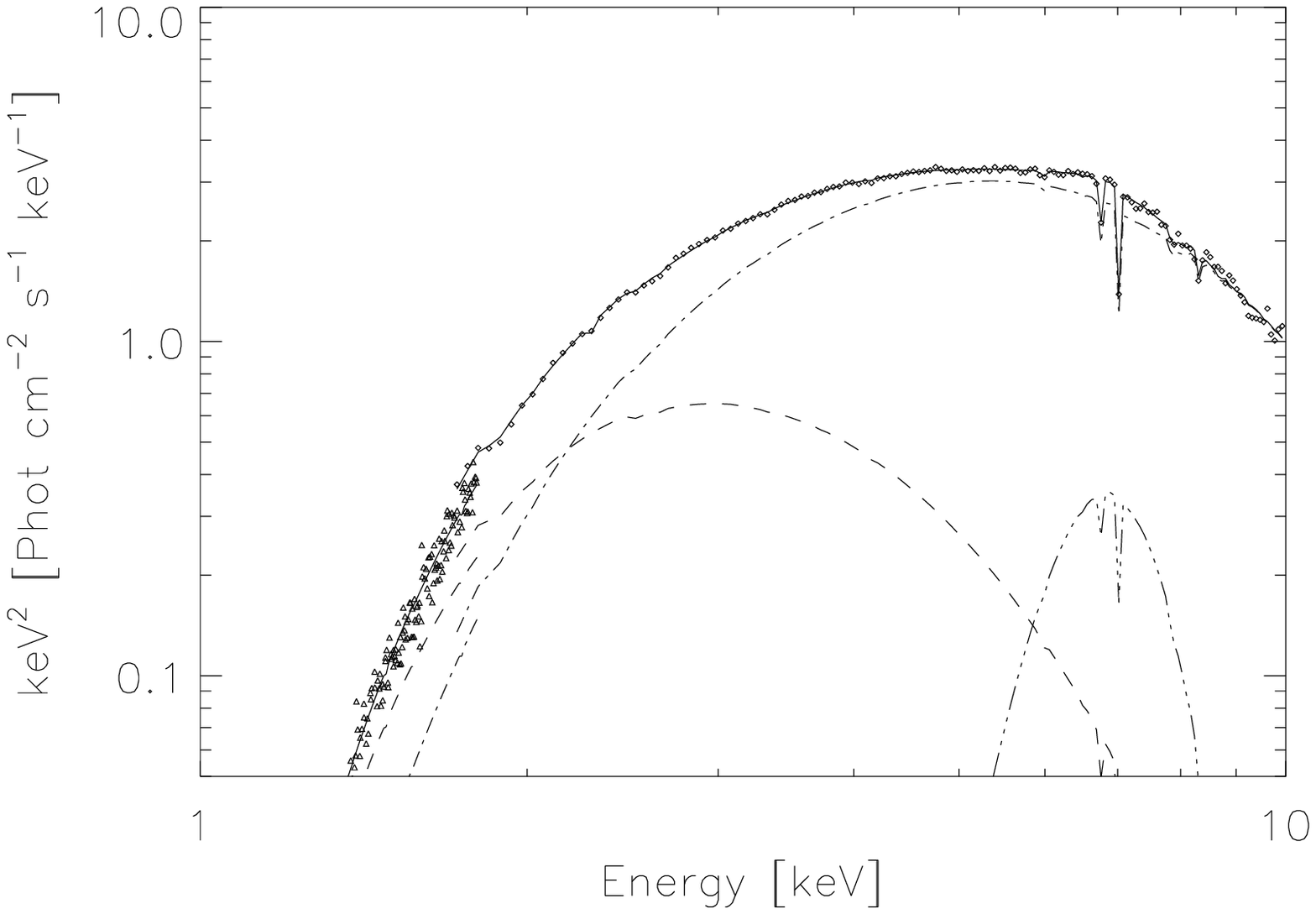}
\caption{0.8--10~keV unfolded spectra. Solid, dashed, dot-dashed and dot-dot-dashed lines represent
the contribution of the total model, disc, blackbody and Gaussian components, respectively.}
\label{fig:eeuf}
\end{figure*}

The fits with Model~2 were acceptable, with \rchisq\ between 1.0 and 1.2
for $\sim$550--700 d.o.f. for obs~4--9. The parameters of the best-fit with
this model are given in Table~\ref{tab:bestfit} and the residuals of
the fit and unfolded spectra are shown in Figs.~\ref{fig:rgspn} and ~\ref{fig:eeuf}.
For each observation, we used the best-fit continuum before adding the
broad emission line and the photo-ionised absorber
as an ionising continuum for the latter. The density of the plasma was set to be $n$\,=\,10$^{13}$\,cm$^{-3}$ following \citet{gx13:ueda04apj}.
The luminosity was defined in the 0.013--13.6~keV band for each spectrum. 
Interestingly, the value of the ionisation parameter for obs~4 in Table~\ref{tab:bestfit} does not reflect the relative
variations in the ionisation of the plasma with respect to obs~6--9 that we inferred in Sect.~\ref{model1}. The most likely reason for this is that
the model that we are using as an ionising continuum for the photoionised absorber is not appropriate. If this is true, we expect to find the larger discrepancy between the ratio of \fetfive\ and \fetsix\ lines and the ionisation parameter in obs~4, since
this observation shows a column density for the ionised plasma larger by more than a factor of two compared to the other observations and this will cause severe obscuration of the continuum. Therefore, we re-fitted again all the spectra but using for all the observations the ionising continuum of obs~6. We chose obs~6 as the more realistic continuum since it has the most highly-ionised plasma and consequently more transparent to the incident radiation. We obtained \logxi\ = 4.18\,$\pm$\,0.15, 4.39\,$\pm$\,0.17, 4.33\,$^{+0.11}_{-0.15}$, 4.00\,$^{+0.10}_{-0.19}$ and 4.09\,$^{+0.11}_{-0.15}$ \xiunit\ for obs~4, 6--9, respectively. The other
parameters and the quality of the fit did not change, as expected. These values are consistent with the line ratios of different species inferred from Table~\ref{tab:bestfit-gau} and Fig.~\ref{fig:rgspndiscrepant} and therefore we use these values rather than the ones in Table~\ref{tab:bestfit} in the remaining sections of this paper. 

Finally, we note that obs~4, 8 and 9 show still after the fit small residuals at the energies of $\sim$2.45 and 2.62~keV (see Fig.~\ref{fig:rgspn}), that we identify with absorption from \sfifteen\ and \ssixteen. Since we account for all the absorption features with the self-consistent model {\tt warmabs}, the existence of residuals indicates that the abundance of Sulphur is most likely larger than the solar value. 
Another possibility would be the presence of a second absorber with a lower ionisation degree, but in this case, we should observe also residuals for other elements in the plasma. We did not attempt to fit the Sulphur abundances due to the small significance of the features, but we note that an overabundance of the S/Fe ratio with respect to the solar value was also pointed out by \citet{gx13:ueda04apj} based on $Chandra$~HETGS data. They argued that this overabundance could indicate an underestimation of  the column density of Fe due to the non-inclusion of reemission lines. Our data could be showing the same scenario, since we also observe small residuals at the energy of the \fetsix\ edge, which could disappear with an increase of the column density of Fe. Alternatively, the remaining residuals at the \fetsix\ edge in obs~4 and 9 could represent the limit of applicability of the {\tt warmabs} model for high column densities ({\tt warmabs} assumes a uniform ionisation for the absorber even if a large column density is present, which is not self-consistent).
A small absorption feature remains also at $\sim$1.88~keV in obs~8 and 9, which could be \sithirteen, or systematic residuals at the Si edge due to a deficient CTI calibration (see Sect.~\ref{sec:observations}).

\begin{table*}
\begin{center}
\caption[]{Best-fits to the 0.8--1.8~keV RGS and 1.7--10~keV EPIC pn
  persistent spectra for all the observations using the {\tt
  tbabs*cabs*warmabs*(diskbb+bbrad+gau)} model. The column density
  of the {\tt cabs} component was tied to the column density of the photo-ionised
  absorber (see text).
  \kbb, \kdbb\ and
  \kgau\ are the normalizations of the blackbody component, disc
  blackbody and Gaussian emission feature, respectively.  \ktbb\ and
  \ktdbb\ are the temperatures of the blackbody and disc blackbody,
  respectively. \egau\ and $\sigma$ represent the energy and width of
  the Gaussian feature. \nhabs\ and \nhwarmabs\ are the column
  densities for the neutral and ionised absorbers,
  respectively. \xil, \sigmav, and $v$ are the ionisation parameter
  (in units of erg cm s$^{-1}$), the turbulent velocity broadening,
  and the average systematic velocity shift of the absorber (negative
  values indicate blueshifts). $F_{tot}$ is the unabsorbed total
  flux, and $F_{gau}$ is the unabsorbed  
  emission Gaussian flux.
The width ($\sigma$) of the Gaussian emission line {\tt
 gau} was constrained to be $\le$1~keV in the fits.
 }
\begin{tabular}{lcccccc}
\hline \hline\noalign{\smallskip}
Observation No. & & 4 & 6 & 7 & 8 & 9 \\
\noalign{\smallskip\hrule\smallskip}
& Comp. & & & & &    \\
Parameter & & & & & & \\
& {\tt dbb} & & & & & \\
\ktdbb\ {\small(keV)} & & 0.86\,$^{+0.59}_{-0.07}$ & 1.16\,$^{+0.19}_{-0.07}$  & 1.15\,$^{+0.14}_{-0.08}$ & 1.80$^{+0.11}_{-0.14}$ & 0.77\,$^{+0.06}_{-0.04}$ \\
\multicolumn{2}{l}{\kdbb\ {\small[(R$_{in}$/D$_{10}$)$^{2}$ cos$\theta$]}} & 492\,$^{+168}_{-369}$ & 226\,$^{+51}_{-54}$ & 201\,$^{+20}_{-44}$ & 38\,$\pm$\,8 & 622\,$^{+188}_{-110}$ \\
& {\tt bb} & & &  & & \\
\ktbb\ {\small(keV)} & & 1.24\,$\pm$\,0.02  & 1.66$^{+0.24}_{-0.03}$ & 1.60\,$^{+0.08}_{-0.02}$ & 1.21$^{+0.08}_{-0.10}$ & 1.28\,$\pm$\,0.01 \\
\multicolumn{2}{l}{\kbb\ {\small [(R$_{in}$/D$_{10}$)$^{2}$]}} & 310\,$^{+37}_{-154}$ & 74\,$^{+8}_{-35}$ & 81\,$^{+7}_{-31}$ & 113$^{+34}_{-38}$ & 272\,$^{+16}_{-21}$ \\
& {\tt gau$_1$}  & & & & & \\  
\multicolumn{2}{l}{ \egau\ {\small(keV)}} & 6.56\,$^{+0.10}_{-0.07}$ & 6.71\,$^{+0.12}_{-0.16}$  & 6.63\,$\pm$\,0.11 & 6.59\,$^{+0.07}_{-0.15}$  & 6.69\,$\pm$\,0.06 \\
\multicolumn{2}{l}{ $\sigma$ {\small(keV)}} & 0.88\,$^{+0.12p}_{-0.07}$ & 0.77$^{+0.23p}_{-0.15}$ & 0.72\,$^{+0.08}_{-0.12}$ & 0.78 $^{+0.17}_{-0.09}$ & 0.77\,$^{+0.06}_{-0.08}$ \\
\multicolumn{2}{l}{ \kgau\ {\small(10$^{-4}$ ph cm$^{-2}$ s$^{-1}$)}} & 0.023\,$\pm$\,0.006  & 0.008\,$^{+0.004}_{-0.002}$ & 0.007\,$^{+0.003}_{-0.001}$  & 0.014\,$^{+0.005}_{-0.003}$ & 0.017\,$\pm$\,0.003 \\
\ew\ (eV) & & 299\,$\pm$\,77 & 91\,$^{+48}_{-23}$ & 88\,$^{+38}_{-13}$ & 195\,$^{+70}_{-42}$ & 235\,$\pm$\,42 \\
& {\tt tbabs} & & &   \\
\multicolumn{2}{l}{\nhabs\ {\small($10^{22}$ cm$^{-2}$)}} & 2.86\,$\pm$\,0.05 & 2.86\,$\pm$\,0.02 & 2.92\,$\pm$\,0.02 & 2.77\,$\pm$\,0.05 & 2.87\,$^{+0.06}_{-0.04}$ \\
& {\tt warmabs} & & & &  & \\
\multicolumn{2}{l}{\nhwarmabs\ {\small($10^{22}$ cm$^{-2}$)}} & 17.2\,$^{+10.3}_{-6.8}$ & 6.6\,$^{+3.2}_{-2.7}$ & 6.6\,$^{+2.9}_{-0.4}$ & 6.0$^{+6.6}_{-2.0}$ & 7.1$^{+0.5}_{-2.3}$ \\
\logxi\ {\small(\xiunit)} &&  4.41\,$^{+0.11}_{-0.19}$& 4.39\,$\pm$\,0.17 & 4.34\,$^{+0.22}_{-0.08}$  & 3.97\,$^{+0.19}_{-0.06}$ & 4.23\,$\pm$\,0.10  \\
\sigmav\ {\small(km s$^{-1}$)} & & 580\,$^{+165}_{-75}$ & 3390\,$^{+1785}_{-2210}$ & 2130\,$^{+1220}_{-920}$ & 480\,$^{+490}_{-175}$ & 975\,$^{+970}_{-710}$ \\
$v$ {\small(km s$^{-1}$)} & & -2310\,$^{+420}_{-360}$ & -3690\,$^{+300}_{-600}$ & -2160\,$^{+270}_{-180}$ & -2130\,$\pm$\,480 & -2400\,$\pm$\,360  \\ 
\noalign {\smallskip}
\noalign {\smallskip}
\hline\noalign {\smallskip}
\multicolumn{2}{l}{\rchisq (d.o.f.)} & 1.11 (586) & 1.12 (701) & 1.04 (686) & 1.09 (548) & 1.23 (575) \\
\hline\noalign {\smallskip}
\hline\noalign {\smallskip}
\multicolumn{2}{l}{$F^{tot}_{2-10~keV}$ \small (10$^{-09}$ \ergcmsec)} & 9.6 & 9.5 & 8.7 & 8.0 & 8.3 \\
\multicolumn{2}{l}{$F^{gau}_{2-10~keV}$ \small (10$^{-11}$ \ergcmsec)} & 24.3 & 7.9 & 7.1 & 14.7 & 17.6 \\
\multicolumn{2}{l}{Exposure (ks)} &  5.2 & 8.9 & 13.5 & 6.5 & 6.7 \\
\noalign{\smallskip\hrule\smallskip}
\label{tab:bestfit}
\end{tabular}
\end{center}
\end{table*}

\subsubsection{Choice of continuum model}
\label{sec:continuum}

In previous section we showed that using the same ionising continuum for all the observations caused a significant change
in the value of \xil\ for obs~4 and 9 and smaller, non-significant, changes in the other observations. From this, we infer that at least
for obs~4 and 9 the continuum may not be unique. Continua that result in a similar quality of the fit in the 2--10~keV band may 
yield significantly different ionising continua when they are extrapolated to the  0.013--13.6~keV energy band.

There are several reasons for obtaining a ``non-realistic" continuum in the limited \xmm\ bandpass. First, different combinations of thermal and non-thermal components can be
used for spectral fitting and the quality of the data does not usually allow to discriminate among them. Second, if the
broad Fe line is caused by reprocessed emission, the ``reprocessed" continuum should be also properly accounted for. A last possibility is that changes in the neutral
and ionised absorber are mimicking a change of continuum, as it is the case in dipping sources \citep{1323:boirin05aa, ionabs:diaz06aa}. We examine these possibilities in detail below and
in next section.

We substituted
the continuum components of the fits reported in
Table~\ref{tab:bestfit} by alternative non-thermal components. We found that the quality 
of the fits did not change. As an example, obs~6 had a
\rchisq\ of 1.12 for 701 d.o.f. when fitted with Model~2. Substituting
the blackbody component by a power-law component in Model~2 ({\tt
tbabs*cabs*warmabs*(diskbb+po+gau)}, hereafter Model~3) we obtained \rchisq\ = 1.12 for 701 d.o.f.. The photon index, $\Gamma$,
and normalisation of the power-law were 2.2\,$\pm$\,0.3 and
1.1\,$\pm$\,0.4, respectively, and the temperature and
normalisation of the disc blackbody component 1.85\,$^{+0.03}_{-0.06}$~keV
and 46\,$^{+8}_{-4}$, respectively. Substituting the disc blackbody component
by a power-law component in Model~2 ({\tt
tbabs*cabs*warmabs*(bbodyrad+po+gau)}, hereafter Model~4) we obtained
\rchisq\ = 1.14 for 701 d.o.f.. The photon index, $\Gamma$, and
normalisation of the power-law were 2.27\,$\pm$\,0.04 and 
3.1\,$\pm$\,0.2, respectively, and
the temperature and normalisation of the blackbody component 
1.12\,$\pm$\,0.02~keV and 281\,$^{+4}_{-9}$, respectively. 
We did not find significant changes in the parameters of the narrow features
when using different continua. In contrast, 
the \ew\ of the Gaussian component
increased from  91\,$^{+48}_{-23}$~eV in Model~2 to 201\,$\pm$\,41
and  348\,$^{+58}_{-16}$~eV in Models~3 and 4,
respectively. 
Finally, substituting the power-law component in
Models~3 or 4 by a cutoff or broken power-law components
gave an unconstrained cutoff or break and a similar fit to the
power-law. 

We also obtained fits of the same quality when using different abundances. 
However, we obtained a change in the \ew\ and breadth of the Gaussian component. As an example, in obs~6 the breadth and \ew\ of  the 
broad emission line increased from 91\,$^{+48}_{-23}$ to 129\,$^{+46}_{-21}$~eV when we substituted the solar abundances of \citet{anders89} by those of \citet{wilms00apj} in Model~2.

In summary, we found that the parameters of the warm absorber were stable
to different choices of continuum components and abundances. In contrast, the 
parameters of the broad Fe line changed significantly, the main reason being
that the quality of the fit before introducing the broad Fe line was worse for 
Models~3 and 4 than for Model~2.

\subsubsection{Modeling of the broad Fe line and continuum variability}
\label{simultaneous_fits}

We examined the possibility that  spectral changes between different observations
are due to variations of the warm absorber and the reprocessed component instead of variations of
the continuum. A motivation for this is Fig.~\ref{fig:colours2}. The resemblance of this figure 
with Fig.~4 of \citet{1323:boirin05aa} indicates that obs~8--9 (7) could be 
 deep (shallow) dipping states with respect to obs~6 (see Sect.~\ref{sec:x-lc}). If this is true we expect 
 the continuum to be relatively stable and the dipping states to have a less ionised absorber and with a larger column density 
 compared to the persistent ones \citep{1323:boirin05aa,ionabs:diaz06aa}. Indeed, obs~8 and 9 show 
a less ionised absorber with visible features of \fetfive, compared to obs~6 and 7.
Similarly, obs~1-3 show a less ionised plasma than obs~4.

In order to test the continuum variability, we first substituted the Gaussian component by the {\tt reflionx} model \citep{ross05mnras}  to 
account for reflection by an ionised, optically thick, illuminated atmosphere of constant density. Since {\tt reflionx} only provides the reflected emission, we added also a power-law component to account for the direct emission ({\tt tbabs*cabs*warmabs*(diskbb+bbodyrad+po+reflionx)}, hereafter Model~5), and we coupled the photon index of the illuminating power-law component to the index of {\tt reflionx}.
If the line is produced by reprocessed emission, Model~5 will yield a more realistic continuum, necessary to perform a simultaneous fit to several observations.
We note that while {\tt reflionx} is the most adequate of the 
available public models for our purposes since it includes self-consistently
line emission and accounts for ionised reflection, more recent refined models
such as those presented in \citet{garcia10apj} point to important differences 
in the resultant spectrum when e.g. new atomic data are considered (see Sect.~\ref{subsec:broadline}).  

For the simultaneous fit we considered obs~6--9 only. We left out obs~4 since it is in a different track in Fig.~\ref{fig:colours2}, together with obs~1--3 and this may
indicate already significant changes in the continuum (a significant increase in Compton scattering due to a larger column density of plasma could be also the reason for the different track of obs~1--4). We imposed the same continuum for obs~6--9 (including index and normalisation of the power-law component) 
and allowed only changes in the neutral absorber and the ionisation degree and normalisation of the reflection component. We kept the parameters of the warm absorber 
fixed to the values shown in Table~\ref{tab:bestfit} but with the values of the ionisation parameter obtained when the ionising continuum of obs~6 was used to fit all observations (see Sect.~\ref{sec:model2}). Finally, we allowed a global difference in normalisation among observations to account for a potential different amount of Compton scattering. 

\begin{figure*}[!ht]
\includegraphics[angle=0,width=0.45\textwidth]{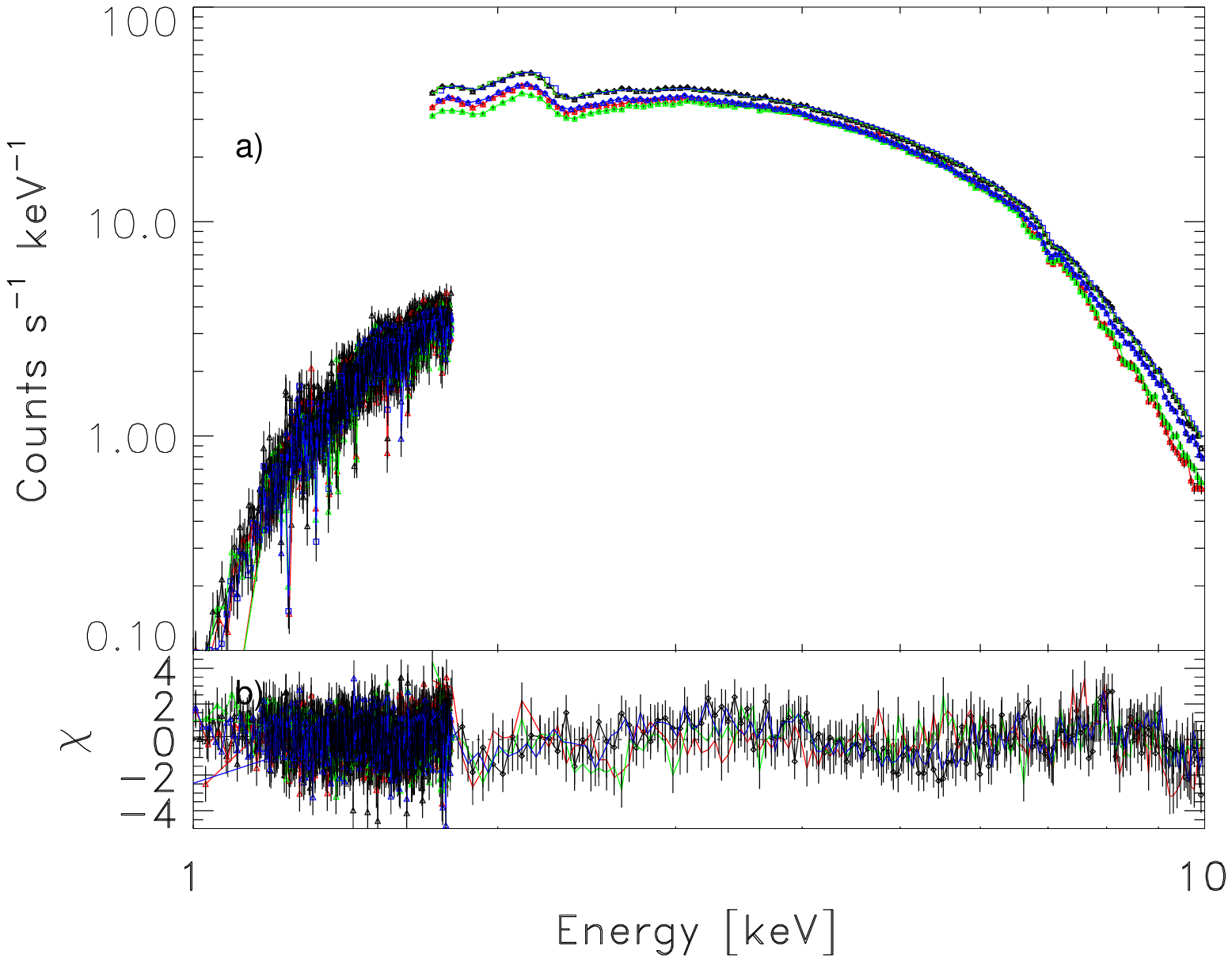} 
\includegraphics[angle=0,width=0.49\textwidth]{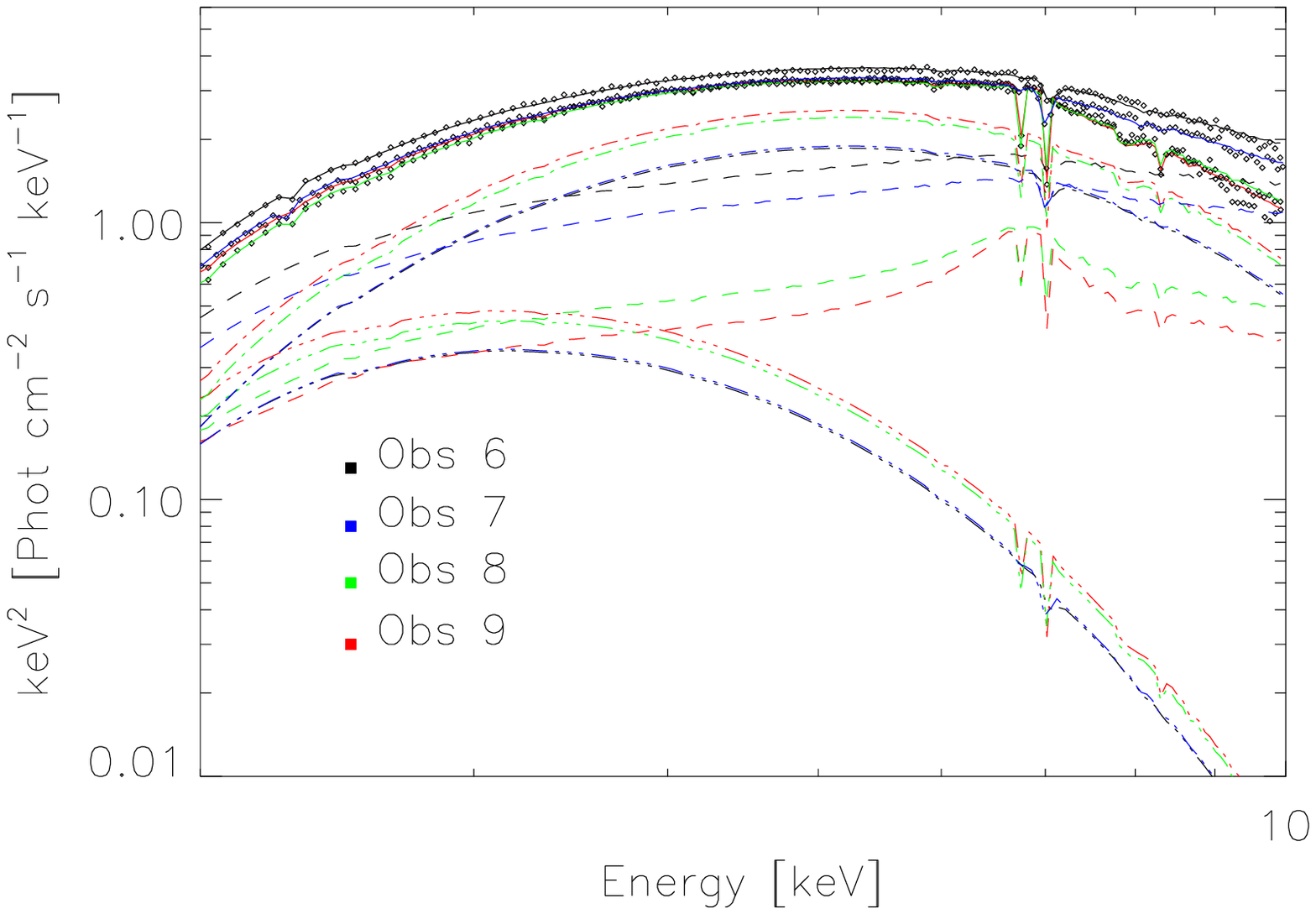}
\caption{{\it Left}: (a) 1.7--10 keV EPIC pn, and 0.8--1.8~keV RGS1
and RGS2 \src\ spectra fitted with Model~5 (see text). (b) Residuals in units of standard
deviation from the above model. {\it Right}: 1--10~keV unfolded spectra. Solid, dot-dot-dashed and dot-dashed and dashed lines represent
the contribution of the total model, disc, blackbody and reflection components, respectively. Obs~6--9 are shown in black, blue, green and red, respectively.}
\label{fig:6-9}
\end{figure*}

We obtained a good quality for the simultaneous fit of obs~6--9 (\rchisq\ = 1.21 for 2537 d.o.f.).
Table~\ref{tab:bestfit_sim} shows the best-fit parameters and the residuals of
the fit and unfolded spectra are shown in Fig.~\ref{fig:6-9}.

We could model the emission line with the reflection component and, as expected, the reprocessed 
component contributes significantly to the continuum emission. Thus, the interpretation of the continuum parameters should
be done with caution. Interestingly, the large difference in the curvature of the spectrum at 7--10~keV between obs~6--7 and 8--9
can be well reproduced by a difference in the ionisation and normalisation of the reflected component. 
In contrast, we were not able to reproduce the broad iron line with reflection models which consider 
a blackbody as an incident spectrum since they predict \ews\ that are too small compared to the observed lines.

Despite the success of Model~5 to reproduce the various spectral features of obs~6--9 while assuming a constant continuum, 
there are still some systematic residuals in the best-fit. The residuals at the
absorption edge at the energy of \fetsix\ in obs~9 were already present in individual fits (see Sect.~\ref{sec:model2}). However, the simultaneous fits show a slight curvature  in the residuals which is not present in the individual fits and could be due to shortcomings of the reflection component, a different origin for the emission line, or an indication for continuum variability.
Finally, if we attribute the differences in the global normalisation constant to Compton scattering of fully ionised plasma, the $\sim$30\%
difference in normalisation between obs~6--7 and 8--9 implies that obs~6--7 are observed through an extra column density of $\approxgt$
10$^{23}$~cm$^{-2}$ of fully ionised plasma.

\begin{table*}
\begin{center}
\caption[]{
Best-fits to the 0.8--1.8~keV RGS and 1.7--10~keV EPIC pn
  persistent spectra for all the observations using the {\tt
  const*tbabs*cabs*warmabs*(diskbb+bbrad+po+reflionx) model} (see caption of Table~\ref{tab:bestfit} for definitions). $\Gamma$ is the index of the incident power-law and is coupled to the index of {\tt reflionx}. $k_{po}$ is the normalisation of the power-law component and \xil\ and $k_{reflionx}$ are the ionisation parameter and the normalisation of the reflection component. The parameters of {\tt warmabs} and {\tt cabs} have been fixed to the values of Table~\ref{tab:bestfit}. }
\begin{tabular}{lccccc}
\hline \hline\noalign{\smallskip}
Observation No. & & 6 & 7 & 8 & 9 \\
\noalign{\smallskip\hrule\smallskip}
& Comp. & & & &    \\
Parameter & & & & & \\
& {\tt dbb} & & & & \\
\ktdbb\ {\small(keV)} & &  \multicolumn{4}{c}{0.86\,$^{+0.08}_{-0.06}$}  \\
\multicolumn{2}{l}{\kdbb\ {\small[(R$_{in}$/D$_{10}$)$^{2}$ cos$\theta$]}} & \multicolumn{4}{c}{189\,$^{+27}_{-55}$}  \\
& {\tt bb} & & &  &  \\
\ktbb\ {\small(keV)} &  & \multicolumn{4}{c}{1.23\,$^{+0.01}_{-0.04}$} \\
\multicolumn{2}{l}{\kbb\ {\small [(R$_{in}$/D$_{10}$)$^{2}$]}} & \multicolumn{4}{c}{199\,$^{+25}_{-13}$}  \\
& {\tt po} & && & \\
$\Gamma$ & & \multicolumn{4}{c}{1.94\,$\pm$\,0.02} \\
$k_{po}$ {\small (ph keV$^{-1}$cm$^{-2}$s$^{-1}$ at 1 keV)} & & \multicolumn{4}{c}{$<$\,0.006} \\
& {\tt reflionx} &&&\\
\logxi\ {\small(\xiunit)} & & 3.94\,$^{+0.06}_{-0.17}$ & 3.83\,$^{+0.05}_{-0.08}$ & 3.41\,$\pm$\,0.04 &  3.30\,$^{+0.01}_{-0.04}$ \\
$k_{reflionx}$ (10$^{-6}$) & & 3.1\,$^{+0.6}_{-0.3}$ & 3.2\,$^{+0.5}_{-0.3}$ & 3.8\,$^{+0.5}_{-0.3}$ & 4.6\,$^{+0.4}_{-0.2}$ \\
& {\tt tbabs} & & &   \\
\multicolumn{2}{l}{\nhabs\ {\small($10^{22}$ cm$^{-2}$)}} & 3.13\,$\pm$\,0.03 & 3.17\,$\pm$\,0.03 & 3.17\,$\pm$\,0.03 & 2.94\,$\pm$\,0.03 \\
& {\tt const} & & & \\
$Const$ & & 1 (f) & 1.02\,$\pm$\,0.03 & 1.29\,$\pm$\,0.03 & 1.37\,$^{+0.02}_{-0.05}$ \\
\noalign {\smallskip}
\noalign {\smallskip}
\hline\noalign {\smallskip}
\multicolumn{2}{l}{\rchisq (d.o.f.)} & \multicolumn{4}{c}{1.21 (2537)} \\
\noalign{\smallskip\hrule\smallskip}
\label{tab:bestfit_sim}
\end{tabular}
\end{center}
\end{table*}

\subsection{RGS spectral analysis}
\label{subsec:RGS}

We examined the 0.8--1.8~keV (9.9--22.3~\ang) first order
RGS spectra to constrain the \nh\ in the direction of the source and to
search for the signature of narrow absorption and
emission features.  We note that the second order spectra have
too few counts to add any value to the fits.

We could fit the RGS spectra of all the
observations with a continuum consisting of a disc blackbody modified by
photo-electric absorption from neutral material. We obtained a C-statistic of between 
450 and 571 for $\sim$434--436 d.o.f. for obs~4 to 9). The values of \nh\ in units of 10$^{22}$\,cm$^{-2}$ were 2.72\,$\pm$\,0.06, 2.73\,$\pm$\,0.05, 2.80\,$^{+0.07}_{-0.03}$, 2.80\,$\pm$\,0.06, 2.76\,$\pm$\,0.07 and 2.75\,$^{+0.07}_{-0.05}$ for obs~4 to 9, respectively.
None of the observations showed significant narrow features. 

To evaluate the effect of the chosen continuum in the value of \nh, we substituted the disc blackbody by a power law. We obtained a C-statistic of between 
448 and 570 for $\sim$434--436 d.o.f. for obs~4 to 9. The values of \nh\ in units of 10$^{22}$\,cm$^{-2}$ were 2.53\,$\pm$\,0.25, 2.53\,$\pm$\,0.19, 2.80\,$\pm$\,0.19, 2.71\,$\pm$0.20, 2.41\,$\pm$0.26 and 2.16\,$^{+0.22}_{-0.16}$  for obs~4 to 9, respectively.

Finally, to evaluate the effect of the absorption component in our fit, we substituted the {\tt tbabs} component by the more recent {\tt tbnew} component and re-fitted the combined spectra. The C-statistic and values of \nh\ did not change. 

In summary, we obtained consistent values of \nh\ for different models of interstellar absorption but significantly different values of \nh\ when using different continuum components. This is expected due to the combination of limited energy band of the RGS and low statistics of these observations. 
Therefore, we conclude that the RGS spectra alone cannot constrain the value of \nh\ for these observations due to the poor statistics and the limited bandwidth, necessary to determine the continuum.

In order to increase the sensitivity to narrow lines and edges, we combined the RGS spectra of several observations. We added the RGS1 and RGS2 spectra of obs~4, 8 and 9 (set 1) and separately of obs~5, 6 and 7 (set 2). The reason to distinguish between sets is that the former set of observations shows possible ``obscuration" events (see Sect.~\ref{sec:spectra}) and, if confirmed, such events indicate a plasma with lower ionisation degree than the second set of observations. We could fit the resultant spectra with the same model as the individual observations. We obtained a C-statistic(d.o.f.) of 318(280) and 390(280) and \nh=2.74\,$\pm$\,0.04 and 2.78\,$^{+0.02}_{-0.04}$\,$\times$10$^{22}$\,cm$^{-2}$ for sets 1 and 2, respectively, when using a disc blackbody continuum and of 292(280) and 385(280) and \nh=2.33\,$\pm$\,0.15 and 2.67\,$\pm$\,0.11\,$\times$10$^{22}$\,cm$^{-2}$  when using a power law.

None of the merged data sets shows any additional narrow features and
including absorption from a photo-ionised plasma in the model does not
improve the fit quality. We calculated upper limits for the existence of the \mgtwelve\ absorption feature 
detected in the $Chandra$ observation with an \ew\ of 2~eV \citep{gx13:ueda04apj}. Assuming that the feature is
narrow (width fixed to 0), we obtained upper limits to the \ew\ between 1 and 3~eV for obs~4--8 for a feature with no velocity shift. For obs~9 we obtain a non-significant
detection of an absorption feature with \ew\,=\,2.5\,$^{+2.1}_{-2.2}$~eV. Then, we assumed velocity blueshifts up to 4000~km/s in steps of 500 km/s and calculated upper limits at the corresponding
wavelengths. We obtained upper limits to the \ew\ between 0.5 and 4.2~eV for obs~4, 7--9. For obs~6 we obtain a non-significant
detection of an absorption feature at a blueshift of 2000~km/s with an \ew\ = 2.0$^{+1.5}_{-1.6}$~eV. 
These results are consistent with the predicted \ew\ for  \mgtwelve, $\approxlt$1~eV,
for an ionised absorber of log \xil $\approxgt$\,4, as fitted in
the EPIC pn exposures.

\subsection{OM data}
\label{sec:om}

We extracted images and light curves from the U, UVW1 and UVM2 filter OM exposures for obs~4--9.

The images show a source in the U filter exposures consistent with the position of \src, which is strongest in obs~4 and weakest in obs~7. The source was not detected in the UVW1 or UVM2 filters, except for obs~5 where there is a detection with the UVW1 filter with a magnitude of 20.7\,$\pm$\,0.7. For obs~4, 5, 7 and 8 we obtained average U optical magnitudes of 19.8\,$\pm$\,0.4, 19.7\,$\pm$\,0.2, 20.8\,$\pm$\,0.7 and 20.5\,$\pm$\,0.5, respectively. We note that \src\ is not detected in the U exposures in obs~6 and 9.

Obs~4 showed a significant variation in the U magnitude with time, from 20.4\,$\pm$\,0.4 to 19.2$\pm$\,0.1 within a 9~ks interval. Unfortunately, we cannot determine the presence of a modulation with orbital phase due to the few detections of the source. Therefore we do not use the OM data any further.

\section{Discussion}
\label{sec:discussion}

We analysed five \xmm\ observations of the LMXB \src\ taken during 2008 to 
investigate the variability and origin of the disc wind present in this source. 

The 0.6--10~keV EPIC pn lightcurves were highly variable with two
observations showing energy-dependent obscuration, similar to the classical dipping sources.

The X-ray continua of the ``least variable" intervals were well fitted by a model consisting
of a blackbody and disc-blackbody components absorbed by neutral and
ionised material. We could substitute the blackbody or disc-blackbody components by
a non-thermal component such as a (cutoff) power-law or a Comptonisation component without 
significantly worsening the quality of the fit.  However, we found a significant change in the parameters of the broad emission line
when such non-thermal components were used (see Sect.~\ref{subsec:pn-RGS} and \ref{subsec:broadline}). 

\citet{1701:lin07apj} showed that the
commonly used models for thermal emission plus Comptonisation were not
favoured for the ``Soft State" of two atoll sources because they failed to satisfy various desirability
criteria, such as the L$_X$ $\propto$ T$^4$ evolution for the multi-colour disc blackbody component and similarity to black holes for correlated timing/spectral behaviour. Similarly, in a systematic analysis of 16 NS LMXBs observed with \xmm, \citet{ng10aa} found that  a continuum consisting of a disc blackbody and blackbody components was favoured in 80\% of the cases. In this model, the blackbody component was interpreted as emission from a radiation pressure supported boundary layer.
The data analysed in this work give further support to a continuum with two thermal components.
However, we note that since  \xmm\ effective 
area is already very low at $\sim$10~keV, we
cannot decide based on this data whether an additional power-law component extending to high energies, 
like the one detected 
by INTEGRAL up to $\sim$40~keV \citep{Paizis2006}, would be requested above 10~keV. If confirmed, the presence of a reprocessed component  representing reflection by an ionised, optically thick, atmosphere illuminated by a power-law component (see Sect.~\ref{simultaneous_fits}) gives indirect evidence of the existence of a power-law component extending to high energies. 

\citet{gx13:ueda04apj} used a blackbody and disc blackbody to fit the 3--25 keV RXTE/PCA spectrum of \src\
and obtained a satisfactory fit but the values for the temperatures of the disc and blackbody components were significantly different than those reported in Table~\ref{tab:bestfit}. Taking into account the flux and the ionisation of the disc wind, the RXTE/PCA observation should be in a similar state to obs~7. However,  \citet{gx13:ueda04apj} report a temperature of 1.52\,$\pm$\,0.03~keV and of 2.56\,$\pm$\,0.02~keV for the disc and blackbody components, much larger than the ones found in this work for obs~7. Clearly, the absolute parameters of the continuum should be taken with caution in fits with limited energy band coverage. 

\subsection{Photoionised absorption}
\label{subsec:discwind}

Absorption from highly-ionised species was present in all the observations of \src\ analysed 
in this work. 

An absorption feature consistent with \fetsix\ was first discovered by
ASCA \citep{gx13:ueda01apjl}. \xmm\ observations in 2000 (obs~1-3)
showed a complex of absorption lines and edges consistent with
absorption from \catwenty, \fetfive\ and \fetsix\
\citep{gx13:sidoli02aa}, similar to the complex found in obs~4, 8 and
9. \citet{gx13:ueda04apj} found additional features of \ssixteen,
\sifourteen, \mgtwelve, \mntfive, \crtfour\ 
and \arseventeen\ in a posterior {\it
Chandra} HETGS observation in 2004 and measured a blueshift of $\sim$400~km 
and an ionisation parameter of \logxi\ $\sim$ 4.1--4.7 for the
plasma responsible for the absorption. 
A
plasma density $\approxgt$10$^{13}$~cm$^{-3}$ and a launching radius for the
wind of $\sim$10$^{10}$-10$^{11}$~cm was inferred.

The depth of the absorption features in the \xmm\ 2008 observations changed significantly in timescales of few days, and more subtle variations are seen in shorter, hours, timescales \citep{gx13:sidoli02aa}.

\begin{figure}[!ht]
\includegraphics[angle=0,width=0.48\textwidth]{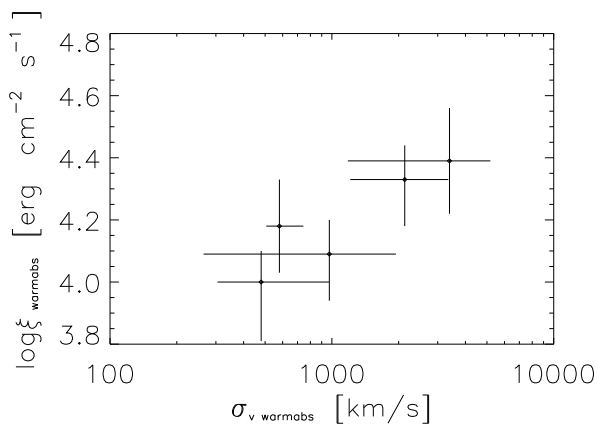}
\caption{Ionisation parameter of the warm absorber with respect to its turbulent velocity.}
\label{fig:warm}
\end{figure}

We found column densities for the absorber between 6 and 17\,$\times$10$^{22}$ cm$^{-2}$ and
ionisation parameters, \logxi, between 4 and 4.4 \xiunit.  
Obs~4 shows a column density larger by more than a factor of two with respect to the other observations, and which is poorly constrained.
This, and the well constrained velocity indicate that the lines lie in the saturated region of the curve of growth, where the \ews\ of the lines do
not depend on the column density anymore but increase with the velocity broadening.
We observed a correlation between the ionisation parameter and the turbulent velocity of the warm absorber (see Fig.~\ref{fig:warm}). 

We found blueshifts between $\sim$2100 and 3700~km s$^{-1}$ for obs~4--9.
Such blueshifts seem too large, compared to the shifts of
$\sim$400~\kms\ obtained by \citet{gx13:ueda04apj} for \src\ and of
$\sim$300--1600~km~s$^{-1}$ in microquasars \citep[e.g.][see however \citet{igr:miller11apj}
for a recently reported blueshift of 3100\,$\pm$\,400 km/s in a $Chandra$ HETGS observation of IGR~J17480--2446]{1655:kallman09apj,1915:ueda09apj}.
Unfortunately, we do not detect any
discrete absorption features from the warm absorber in the RGS, which
has a much better energy resolution than EPIC. The lack of spectral
features in the RGS wavelength range is consistent with the high
degree of ionisation of the absorber (see Sect.~\ref{subsec:RGS}). Thus, a more accurate
measurement of the blueshifts is not possible for these
observations. However, it seems plausible that the shifts detected with the pn
camera are overestimated compared to the real blueshifts due to
residual calibration uncertainties (see Sect.~\ref{model1}).

We can calculate  the distance between the ionising source and the
slab, $r$. Since $\xi = L /n_{\rm e} \, r^{2}$ and $n_{\rm
e}\sim$ \nhwarmabs/$d$ (where $d$ represents the thickness of the slab of ionised absorbing
material), we can calculate $r$ as
($L$/\xil~\nhwarmabs)$(d/r)$. Considering plausible values of $d/r$ range between 0.1 and 1,
we obtain values of $r$ between 3$\times$10$^9$~cm and 1$\times$10$^{11}$~cm. Further, assuming that
the value of $r$ should not change significantly between observations, we estimate $r$ as 1--3\,$\times$\,10$^{10}$~cm, in excellent 
agreement with the value obtained by \citet{gx13:ueda04apj} of 2\,$\times$\,10$^{10}$~cm.

The characteristics of the wind allow us to discriminate among the possible launching mechanisms, namely thermal, radiative,
magnetic or a combination of them. Thermal driving is effective at large distances 
from the central compact object, black hole or NS,
where the thermal velocity exceeds the local escape velocity. Thermally driven models
are thus most effective in producing slow winds and at large radii. The distance at which the wind is launched in \src, 
1--3\,$\times$\,10$^{10}$~cm, points to a thermal origin, as discussed by \citet{gx13:ueda04apj}.
Moreover, the fact that the only known NS LMXBs showing winds are \src\ \citep{gx13:ueda04apj} and  IGR~J17480--2446 \citep{igr:miller11apj}
is a logical consequence of the wind launching mechanism being thermal. As X-rays heat low
density gas to a temperature T\,$\sim$\,10$^7$~K, the surface of a
heated disc is expected to either puff up and form a static corona or
emit a thermal wind, depending whether the thermal velocity exceeds
the local escape velocity \citep{begelman83apj,woods96apj,proga02apj}. \citet{woods96apj} estimated that
for a radius of $\sim$2.6\,$\times$\,10$^{10}$~cm, an isothermal wind would develop for
luminosities above $\sim$\,3$\times$10$^{37}$~erg s$^{-1}$, while such a wind
would be inhibited by gravity at lower luminosities (see their
Fig. 17). For larger radii, a steadily heated free wind could develop already at luminosities below $\sim$\,3$\times$10$^{37}$ erg s$^{-1}$. \citet{proga02apj} found that including the radiation force due to electron scattering, the radius at which the maximum of mass flux density occurred could be reduced by a factor of 20 compared to the calculations of  \citet{woods96apj}. Taking these values into account, the luminosity, and not the radius, seems to be the critical parameter for the existence of a wind in the classical dipping sources. Indeed, all the dipping sources
showing ionised absorption for which no significant blueshifts have been detected \citep{0748:cottam01aa,1658:sidoli01aa,1916:juett06apj,1323:boirin05aa,1254:iaria07aa}  have a luminosity below the one needed for the wind to overcome gravity in these models \citep[see Table~1 of][]{ionabs:diaz06aa}, with the possible exception of \mxb, which has a luminosity close to the critical value. In contrast, \bigdip, \src\ and IGR~J17480--2446 have luminosities of 4.7\,$\times$\,10$^{37}$, 6--9\,$\times$\,10$^{37}$ and 3.7\,$\times$\,10$^{37}$~erg/s, respectively, close to or above the critical luminosity. From these sources, two show clearly outflowing winds (see above) and the third one, \bigdip, shows  a significant blueshift of 607\,$^{+354}_{-342}$~km/s for the most highly ionised absorbing component in $Chandra$ HETGS observations \citep{1624:xiang09apj}. In summary, a thermal wind launching mechanism can explain the appearance of hot static atmospheres or slow disc winds in all high inclination NS LMXBs and its absence in low inclination NS LMXBs.
We caution that Compton scattering in a hot atmosphere or disc wind with column densities above 10$^{22}$~cm$^{-2}$ can reduce significantly the flux of the central source seen by a distant observer. Therefore, a potential larger luminosity than ``observed" (as the luminosity of the system could be significantly underestimated due to Compton scattering) and the radiation force due to electron scattering when a large column density of highly ionised plasma is present, should be always considered before ruling out the thermal wind launching mechanism.

Finally, if the thermal wind launching mechanism is in place for all NS LMXBs, we expect that a disc wind will develop for systems brighter than \src, known
as Z-sources , since their luminosity is well above the critical luminosity. \citet{gx13:ueda04apj} estimated the mass outflow rate in \src\ to be $\approxgt$\,0.7--2.7\,$\times$\,10$^{18}$ g s$^{-1}$. They assumed that the plasma had an equatorial geometry and was outflowing from both sides of the disc to define a solid angle of $\Omega/4\pi$\,=\,0.4. In our observations, taking a similar solid angle into account we calculate outflow rates of 0.1--0.2\,$\times$\,10$^{18}$ g s$^{-1}$ for an outflow velocity of 400 km s$^{-1}$, as found by  \citet{gx13:ueda04apj}, and of 0.5--1\,$\times$\,10$^{18}$ g s$^{-1}$ for an outflow velocity of 2000-3500 km s$^{-1}$, as found in these observations. This outflow rate is comparable with the mass accretion rate of 10$^{18}$ g s$^{-1}$, as already pointed out by \citet{gx13:ueda04apj}, and implies that  
the disc wind could play an important role in the whole dynamics of the accretion disc for such bright sources. 
 \citet{proga02apj} calculated the mass-outflow rate for a NS accreting at a rate of 1.3\,$\times$\,10$^{18}$~g/s and found mass-outflow rates between 3.9\,$\times$\,10$^{15}$ and 0.65\,$\times$\,10$^{18}$~g/s for different values of X-ray attenuation (see their Table~1). Unfortunately, from their simulations we cannot estimate what is the effect of increasing the mass accretion rate in the mass-outflow rate for a given value of X-ray attenuation. However, the fact that for the same attenuation, an increase in accretion rate results in an increase of column density of the ionised plasma (see e.g. models L2 and H2 in their Table~1) may indicate that the mass-loss rate will be larger for sources with higher accretion rate, since the radiation pressure due to electron scattering will also increase for such sources. 

Although narrow absorption features are not detected in the Z-sources, most likely due to a lower inclination or to fully ionised winds, if we consider other signatures of a disc wind such as broad iron line emission (see Sect.~\ref{subsec:self}) a study of the strength of the wind in such sources may become possible and of utmost importance to evaluate the balance between inflowing and outflowing matter in such systems.

\subsection{Broad iron line emission}
\label{subsec:broadline}

We found a broad, $\sigma\sim$\,0.7--0.9~keV, emission line in all the observations analysed in this work.
The energy centroid varied between 6.55 and 6.7~keV, consistent with emission of highly ionised species 
of iron (\fetone--\fetfive). The \ew\ of the line increased from $\sim$\,100~eV in obs~6 and 7 to $\sim$300~eV in obs~4.

The absolute breadth and \ew\ of the lines was strongly dependent on the continuum model.
As an example, the \ew\ of the line in obs~6 increased from 91\,$^{+48}_{-23}$ to 201\,$\pm$\,41 (348\,$^{+58}_{-16}$)~eV when we substituted 
the disc blackbody (blackbody) component by a power-law component in Model~2 (see Sect.~\ref{sec:model2}).  
For this highly absorbed source we also found a dependency of the breadth and \ew\ of the line on 
the abundances used, which caused a significant change in the continuum parameters, 
especially in the soft band. For obs~6, we obtained an \ew\ of 91\,$^{+48}_{-23}$~eV using solar abundances from \citet{anders89} and of 129\,$^{+46}_{-21}$~eV with abundances from \citet{wilms00apj}.
In general, continuum models that yielded a higher \rchisq\ for the fit before adding the line resulted in broader lines, since the Gaussian component
absorbed the deficiencies of the continuum model. 
Hence, it is challenging to investigate the origin of the broad iron line 
based on its absolute parameters. In contrast, we can examine the relative changes of the line 
among observations, since any systematic effect caused by our choice of continuum, abundances
or spectral analysis should be the same for all observations. Further, since we are comparing observations of one source and taken with the
same instrument, residual calibration effects or properties of the source such as inclination should not affect our conclusions. 

The presence of broad Fe lines in NS LMXBs has been extensively studied \citep[see][for systematic studies with different X-ray observatories]{white86mnras,hirano87pasj,asai00apjs,ng10aa}.
In these systems, broad lines could arise in the inner accretion disc by fluorescence following illumination by an external source of X-rays (e.g. the boundary layer where the accretion disc meets the star), and be broadened by relativistic effects near the compact object \citep[e.g.][]{reynolds03ps, fabian05, matt06an}. Alternatively, they could originate in the inner
part of the so-called accretion disc corona, formed by
evaporation of the outer layers of the disc illuminated by the
emission of the central object \citep[e.g.,][]{white82apj, kallman89apj} and be broadened predominantly by Compton scattering \citep[e.g.][]{pozdnyakov79aa, sunyaev80aa}. A third possibility, is that they originate in a partially ionised wind as a 
result of illumination by the central source continuum photons and are
broadened by electron downscattering in the wind environment \citep{laurent07apj}. The first interpretation differs from the other two in that Compton scattering does not suffice to broaden the lines to the observed values, hence relativistic broadening has to be invoked. Therefore, to discriminate between the different interpretations it is key to determine the characteristics of the plasma where the lines originate, mainly its temperature and state of ionisation, by comparison with detailed reflection models. 

Recently, \citet{garcia10apj} presented new models for illuminated accretion discs and their structure  implementing state-of-the-art atomic data for the isonuclear sequences of iron and oxygen and increasing the energy, spatial and angular resolution compared to previous works. 
They found \ews\ of the Fe K emission line between 400 and 800 eV for the cases with \logxi\ between 1 and 3. The line \ews\ decreased rapidly for higher values of \xil, with \ew\ $\sim$ 40 eV for \logxi\,= 3.8. The \ews\ of the lines found in this work range between $\sim$90 and $\sim$300~eV, indicating a plasma with \logxi\ between $\sim$3.7 and 3\footnote[6]{We note that these values of \logxi\ cannot be directly compared with those of the warm absorber, since the models shown in \citet{garcia10apj} have an ionising continuum consistent of a power law with index 2.} \citep[see also Fig.~2 of][]{garcia11apj}, in broad agreement with the values of \logxi\ $\sim$ 3.3--3.9 found for the reflection component used to model the emission line in Sect.~\ref{simultaneous_fits}. For these values of the ionisation parameter \citet{garcia11apj} found that the most prominent lines had centroids between 6.65 and 6.74~keV, slightly larger than the values found in this work, 6.56--6.71~keV, but still consistent within the errors. 

In summary, we found that the lines observed in this work can be explained by ionised reflection of the disc. In this model, the breadth of the lines results primarily from the Compton scattering in the hot atmosphere or corona, and relativistic effects are not required as a broadening mechanism. 
 
However, the contemporaneous evolution of warm absorption and broad Fe line emission for the observations analysed in this work indicates that they probably form in the same environment, namely the disc wind (see sect.~\ref{subsec:self}). As pointed out by \citet{sim10mnras}, in the presence of a disc wind we expect a scattered/reprocessed component that is qualitatively similar to that produced by standard disc reflection models. In particular, a moderately strong Fe line is present in both models.  In contrast, it is difficult to justify the contemporaneous changes of the broad iron line and the absorption features if the former is produced at the inner disc or the boundary layer, as expected in the classical reflection scenarios, and the latter at the outer disc, as shown in Sect.~\ref{subsec:discwind}.  We note that strong broad Fe lines were predicted for LMXBs which had a wind with a Thomson scattering optical depth close to unity far from the compact object and which is illuminated by a hard component \citep{laming04apjl}. In that model, the optical depth in the Fe K continuum is about 1-3 times higher than that due to electron scattering. The wind found in obs~4, for which the Fe line is strongest, has a column density of 2\,$\times$\,10$^{23}$~cm$^{-2}$, but this may be only a lower limit to the real column density. The fact that the absorption lines are saturated and that the absorption edges are not properly fitted points to a failure of the {\tt warmabs} model when the optically thick limit is approached. Therefore the conditions necessary for a wind origin of the broad line are most likely fulfilled.

Finally, while we cannot assure that a hard illuminating component, as the one assumed by \citet{laming04apjl}, is present in this data due to the limited energy coverage of \xmm, if such component is absent, disc reflection would not be an alternative mechanism, since strong lines cannot be produced with a soft illuminating component \citep[see Fig.~3 of][]{ballantyne04mnras}.
 
Therefore, we favour reprocessing in a hot atmosphere or disc wind as the origin for the broad Fe line in NS LMXBs and justify this in detail in next section.
 
\subsection{A disc wind as a self-consistent scenario for photoionised absorption and broad iron line emission}
\label{subsec:self}

\begin{figure*}[!ht]
\includegraphics[angle=0,width=0.48\textwidth]{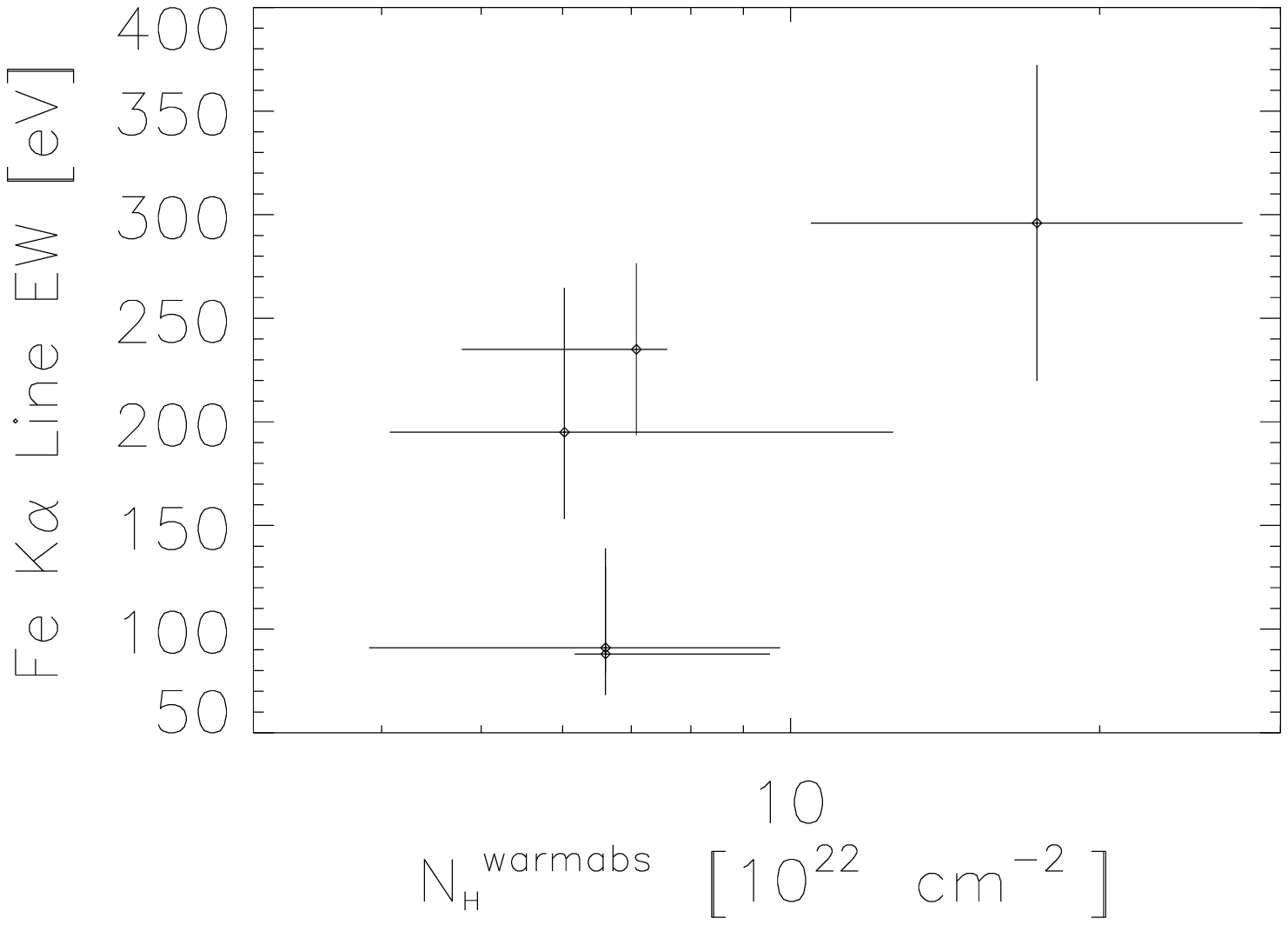}
\includegraphics[angle=0,width=0.48\textwidth]{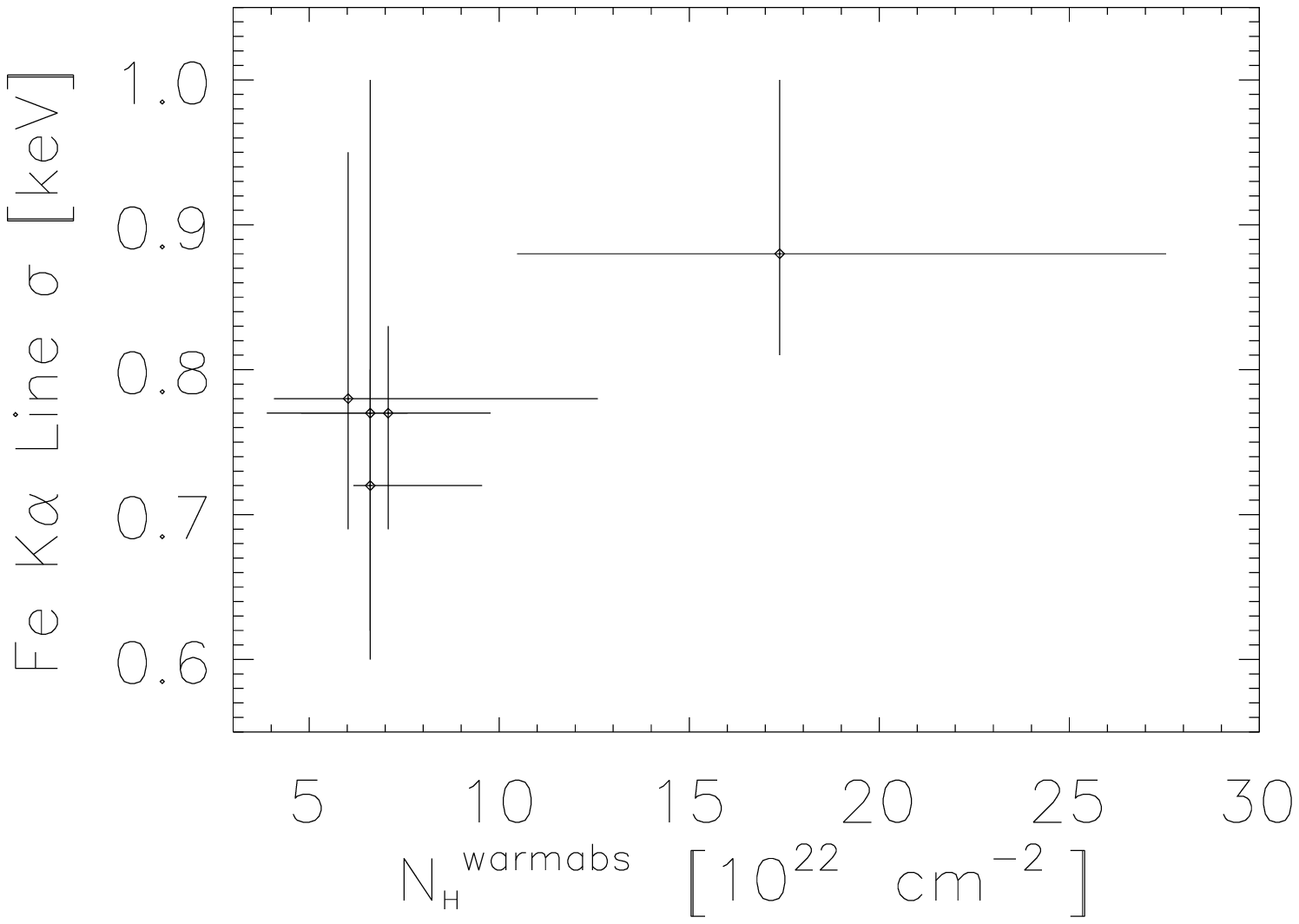} 
\includegraphics[angle=0,width=0.48\textwidth]{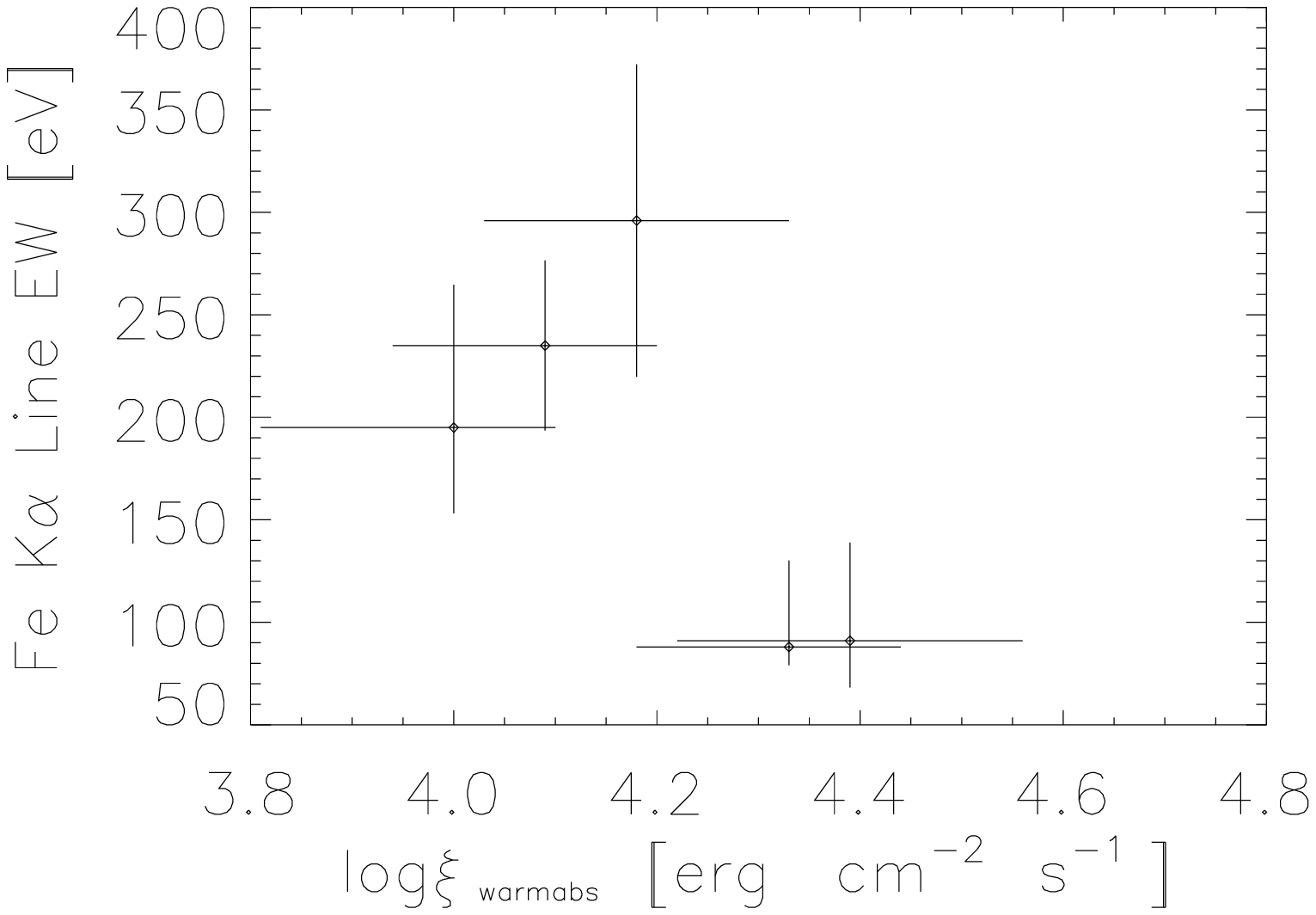}
\hspace{0.45cm}
\includegraphics[angle=0,width=0.48\textwidth]{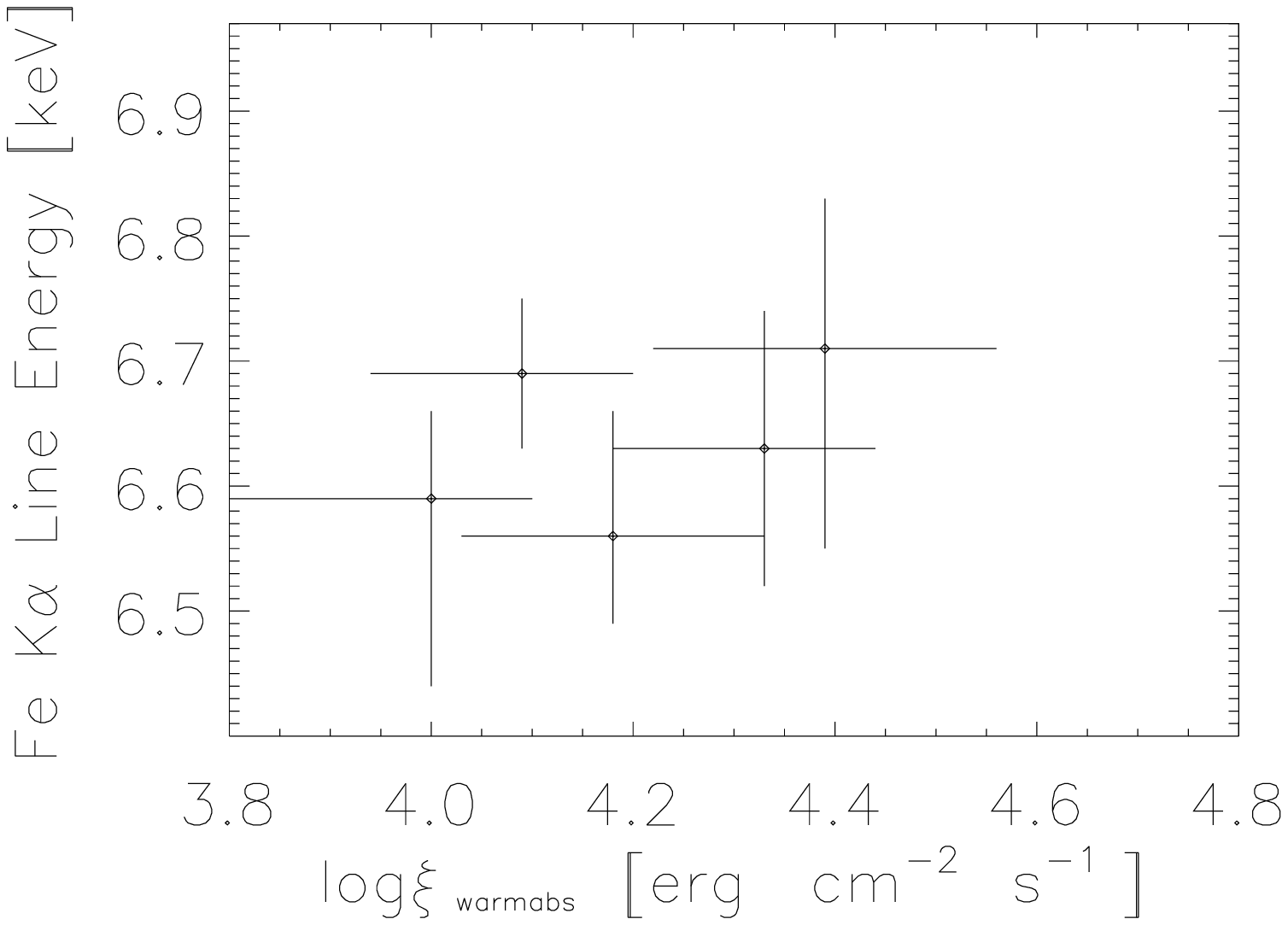}

\caption{{\it Upper panels}: \ew\ (upper-left) and width (upper-right) of the broad iron line with respect to the column density of the warm absorber. {\it Lower panels}: \ew\ (lower-left) and energy (lower-right) of the broad iron line with respect to the ionisation of the warm absorber.}
\label{fig:wind}
\end{figure*}

\citet{1915:neilsen09nat} reported the presence of a broad emission line in the faint, hard states and narrow absorption lines in the bright, soft states of the microquasar \grs. 
Since the hard states exhibit prominent radio jets, which are absent during the soft states, they argued that the broad emission line arises when the jet illuminates the inner accretion disc and the absorption lines originate when the powerful radiation field around the black hole drives a hot wind off the accretion disc. 
Since \src\ shows also jets, which are associated with hard states \citep{Grindlay1986}, we could expect a similar behaviour regarding the appearance of broad emission lines and narrow absorption lines as for \grs. 

In contrast, we observe contemporaneously broad emission line  and narrow absorption lines similarly to most of the classical dippers, and the emission line becomes stronger as the hard flux decreases. Moreover, we find that the \ew\ and the width of the broad Fe line increase with the column density of the absorber (see Fig.~\ref{fig:wind}, upper panels), a decrease of the \ew\ of the broad line with the increase of the ionisation parameter of the absorber (Fig.~\ref{fig:wind}, lower-left panel) and a simultaneous increase in the ionisation of the absorber and the energy centroid of the broad line (Fig.~\ref{fig:wind}, lower-right panel) .
We can naturally explain all these correlations in the frame of a disc wind scenario in which both the absorption and the emission are produced in the wind \citep[][]{laurent07apj,sim10mnras}. In this scenario, we observe absorption features as we look through the wind (predominantly in high inclination sources), while the radiation scattered in the outflow produces the broad iron line emission and other distinct features such as a Compton hump, which will be more or less visible depending on their contribution with respect to the incident radiation. Higher column densities make scattered/reprocessed radiation more dominant and so we observe a most prominent iron line whenever the column density of the absorbing plasma is larger and the ionisation smaller. 
We refer to \citet{sim10mnras,sim10mnras2} for a detailed explanation of the total spectra expected at different inclinations and for different characteristics of the wind and summarise the relevant results in what follows. 

At low, $\approxlt$45\deg, inclinations the wind does not obscure the X-ray source. However, a significant component of scattered/reprocessed radiation in the outflow is present in addition to the direct emission and responsible for the broad iron line emission. The emission is predominantly formed by \fetfive\ and \fetsix, since the part of the wind seen at low inclinations is the inner, highly ionised, surface of the outflow. 
\citet{sim10mnras2} report emission features that are moderately strong, $\sim$140~eV in their example, and which are significantly broadened by the Doppler motions of the material to a width of $\sigma$\,$\sim$\,0.3~keV. This is consistent with what \citet{ng10aa} report in a systematic study of low-inclination (non-dipping) NS LMXBs with \xmm. 
First, they report an emission line for 80\% of the sources, i.e. the line is almost ubiquitous. Second, they report a weighted average energy for the sample of 6.67\,$\pm$\,0.02~keV, consistent with \fetfive, and a width of $\sigma$\,=\,0.33\,$\pm$\,0.02~keV. Finally they find \ews\ between 17 and 189~eV. Indeed, a qualitative comparison of Fig.~4 of \citet{sim10mnras} with Fig.~3 of \citet{ng10aa} indicates that even the shape of the lines looks similar. In particular, the line of 4U~1705-44 shown in Fig.~3 (third row, third column) of \citet{ng10aa} is strikingly similar to the one shown in Fig.~4 (upper-left panel) of \citet{sim10mnras}. 

As the inclination increases to $\sim$60\deg, the hottest and most ionised layers of the outflow are first intercepted. At this inclination the material is almost fully ionised and the only important opacity source is Compton scattering. Due to the energy dependence of the relativistic Klein-Nishina formula for Compton scattering, the opacity will be lower and the spectrum will curve upwards at high energies. Compared to lower inclinations, the emission feature will be slightly broader since a larger column density and gradient in velocity fields will be traversed. \citet{sim10mnras2} report widths of around 0.35~keV and appearance of blueshifted \fetsix\ narrow absorption lines. 
At even higher inclinations, the part of the wind seen by the observer is less ionised and has a larger column density, and consequently the absorption features of less ionised ions, such as \fetfive, become deeper. The direct component of the spectrum becomes increasingly
dominated by the scattered/reprocessed component resulting in a prominent emission lines with \ews\ of up to 400~eV  \citep{sim10mnras2}. 
This is consistent with the results of this work and of previous works on classical dippers (i.e. high inclination sources), in which narrow absorption lines and broad iron lines are observed simultaneously \citep[e.g.][]{1658:sidoli01aa,1624:parmar02aa,1323:boirin05aa,1254:diaz09aa}. \citet[][]{1323:boirin05aa} and \citet{ionabs:diaz06aa} found that 
a highly ionised absorption plasma is ubiquitous for these sources and inferred that such an equatorial plasma must be present in all LMXBs. Since the plasma will develop as a wind or not depending on the size of the system and on its luminosity (see Sect.~\ref{subsec:discwind}), in cases where the thermal wind cannot develop due to the low luminosity of the central source the scattering will be produced in the hot, static, atmosphere that is revealed through absorption lines. 

A caveat of the scenario outlined above may be the relevance of radiation pressure in the disc wind. \citet{sim10mnras,sim10mnras2} considered parameters typical of AGN for the disc wind simulations, for which radiation pressure and line-driving is considered to be more important than for LMXBs \citep[see e.g.][]{proga02apj}. For example, the inner launching radius is expected to be smaller for AGN and the blueshifts of the absorption lines larger. However, the large column density of the absorber in obs~4 indicates that electron scattering and therefore radiation pressure is non-negligible and that shielding could be even relevant for some parts of the wind. An extrapolation of the region of emission line formation found by \citet{sim10mnras2} of 4.7\,$\times$\,10$^3$\,r$_g$ to \src\ yields a radius of $\sim$10$^9$\,cm, very close to the estimated launching radius of the wind of $\sim$10$^{10}$\,cm. Therefore, dedicated simulations for disc winds with parameters appropriate to LMXBs are now of utmost importance to allow spectral modeling and confirm this scenario. 

\subsection{Orbital period dependence of the wind}

We found significant variability both within each observation and among all the observations
analysed in this work. 
If the disc wind scenario outlined in the previous section is correct, we can naturally explain the spectral differences 
as different parts of the wind being probed at different phases, i.e. changes in the column density and ionisation of 
the warm absorber/wind in the line of sight and of the reprocessed component drive 
the spectral variability among observations.

\citet{gx13:corbet10apj} reported a modulation of $\sim$\,24~d based on RXTE/ASM data spanning 14.2~years. They also found that the modulation is not constant. For example, their Fig.~6 indicates that for some time intervals the dominant peak is $\sim$~24~d but for others the
peak lies below (third panel from the bottom) or above (second panel from the bottom). 
The fact that a modulation is observed in X-rays \citep{gx13:corbet10apj} indicates most likely a high inclination 
for the source. Although regular dipping has not been reported for \src, this could be due to the long
orbital period of $\sim$24~d and the lack of a precise ephemeris. While modulation in NIR can be 
explained e.g. by a modulation of the companion (but note that  \citet{Froning2008} attribute 3/4 of the NIR flux to the disc), it is difficult to explain a periodicity in X-rays 
without invoking structure in the accretion disc or its atmosphere. Further, the detection of drops of intensity of up to  80\%
in the light curves of \src\ which are energy-dependent strongly resembles the so-called dipping behaviour
of high-inclination sources \citep[see e.g. Fig.~1 of][]{ionabs:diaz06aa}. Therefore, interpretations
of the spectral/timing behaviour of the source in the context of Doppler
boosting of a jet pointing directly towards us \citep{gx13:schnerr03aa} are very unlikely
at the light of the 2008 \xmm\ observations. Instead we favour an inclination for the system of $\sim$60--80\deg.

\citet{gx13:corbet10apj} concluded that the X-ray modulation could be caused by dipping behaviour and that the structure
causing the modulation should not be phase-locked. In classical dippers, the dips recur at the
orbital period of the system and are believed to be caused by
periodic obscuration of a central X-ray source by structure
located in the outer regions of a disc resulting from the impact
of the accretion flow from the companion star into the disc \citep{1916:white82apjl}. However, the structure is locked at
phase~$\sim$\,0.8 (considering phase~0 as the phase at which the disc is eclipsed by the companion). 

We therefore propose a warped, tilted and precessing disc as the origin for the variability of the X-ray modulation.  
\citet{1254:diaz09aa} explained the appearance and disappearance of dips in \twelve\ as well as the changes
in the optical emission by invoking a precessing warped disc, similar to the one in Her~X-1 \citep{gerend76apj}. 
The existence of a warped disc in \src\ together with a high inclination for the system would explain both the 
detection of a modulation in X-rays and the fact that the structure causing it is not phase-locked. Further, the warp
would imply that the line of sight is slightly different at different phases, explaining the fact that we see different
parts of the wind along the orbital period. The warp could have been caused by the disc wind, as in the case of Her~X-1 \citep{herx1:schandl96aa}
or simply by a critical illumination of the disc \citep{foulkes10mnras}.

\section{Conclusions}

We find that:

- the disc wind in \src\ originates at large radii, $\sim$10$^{10}$~cm, and is consistent with being thermally driven, consistent with work by \citet{gx13:ueda04apj}

- the broad iron line can be explained by reprocessed emission in a hot atmosphere, and Compton scattering is the main source of its breadth

- contemporaneous changes of the broad iron line and the warm absorber point to an origin in the same
disc wind. In particular, the broad iron line is too strong to be reproduced by reflection at
the inner disc if we assume illumination from the boundary layer and reprocessing in a hot atmosphere, 
corona or wind illuminated by a harder, power-law-like, continuum is needed 

- the presence of strong energy-dependent obscuration, found for first time in \src, and the absence of 
eclipses combined with the spectral type of the companion, indicate an 
inclination of 60--80\deg. The emission and absorption features expected in the disc wind 
scenario for such an inclination are consistent with the observed ones. This gives additional support
to such scenario.


\begin{acknowledgements}
We thank an anonymous referee for useful comments.
We thank Robin Corbet for providing us with the NIR light curves
presented in \citet{gx13:corbet10apj}. M. D{\'i}az Trigo thanks Chris Done
for useful discussions regarding optically thick winds and reflection models.
Based on observations obtained with XMM-Newton, an ESA science
mission with instruments and contributions directly funded by ESA
member states and the USA (NASA).
This research has made use of 
data obtained from the High Energy Astrophysics Science Archive 
Research Center (HEASARC), provided by NASA's Goddard 
Space Flight Center. 
We acknowledge support from the Faculty of the 
European Space Astronomy Centre (ESAC).
\end{acknowledgements}


\bibliographystyle{aa}
\bibliography{19049}

\begin{thebibliography}{70}
\expandafter\ifx\csname natexlab\endcsname\relax\def\natexlab#1{#1}\fi

\bibitem[{{Anders} \& {Grevesse}(1989)}]{anders89}
{Anders}, E. \& {Grevesse}, N. 1989, \gca, 53, 197

\bibitem[{{Arnaud}(1996)}]{arnaud96conf}
{Arnaud}, K.~A. 1996, in ASP Conf. Ser. 101: Astronomical Data Analysis
  Software and Systems V, 17

\bibitem[{{Asai} {et~al.}(2000){Asai}, {Dotani}, {Nagase}, \&
  {Mitsuda}}]{asai00apjs}
{Asai}, K., {Dotani}, T., {Nagase}, F., \& {Mitsuda}, K. 2000, \apjs, 131, 571

\bibitem[{{Ballantyne}(2004)}]{ballantyne04mnras}
{Ballantyne}, D.~R. 2004, \mnras, 351, 57

\bibitem[{{Bandyopadhyay} {et~al.}(2002){Bandyopadhyay}, {Charles}, {Shahbaz},
  \& {Wagner}}]{gx13:bandyopadhyay02apj}
{Bandyopadhyay}, R.~M., {Charles}, P.~A., {Shahbaz}, T., \& {Wagner}, R.~M.
  2002, \apj, 570, 793

\bibitem[{{Bandyopadhyay} {et~al.}(1999){Bandyopadhyay}, {Shahbaz}, {Charles},
  \& {Naylor}}]{Bandyopadhyay1999}
{Bandyopadhyay}, R.~M., {Shahbaz}, T., {Charles}, P.~A., \& {Naylor}, T. 1999,
  \mnras, 306, 417

\bibitem[{{Begelman} {et~al.}(1983){Begelman}, {McKee}, \&
  {Shields}}]{begelman83apj}
{Begelman}, M.~C., {McKee}, C.~F., \& {Shields}, G.~A. 1983, \apj, 271, 70

\bibitem[{{Boirin} {et~al.}(2005){Boirin}, {M\'endez}, {Parmar}, \&
  {Kaastra}}]{1323:boirin05aa}
{Boirin}, L., {M\'endez}, M.~{D{\'i}az Trigo}, M., {Parmar}, A.~N., \&
  {Kaastra}, J. 2005, \aap, 436, 195

\bibitem[{{Boirin} {et~al.}(2004){Boirin}, {Parmar}, {Barret}, {Paltani}, \&
  {Grindlay}}]{1916:boirin04aa}
{Boirin}, L., {Parmar}, A.~N., {Barret}, D., {Paltani}, S., \& {Grindlay},
  J.~E. 2004, \aap, 418, 1061

\bibitem[{{Cash}(1979)}]{cash79apj}
{Cash}, W. 1979, \apj, 228, 939

\bibitem[{{Corbet}(2003)}]{gx13:corbet03apj}
{Corbet}, R.~H.~D. 2003, \apj, 595, 1086

\bibitem[{{Corbet} {et~al.}(2010){Corbet}, {Pearlman}, {Buxton}, \&
  {Levine}}]{gx13:corbet10apj}
{Corbet}, R.~H.~D., {Pearlman}, A.~B., {Buxton}, M., \& {Levine}, A.~M. 2010,
  \apj, 719, 979

\bibitem[{{Cottam} {et~al.}(2001){Cottam}, {Kahn}, {Brinkman}, {den Herder}, \&
  {Erd}}]{0748:cottam01aa}
{Cottam}, J., {Kahn}, S.~M., {Brinkman}, A.~C., {den Herder}, J.~W., \& {Erd},
  C. 2001, \aap, 365, L277

\bibitem[{{Den Herder} {et~al.}(2001){Den Herder}, {Brinkman}, {Kahn},
  {Branduardi-Raymont}, {Thomsen}, {Aarts}, {Audard}, {Bixler}, {den Boggende},
  {Cottam}, {Decker}, {Dubbeldam}, {Erd}, {Goulooze}, {G{\" u}del},
  {Guttridge}, {Hailey}, {Janabi}, {Kaastra}, {de Korte}, {van Leeuwen},
  {Mauche}, {McCalden}, {Mewe}, {Naber}, {Paerels}, {Peterson}, {Rasmussen},
  {Rees}, {Sakelliou}, {Sako}, {Spodek}, {Stern}, {Tamura}, {Tandy}, {de
  Vries}, {Welch}, \& {Zehnder}}]{xmm:denherder01aa}
{Den Herder}, J.~W., {Brinkman}, A.~C., {Kahn}, S.~M., {et~al.} 2001, \aap,
  365, L7

\bibitem[{{D{\'i}az Trigo} {et~al.}(2006){D{\'i}az Trigo}, {Parmar}, {Boirin},
  {M\'endez}, \& {Kaastra}}]{ionabs:diaz06aa}
{D{\'i}az Trigo}, M., {Parmar}, A.~N., {Boirin}, L., {M\'endez}, M., \&
  {Kaastra}, J. 2006, \aap, 445, 179

\bibitem[{{D{\'{\i}}az Trigo} {et~al.}(2009){D{\'{\i}}az Trigo}, {Parmar},
  {Boirin}, {Motch}, {Talavera}, \& {Balman}}]{1254:diaz09aa}
{D{\'{\i}}az Trigo}, M., {Parmar}, A.~N., {Boirin}, L., {et~al.} 2009, \aap,
  493, 145

\bibitem[{{Done} \& {D\'iaz Trigo}(2010)}]{gx339:done10mnras}
{Done}, C. \& {D\'iaz Trigo}, M. 2010, \mnras, 407, 2287

\bibitem[{{Fabian} \& {Miniutti}(2005)}]{fabian05}
{Fabian}, A.~C. \& {Miniutti}, G. 2005, ArXiv Astrophysics e-prints

\bibitem[{{Fleischman}(1985)}]{gx13:fleischman85aa}
{Fleischman}, J.~R. 1985, \aap, 153, 106

\bibitem[{{Foulkes} {et~al.}(2010){Foulkes}, {Haswell}, \&
  {Murray}}]{foulkes10mnras}
{Foulkes}, S.~B., {Haswell}, C.~A., \& {Murray}, J.~R. 2010, \mnras, 401, 1275

\bibitem[{{Froning} {et~al.}(2008){Froning}, {Robinson}, \&
  {Bitner}}]{Froning2008}
{Froning}, C.~S., {Robinson}, E.~L., \& {Bitner}, M.~A. 2008, in American
  Institute of Physics Conference Series, Vol. 1010, A Population Explosion:
  The Nature \& Evolution of X-ray Binaries in Diverse Environments, ed. R.~M.
  {Bandyopadhyay}, S.~{Wachter}, D.~{Gelino}, \& C.~R. {Gelino}, 192--194

\bibitem[{{Garc\'ia} \& {Kallman}(2010)}]{garcia10apj}
{Garc\'ia}, J. \& {Kallman}, T.~R. 2010, \apj, 718, 695

\bibitem[{{Garc{\'{\i}}a} {et~al.}(2011){Garc{\'{\i}}a}, {Kallman}, \&
  {Mushotzky}}]{garcia11apj}
{Garc{\'{\i}}a}, J., {Kallman}, T.~R., \& {Mushotzky}, R.~F. 2011, \apj, 731,
  131

\bibitem[{{Gerend} \& {Boynton}(1976)}]{gerend76apj}
{Gerend}, D. \& {Boynton}, P.~E. 1976, \apj, 209, 562

\bibitem[{{Grindlay} \& {Seaquist}(1986)}]{Grindlay1986}
{Grindlay}, J.~E. \& {Seaquist}, E.~R. 1986, \apj, 310, 172

\bibitem[{{Hirano} {et~al.}(1987){Hirano}, {Hayakawa}, {Nagase}, \&
  {Masai}}]{hirano87pasj}
{Hirano}, T., {Hayakawa}, S., {Nagase}, F., \& {Masai}, K. a nd~{Mitsuda}, K.
  1987, \pasj, 39, 619

\bibitem[{{Homan} {et~al.}(1998){Homan}, {van der Klis}, {Wijnands}, {Vaughan},
  \& {Kuulkers}}]{Homan1998}
{Homan}, J., {van der Klis}, M., {Wijnands}, R., {Vaughan}, B., \& {Kuulkers},
  E. 1998, \apjl, 499, L41+

\bibitem[{{Iaria} {et~al.}(2007){Iaria}, {di Salvo}, {Lavagetto},
  {D'A{\'{\i}}}, \& {Robba}}]{1254:iaria07aa}
{Iaria}, R., {di Salvo}, T., {Lavagetto}, G., {D'A{\'{\i}}}, A., \& {Robba},
  N.~R. 2007, \aap, 464, 291

\bibitem[{{Jansen} {et~al.}(2001){Jansen}, {Lumb}, {Altieri}, {Clavel}, {Ehle},
  {Erd}, {Gabriel}, {Guainazzi}, {Gondoin}, {Much}, {Munoz}, {Santos},
  {Schartel}, {Texier}, \& {Vacanti}}]{xmm:jansen01aa}
{Jansen}, F., {Lumb}, D., {Altieri}, B., {et~al.} 2001, \aap, 365, L1

\bibitem[{{Juett} \& {Chakrabarty}(2006)}]{1916:juett06apj}
{Juett}, A. \& {Chakrabarty}, D. 2006, \apj, 646, 493

\bibitem[{{Kaastra} {et~al.}(1996){Kaastra}, {Mewe}, \&
  {Nieuwenhuijzen}}]{kaastra96}
{Kaastra}, J.~S., {Mewe}, R., \& {Nieuwenhuijzen}, H. 1996, in UV and X-ray
  Spectroscopy of Astrophysical and Laboratory Plasmas, ed. {K.~Yamashita \&
  T.~Watanabe}, 411--414

\bibitem[{{Kallman} \& {White}(1989)}]{kallman89apj}
{Kallman}, T. \& {White}, N.~E. 1989, \apj, 341, 955

\bibitem[{{Kallman} {et~al.}(2009){Kallman}, {Bautista}, {Goriely}, {Mendoza},
  {Miller}, {Palmeri}, {Quinet}, \& {Raymond}}]{1655:kallman09apj}
{Kallman}, T.~R., {Bautista}, M.~A., {Goriely}, S., {et~al.} 2009, \apj, 701,
  865

\bibitem[{{Laming} \& {Titarchuk}(2004)}]{laming04apjl}
{Laming}, J.~M. \& {Titarchuk}, L. 2004, \apjl, 615, L121

\bibitem[{{Laurent} \& {Titarchuk}(2007)}]{laurent07apj}
{Laurent}, P. \& {Titarchuk}, L. 2007, \apj, 656, 1056

\bibitem[{{Lin} {et~al.}(2007){Lin}, {Remillard}, \& {Homan}}]{1701:lin07apj}
{Lin}, D., {Remillard}, R.~A., \& {Homan}, J. 2007, \apj, 667, 1073

\bibitem[{{Mason} {et~al.}(2001){Mason}, {Breeveld}, {Much}, {Carter},
  {Cordova}, {Cropper}, {Fordham}, {Huckle}, {Ho}, {Kawakami}, {Kennea},
  {Kennedy}, {Mittaz}, {Pandel}, {Priedhorsky}, {Sasseen}, {Shirey}, {Smith},
  \& {Vreux}}]{xmm:mason01aa}
{Mason}, K.~O., {Breeveld}, A., {Much}, R., {et~al.} 2001, \aap, 365, L36

\bibitem[{{Matsuba} {et~al.}(1995){Matsuba}, {Dotani}, {Mitsuda}, {Asai},
  {Lewin}, {van Paradijs}, \& {van der Klis}}]{gx13:matsuba95pasj}
{Matsuba}, E., {Dotani}, T., {Mitsuda}, K., {et~al.} 1995, \pasj, 47, 575

\bibitem[{{Matt}(2006)}]{matt06an}
{Matt}, G. 2006, Astronomische Nachrichten, 327, 949

\bibitem[{{Miller} {et~al.}(2011){Miller}, {Maitra}, {Cackett},
  {Bhattacharyya}, \& {Strohmayer}}]{igr:miller11apj}
{Miller}, J.~M., {Maitra}, D., {Cackett}, E.~M., {Bhattacharyya}, S., \&
  {Strohmayer}, T.~E. 2011, \apjl, 731, L7

\bibitem[{{Neilsen} \& {Lee}(2009)}]{1915:neilsen09nat}
{Neilsen}, J. \& {Lee}, J.~C. 2009, \nat, 458, 481

\bibitem[{{Ng} {et~al.}(2010){Ng}, {D\'iaz Trigo}, {Cadolle Bel}, \&
  {Migliari}}]{ng10aa}
{Ng}, C., {D\'iaz Trigo}, M., {Cadolle Bel}, M., \& {Migliari}, S. 2010, \aap,
  522, id.A96

\bibitem[{{Paizis} {et~al.}(2006){Paizis}, {Farinelli}, {Titarchuk},
  {Courvoisier}, {Bazzano}, {Beckmann}, {Frontera}, {Goldoni}, {Kuulkers},
  {Mereghetti}, {Rodriguez}, \& {Vilhu}}]{Paizis2006}
{Paizis}, A., {Farinelli}, R., {Titarchuk}, L., {et~al.} 2006, \aap, 459, 187

\bibitem[{{Parmar} {et~al.}(2002){Parmar}, {Oosterbroek}, {Boirin}, \&
  {Lumb}}]{1624:parmar02aa}
{Parmar}, A.~N., {Oosterbroek}, T., {Boirin}, L., \& {Lumb}, D. 2002, \aap,
  386, 910

\bibitem[{{Pozdnyakov} {et~al.}(1979){Pozdnyakov}, {Sobol}, \&
  {Sunyaev}}]{pozdnyakov79aa}
{Pozdnyakov}, L.~A., {Sobol}, I.~M., \& {Sunyaev}, R.~A. 1979, \aap, 75, 214

\bibitem[{{Proga} \& {Kallman}(2002)}]{proga02apj}
{Proga}, D. \& {Kallman}, T.~R. 2002, \apj, 565, 455

\bibitem[{{Reig} {et~al.}(2003){Reig}, {Papadakis}, \& {Kylafis}}]{Reig2003}
{Reig}, P., {Papadakis}, I., \& {Kylafis}, N.~D. 2003, \aap, 398, 1103

\bibitem[{{Reynolds} \& {Nowak}(2003)}]{reynolds03ps}
{Reynolds}, C.~S. \& {Nowak}, M.~A. 2003, Physics Reports, 377, 389

\bibitem[{{Ross} \& {Fabian}(2005)}]{ross05mnras}
{Ross}, R.~R. \& {Fabian}, A.~C. 2005, \mnras, 358, 211

\bibitem[{{Schandl}(1996)}]{herx1:schandl96aa}
{Schandl}, S. 1996, \aap, 307, 95

\bibitem[{{Schnerr} {et~al.}(2003){Schnerr}, {Reerink}, {van der Klis},
  {Homan}, {M{\' e}ndez}, {Fender}, \& {Kuulkers}}]{gx13:schnerr03aa}
{Schnerr}, R.~S., {Reerink}, T., {van der Klis}, M., {et~al.} 2003, \aap, 406,
  221

\bibitem[{{Sidoli} {et~al.}(2001){Sidoli}, {Oosterbroek}, {Parmar}, {Lumb}, \&
  {Erd}}]{1658:sidoli01aa}
{Sidoli}, L., {Oosterbroek}, T., {Parmar}, A.~N., {Lumb}, D., \& {Erd}, C.
  2001, \aap, 379, 540

\bibitem[{{Sidoli} {et~al.}(2002){Sidoli}, {Parmar}, {Oosterbroek}, \&
  {Lumb}}]{gx13:sidoli02aa}
{Sidoli}, L., {Parmar}, A.~N., {Oosterbroek}, T., \& {Lumb}, D. 2002, \aap,
  385, 940

\bibitem[{{Sim} {et~al.}(2010{\natexlab{a}}){Sim}, {Miller}, {Long}, {Turner},
  \& {Reeves}}]{sim10mnras}
{Sim}, S.~A., {Miller}, L., {Long}, K.~S., {Turner}, T.~J., \& {Reeves}, J.~N.
  2010{\natexlab{a}}, \mnras, 404, 1369

\bibitem[{{Sim} {et~al.}(2010{\natexlab{b}}){Sim}, {Proga}, {Miller}, {Long},
  \& {Turner}}]{sim10mnras2}
{Sim}, S.~A., {Proga}, D., {Miller}, L., {Long}, K.~S., \& {Turner}, T.~J.
  2010{\natexlab{b}}, \mnras, 408, 1396

\bibitem[{{Stella} {et~al.}(1985){Stella}, {White}, \&
  {Taylor}}]{gx13:stella85conf}
{Stella}, L., {White}, N.~E., \& {Taylor}, B.~G. 1985, in Proc. ESA Workshop:
  Recent Results on Cataclysmic Variables, Vol. ESA SP-236, 125

\bibitem[{{Stelzer} {et~al.}(1999){Stelzer}, {Wilms}, {Staubert}, {Gruber}, \&
  {Rothschild}}]{herx1:stelzer99}
{Stelzer}, B., {Wilms}, J., {Staubert}, R., {Gruber}, D., \& {Rothschild}, R.
  1999, \aap, 342, 736

\bibitem[{{Str{\" u}der} {et~al.}(2001){Str{\" u}der}, {Briel}, {Dennerl},
  {Hartmann}, {Kendziorra}, {Meidinger}, {Pfeffermann}, {Reppin}, {Aschenbach},
  {Bornemann}, {Br{\" a}uninger}, {Burkert}, {Elender}, {Freyberg}, {Haberl},
  {Hartner}, {Heuschmann}, {Hippmann}, {Kastelic}, {Kemmer}, {Kettenring},
  {Kink}, {Krause}, {M{\" u}ller}, {Oppitz}, {Pietsch}, {Popp}, {Predehl},
  {Read}, {Stephan}, {St{\" o}tter}, {Tr{\" u}mper}, {Holl}, {Kemmer},
  {Soltau}, {St{\" o}tter}, {Weber}, {Weichert}, {von Zanthier},
  {Carathanassis}, {Lutz}, {Richter}, {Solc}, {B{\" o}ttcher}, {Kuster},
  {Staubert}, {Abbey}, {Holland}, {Turner}, {Balasini}, {Bignami}, {La
  Palombara}, {Villa}, {Buttler}, {Gianini}, {Lain{\' e}}, {Lumb}, \&
  {Dhez}}]{xmm:struder01aa}
{Str{\" u}der}, L., {Briel}, U., {Dennerl}, K., {et~al.} 2001, \aap, 365, L18

\bibitem[{{Sunyaev} \& {Titarchuk}(1980)}]{sunyaev80aa}
{Sunyaev}, R.~A. \& {Titarchuk}, L.~G. 1980, \aap, 86, 121

\bibitem[{{Tombesi} {et~al.}(2011){Tombesi}, {Cappi}, {Reeves}, {Palumbo},
  {Braito}, \& {Dadina}}]{tombesi11apj}
{Tombesi}, F., {Cappi}, M., {Reeves}, J.~N., {et~al.} 2011, \apj, 742, 44

\bibitem[{{Turner} {et~al.}(2001){Turner}, {Abbey}, {Arnaud}, {Balasini},
  {Barbera}, {Belsole}, {Bennie}, {Bernard}, {Bignami}, {Boer}, {Briel},
  {Butler}, {Cara}, {Chabaud}, {Cole}, {Collura}, {Conte}, {Cros}, {Denby},
  {Dhez}, {Di Coco}, {Dowson}, {Ferrando}, {Ghizzardi}, {Gianotti}, {Goodall},
  {Gretton}, {Griffiths}, {Hainaut}, {Hochedez}, {Holland}, {Jourdain},
  {Kendziorra}, {Lagostina}, {Laine}, {La Palombara}, {Lortholary}, {Lumb},
  {Marty}, {Molendi}, {Pigot}, {Poindron}, {Pounds}, {Reeves}, {Reppin},
  {Rothenflug}, {Salvetat}, {Sauvageot}, {Schmitt}, {Sembay}, {Short},
  {Spragg}, {Stephen}, {Str{\" u}der}, {Tiengo}, {Trifoglio}, {Tr{\" u}mper},
  {Vercellone}, {Vigroux}, {Villa}, {Ward}, {Whitehead}, \&
  {Zonca}}]{xmm:turner01aa}
{Turner}, M.~J.~L., {Abbey}, A., {Arnaud}, M., {et~al.} 2001, \aap, 365, L27

\bibitem[{{Ueda} {et~al.}(2001){Ueda}, {Asai}, {Yamaoka}, {Dotani}, \&
  {Inoue}}]{gx13:ueda01apjl}
{Ueda}, Y., {Asai}, K., {Yamaoka}, K., {Dotani}, T., \& {Inoue}, H. 2001,
  \apjl, 556, L87

\bibitem[{{Ueda} {et~al.}(2004){Ueda}, {Murakami}, {Yamaoka}, {Dotani}, \&
  {Ebisawa}}]{gx13:ueda04apj}
{Ueda}, Y., {Murakami}, H., {Yamaoka}, K., {Dotani}, T., \& {Ebisawa}, K. 2004,
  \apj, 609, 325

\bibitem[{{Ueda} {et~al.}(2009){Ueda}, {Yamaoka}, \&
  {Remillard}}]{1915:ueda09apj}
{Ueda}, Y., {Yamaoka}, K., \& {Remillard}, R. 2009, \apj, 695, 888

\bibitem[{{White} \& {Holt}(1982)}]{white82apj}
{White}, N.~E. \& {Holt}, S.~S. 1982, \apj, 257, 318

\bibitem[{{White} {et~al.}(1986){White}, {Peacock}, {Hasinger}, {Mason},
  {Manzo}, {Taylor}, \& {Branduardi-Raymont}}]{white86mnras}
{White}, N.~E., {Peacock}, A., {Hasinger}, G., {et~al.} 1986, \mnras, 218, 129

\bibitem[{{White} \& {Swank}(1982)}]{1916:white82apjl}
{White}, N.~E. \& {Swank}, J.~H. 1982, \apjl, 253, L61

\bibitem[{{Wilms} {et~al.}(2000){Wilms}, {Allen}, \& {McCray}}]{wilms00apj}
{Wilms}, J., {Allen}, A., \& {McCray}, R. 2000, \apj, 542, 914

\bibitem[{{Woods} {et~al.}(1996){Woods}, {Klein}, {Castor}, {McKee}, \&
  {Bell}}]{woods96apj}
{Woods}, D.~T., {Klein}, R.~I., {Castor}, J.~I., {McKee}, C.~F., \& {Bell},
  J.~B. 1996, \apj, 461, 767

\bibitem[{{Xiang} {et~al.}(2009){Xiang}, {Lee}, {Nowak}, {Wilms}, \&
  {Schulz}}]{1624:xiang09apj}
{Xiang}, J., {Lee}, J.~C., {Nowak}, M.~A., {Wilms}, J., \& {Schulz}, N.~S.
  2009, \apj, 701, 984

\end{thebibliography}

\end{document}